\documentclass{statsoc}
\usepackage[a4paper]{geometry}
\usepackage{graphicx}
\usepackage[textwidth=8em,textsize=small]{todonotes}
\usepackage{amsmath}
\usepackage{amssymb}
\usepackage{natbib}
\usepackage{bibunits}
\usepackage{xcolor}
\usepackage{lineno}
\usepackage{slashbox}
\usepackage{graphicx}
\usepackage{multirow}
\usepackage{booktabs}
\usepackage{makecell}
\usepackage{enumitem}
\usepackage{float}
\usepackage{placeins}
\usepackage{url}
\usepackage{dsfont}
\usepackage{comment}
\usepackage{lscape}
\usepackage{longtable}
\usepackage{footnote}
\usepackage{threeparttable}
\usepackage{tablefootnote}
\usepackage{array,mathtools,tabulary}
\usepackage{colortbl}
\usepackage[T1]{fontenc}
\usepackage[utf8]{inputenc}
\usepackage{siunitx}
 
\usepackage{adjustbox}
\usepackage{xr}

\setlist[2]{noitemsep} 
\setenumerate{noitemsep}
\setitemize{noitemsep}

\usepackage{tikz}
\usetikzlibrary{arrows, decorations.markings, matrix}

\newcommand{\Indicator}[1]{\mathds{1}\left\{#1\right\}}

\makeatletter
\newcommand*{\addFileDependency}[1]{
  \typeout{(#1)}
  \@addtofilelist{#1}
  \IfFileExists{#1}{}{\typeout{No file #1.}}
}
\makeatother

\newcommand*{\myexternaldocument}[1]{%
    \externaldocument{#1}%
    \addFileDependency{#1.tex}%
    \addFileDependency{#1.aux}%
}

\myexternaldocument{supp}

\makeatother

\title[]{Inferring the sources of HIV infection in Africa from deep-sequence data with semi-parametric Bayesian Poisson flow models}
\author[Author 1 {\it et al.}]{Xiaoyue Xi}
\address{Department of Mathematics, Imperial College London, London SW72AZ, United Kingdom.}
\author{Simon EF Spencer}
\address{Department of Statistics, University of Warwick, Coventry CV47AL, United Kingdom.}
\author{Matthew Hall}
\address{Big Data Institute, University of Oxford, Oxford OX3 7LF, United Kingdom.}
\author{M Kate Grabowski}
\address{Department of Pathology, Johns Hopkins University, Baltimore, MD, USA; Rakai Health Sciences Program, Entebbe, Uganda.}
\author{Joseph Kagaayi}
\address{Rakai Health Sciences Program, Entebbe, Uganda.}
\author{Oliver Ratmann}
\address{Department of Mathematics, Imperial College London, London SW72AZ, United Kingdom.}
\author{}
\email{oliver.ratmann@imperial.ac.uk}
\author{on behalf of Rakai Health Sciences Program and PANGEA-HIV}

\begin{document}

\setlength{\parskip}{0pt}
\linespread{0}

\begin{abstract}

Pathogen deep-sequencing is an increasingly routinely used technology in infectious disease surveillance. We present a semi-parametric Bayesian Poisson model to exploit these emerging data for inferring infectious disease transmission flows and the sources of infection at the population level. The framework is computationally scalable in high-dimensional flow spaces thanks to Hilbert Space Gaussian process approximations, allows for sampling bias adjustments, and estimation of gender- and age-specific transmission flows at finer resolution than previously possible. We apply the approach to densely sampled, population-based HIV deep-sequence data from Rakai, Uganda, and find substantive evidence that adolescent and young women are predominantly infected through age-disparate relationships.
\end{abstract}

\begin{bibunit}[rss]
\section{Introduction}\label{sec:intro}

\subsection{Inferring the sources, sinks and hubs of transmission flows to aid the design of HIV prevention interventions}

HIV remains one of the largest public health threats, especially in sub-Saharan Africa where approximately 61\% of all new cases worldwide occur \citep{unaids2019}. 
In recent years, rates of incident cases have overall dropped considerably with the widespread adoption of prevention interventions such as voluntary medical male circumcision (VMMC) to reduce the risk of HIV acquisition in men, or immediate provision of antiretroviral therapy (ART) to suppress the virus in infected individuals and thereby stop onward transmission \citep{ cohen2011prevention, grabowski2017hiv, Hayes2019}, although they remain well above UNAIDS thresholds for elimination \citep{unaids2018}. 

\begin{figure}[!tb]
    \centering
    \includegraphics[width=0.8\textwidth]{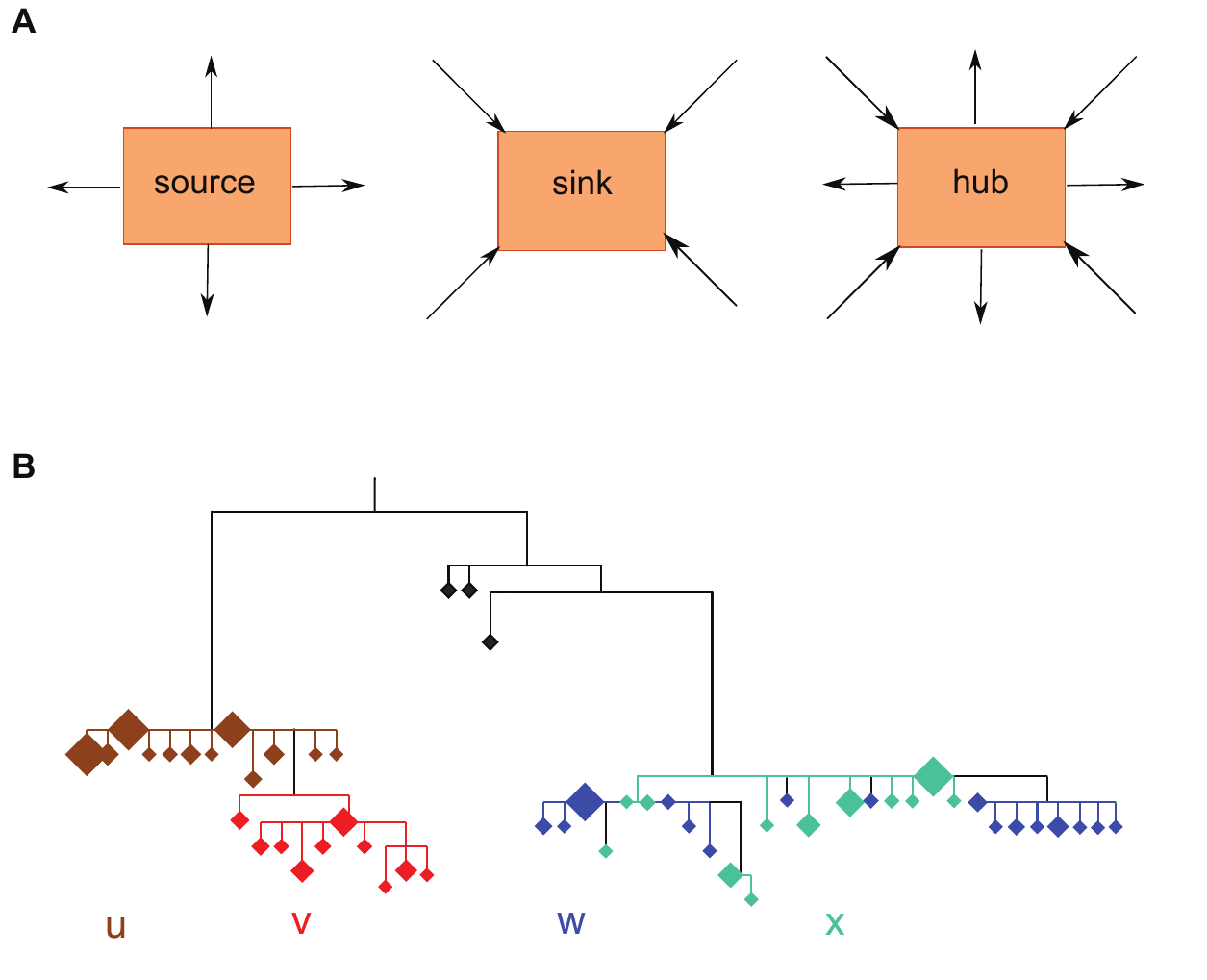}
    \caption{\textbf{Analysis aims and sketch of deep-sequence phylogenetic data to address these aims.} (\textbf{A}) The overall aim of phylogenetic source attribution analyses is to infer how pathogens are passed on between population groups. Conceptually populations can be divided into source populations that predominantly transmit disease, sink populations that predominantly receive infection, and hub populations that both disproportionally transmit and receive infections. \textbf{(B}) Viral deep-sequencing generates many sequence samples per host, which can be used to establish phylogenetic orderings between individuals, and thereby estimate the   direction of pathogen spread among sampled individuals. The figure sketches a deep-sequence phylogeny of pathogen sequences from individuals $u$, $v$, $w$, and $x$. Each tip (diamonds) represents a unique sequence, and the size of the tip copy number. Black tips correspond to out-of-sample reference sequences. Phylogenetic lineages are attributed to individuals (colours) using ancestral state reconstruction. Black lineages cannot be attributed to individuals. The subgraph of the tree associated with individual $u$ is ancestral to that of $v$, suggesting that infection spread from $u$ to $v$ potentially via unsampled intermediates. Individual $w$ has five subgraphs, some of which are ancestral to those of $x$ and some of which are descendent from those of $x$, indicating a complex ordering from which the direction of infection spread cannot be inferred. Subgraphs of $v$ and $w$ are not phylogenetically adjacent (disconnected), suggesting that one did not infect the other. With such information from a population-based sample of infected individuals, it is possible to quantify population-level transmission flows, sources, sinks, and hubs.
    }
    \label{fig:sourcesinkhub}
\end{figure}

Within Africa, there is increasing focus on identifying groups of individuals that are at high risk of acquiring HIV and at high risk of spreading the virus with the goal of targeted control interventions to these groups~\citep{abeler2019pangea}. Conceptually, the first step in this strategy is to break down the epidemic into source, sink and hub populations, according to the transmission flows that occur between them (Figure \ref{fig:sourcesinkhub}). Sources are population groups that disproportionately pass on infection, sinks are groups that disproportionately acquire infection, and hubs are both sources and sinks. The population groups can be defined in various ways. 

For example \cite{dwyer2019mapping} provided sub-national estimates of HIV prevalence across Africa, adding to data showing that the epidemic is highly heterogeneous across Africa, with small areas of very high prevalence (i.e., hotspots) that are surrounded by neighbouring areas with substantially lower prevalence. 
Although often assumed, it is unclear if hotspots are also sources of epidemic spread to neighbouring lower-prevalence communities \citep{ratmann2020quantifying}. 
Here, study populations are divided into individuals living in high-prevalence areas ($h$) and low-prevalence areas ($l$), and then transmission flows are estimated within and between them,
\begin{linenomath*}
\begin{equation}\label{eq:highlowflow}
\boldsymbol{\pi}=
\begin{pmatrix}
\pi_{hh} & \pi_{hl} \\
\pi_{lh} & \pi_{ll} \\
\end{pmatrix},
\end{equation}
\end{linenomath*}
where $\pi_{ab}$ is the proportion of transmission flows from group $a$ to group $b$ subject to $\sum_{ab}\pi_{ab}=1$.

Another prominent application concerns the interruption of infection cycles between men and women of different ages. \citet{de2017transmission}~proposed the scenario that young women aged $<$25 years are predominantly infected by older men aged 25-40 years, and later spread the virus to similarly aged men in their late twenties and early thirties. 
Here, study populations are divided into sex-specific age groups (we consider 1-year age groups between 15 to 49 years), and then transmission flows between and within age groups are estimated,
\begin{linenomath*}
\begin{equation}\label{eq:agesexflow}
\boldsymbol{\pi}=
\begin{pmatrix}
\boldsymbol{\pi}^{mf} & \boldsymbol{0} \\
\boldsymbol{0} & \boldsymbol{\pi}^{fm}
\end{pmatrix},
\quad
\boldsymbol{\pi}^{mf}=
\begin{pmatrix}
\pi^{mf}_{11} & \cdots & \pi^{mf}_{1K}\\
\vdots & \ddots & \vdots\\
\pi^{mf}_{K1} & \cdots & \pi^{mf}_{KK}
\end{pmatrix}
\text{ }
\boldsymbol{\pi}^{fm}=
\begin{pmatrix}
\pi^{fm}_{11} & \cdots & \pi^{fm}_{1K}\\
\vdots & \ddots & \vdots\\
\pi^{fm}_{K1} & \cdots & \pi^{fm}_{KK}
\end{pmatrix},
\end{equation}
\end{linenomath*}
where $\pi^{mf}_{ab}$ is the proportion of transmissions from men in age band $a$ to women in age band $b$, and similarly for $\pi^{fm}_{ab}$. We consider here only male-female transmission flows because we found no evidence of male-male transmission in previous analyses~\citep{ratmanninfer} and sexual transmission between women is extremely rare. The flow matrix \eqref{eq:agesexflow} has $2K^2$ non-zero entries to estimate, which for 1-year age bands amounts to $2450$ variables. Important summary statistics are the vector of sources of infection in group $b$ individuals ($\boldsymbol{\delta}^b$), for example in young women aged 20 years; the vector of recipients of infection from group $a$ individuals ($\boldsymbol{\omega}^a$), for example from men aged 25 years; and flow ratios from $a$ to $b$ ($\gamma_{ab}$), for example the ratio of transmissions from high-prevalence to low-prevalence areas compared to transmissions from low-prevalence to high-prevalence areas.  Respectively these quantities are defined by
\begin{linenomath*}
\begin{subequations}\label{eq:sourcesrecipientsratios}
\begin{align}
\boldsymbol{\delta}^b=(\delta^b_a)_{a\in\mathcal{A}} 
&,\quad \delta^b_a=\pi_{ab}/\sum_c \pi_{cb};
\label{eq:sources}
\\
\boldsymbol{\omega}^a=(\omega^a_b)_{b\in\mathcal{A}} 
&,\quad \omega^a_b=\pi_{ab}/\sum_c\pi_{ac};
\label{eq:recipients}\\
&\hphantom{,}\quad\gamma_{ab}=\pi_{ab}/\pi_{ba}.
\label{eq:ratios}
\end{align}
\end{subequations}
\end{linenomath*}
Flow matrices of the form (\ref{eq:highlowflow}-\ref{eq:agesexflow}) are also of central interest to characterise human migration flows between countries \citep{raymer2013integrated}, transport flows between locations \citep{tebaldi1998bayesian},  bacterial migration in humans \citep{ailloud2019within}, or human contact intensities~\citep{van2017efficient}, and are alternatively referred to as origin-destination matrices \citep{hazelton2001inference, miller2019towards}.

\subsection{Inference from pathogen sequence data}
Transmission flow matrices have been estimated from contact tracing or survey data on partner characteristics, though these data are often subject to reporting and/or social desirability biases, especially for sexual diseases that are associated with stigma or remain criminalised in many countries \citep{barre2018expert}.
Here, we are concerned in estimating the quantities (\ref{eq:highlowflow}-\ref{eq:sourcesrecipientsratios}) from pathogen sequences, which are considered an objective marker of disease flow. 
For fast-evolving pathogens like HIV, mutations accrue quickly and the phylogenetic relationship of pathogen sequences can be used to evaluate many aspects of transmission dynamics such as the origins of HIV \citep{faria2014early}, the contribution of different disease phases to onward spread \citep{volz2013hiv, ratmann2016sources}, or outbreak detection \citep{poon2016near}.

Traditionally, HIV sequences are obtained through Sanger sequencing, which returns for each sample one consensus nucleotide sequence that captures the entire viral diversity in the sample from one individual. The genetic distance between two consensus sequences can be used to estimate if the corresponding two individuals are epidemiologically closely related, however the data are insufficient to estimate the direction of transmission between any two sampled individuals \citep{leitner2018phylogenetic}. For this reason most methods infer transmission flows indirectly from statistics of the entire phylogeny, usually the coalescent times (i.e.~the times when two lineages coalesce into one, backwards in time) and the disease states of infected individuals at time of sampling (such as location or age in the two applications discussed above). In the mugration model \citep{lemey2009bayesian}, the states of viral lineages at any time are described with a continuous-time Markov chain (CTMC) that is independent of the evolutionary process. Flow estimates between groups can be obtained from the posterior distribution of the transition rates of the CTMC model via MCMC sampling, as well as posterior estimates of the phylogeny and the states of its lineages, which are latent variables in the model \citep{lemey2009bayesian}. The MultiTypeTree model \citep{vaughan2014efficient} removes the independence assumption that the evolutionary history of the genealogy is independent of population structure, however sampling correlated latent phylogenies and state histories is often computationally infeasible. This limitation is addressed with the structured coalescent of \cite{volz2009phylodynamics}, which integrates over the state histories of the phylogeny and describes the marginal probabilities of each viral lineage to be in a particular state at a particular time. The changes in the state probabilities along lineages and through coalescent events are derived under ordinary differential equations (ODE) models of disease spread. The flow parameters are obtained as by-products of the estimated latent states and parameters of the compartmental model, and in general vary in time over the phylogenetic history. 
Adopting the marginalisation approach, a greater range of flow models and data sets can be analysed, though computational run-times often remain on the order of several weeks for data from hundreds of individuals~\citep{vaughan2014efficient, volz2013hiv}.

An emerging strategy for estimating transmission flows involves phylogenetic analysis of multiple distinct pathogen sequences per infected host, because such data make possible to attribute sets of viral lineages to individuals and infer the ancestral relationships between them, which can provide direct evidence into the direction of transmission between two individuals \citep{ leitner2018phylogenetic}. Such analyses are becoming broadly applicable, because deep sequencing technology now allows generating thousands to millions of distinct pathogen sequence fragments per sample \citep{gall2012universal, zhang2020evaluation}. 
Prior work focused on software development \citep{wymant2017phyloscanner, skums2018quentin}, validation of the bioinformatics protocol for inferring the direction of transmission \citep{ratmanninfer, zhang2020evaluation}, and reconstruction of partially observed transmission networks at the population level \citep{ratmanninfer}. 

\subsection{Semi-parametric Poisson flow models}

The starting point of this paper is the output of a typical deep-sequence phylogenetic analysis \citep{wymant2017phyloscanner}, which includes, for each ordered pair of sampled individuals, a viral phylogenetic measure in $[0,1]$ giving a score that transmission occurred from the first to the second individual, possibly via unsampled intermediate individuals (phylogenetic direction scores). This is advantageous, because first, in this canonical form the data enable us to present the estimation problem in terms of a class of Bayesian Poisson models for non-Gaussian flow data that can flexibly describe a range of epidemiological questions including transmission in space \eqref{eq:highlowflow}, or by age and sex \eqref{eq:agesexflow}, similar to the Poisson models used for estimating transport or migration flows~\citep{tebaldi1998bayesian, raymer2013integrated}. 
Second, the models can account for multi-level sampling heterogeneity, which is typically present in population-based disease occurrence data but was not emphasised in phylogenetic analysis. We leverage Bayesian data augmentation to adjust for sampling heterogeneity \citep{givens1997publication}, and exploit the fact that the additional latent variables can be integrated out in our framework, so that computational inference remains inexpensive. Third, while typical phylodynamic approaches are limited to estimating transmission flows between coarse population strata, for example by age brackets $15-24$ years and $25-40$ years \citep{de2017transmission, le2019hiv}, we can employ Gaussian-process-based regularisation techniques to capture fine detail in transmission flows by annual age increments.  
Specifically, we propose using recently developed Hilbert Space Gaussion Process (HGSP) approximations to ensure the regularisation priors remain computationally tractable~\citep{solin2014hilbert}. 
This brings our approach into the form of semi-parametric Bayesian Poisson models, which enable inference of high-resolution transmission flows similar to the Integrated Nested Laplace Approximations used for inferring high-resolution human contact matrices~\citep{van2017efficient}.

In Section~\ref{sec:methods}, we introduce our notation and develop the semi-parametric Bayesian Poisson flow model. 
In Sections~\ref{sec:result:noadj}-\ref{sec:performance_hsgp}, we  assess the performance of the Poisson model in estimating transmission flows from sampling-biased data, and identify suitable HGSP approximations. Sections~\ref{sec:deepsamplinghet}-\ref{sec:youngwomen} illustrate our approach on HIV deep sequence data from the Rakai Community Cohort Study (RCCS) of the Rakai Health Sciences program, situated in south-eastern Uganda \citep{grabowski2017hiv}. 
Between August 10 2011 to January 30 2015, virus from 2652 HIV-infected individuals could be deep-sequenced, and 293 pairs of individuals with phylogenetically strong support for the direction of transmission were identified. 
We demonstrate that the new type of phylogenetic data and our statistical model enable estimation of age- and gender-specific transmission flows at finer detail than previously possible while remaining computationally scalable. Particular attention is given to potential sampling biases, and we propose a hierarchical model of the sequence sampling cascade for analysis of transmission flows. Section~\ref{sec:discussion} closes with a discussion.

\begin{figure}[hbt!]
    \centering
    \includegraphics[width=0.8\textwidth]{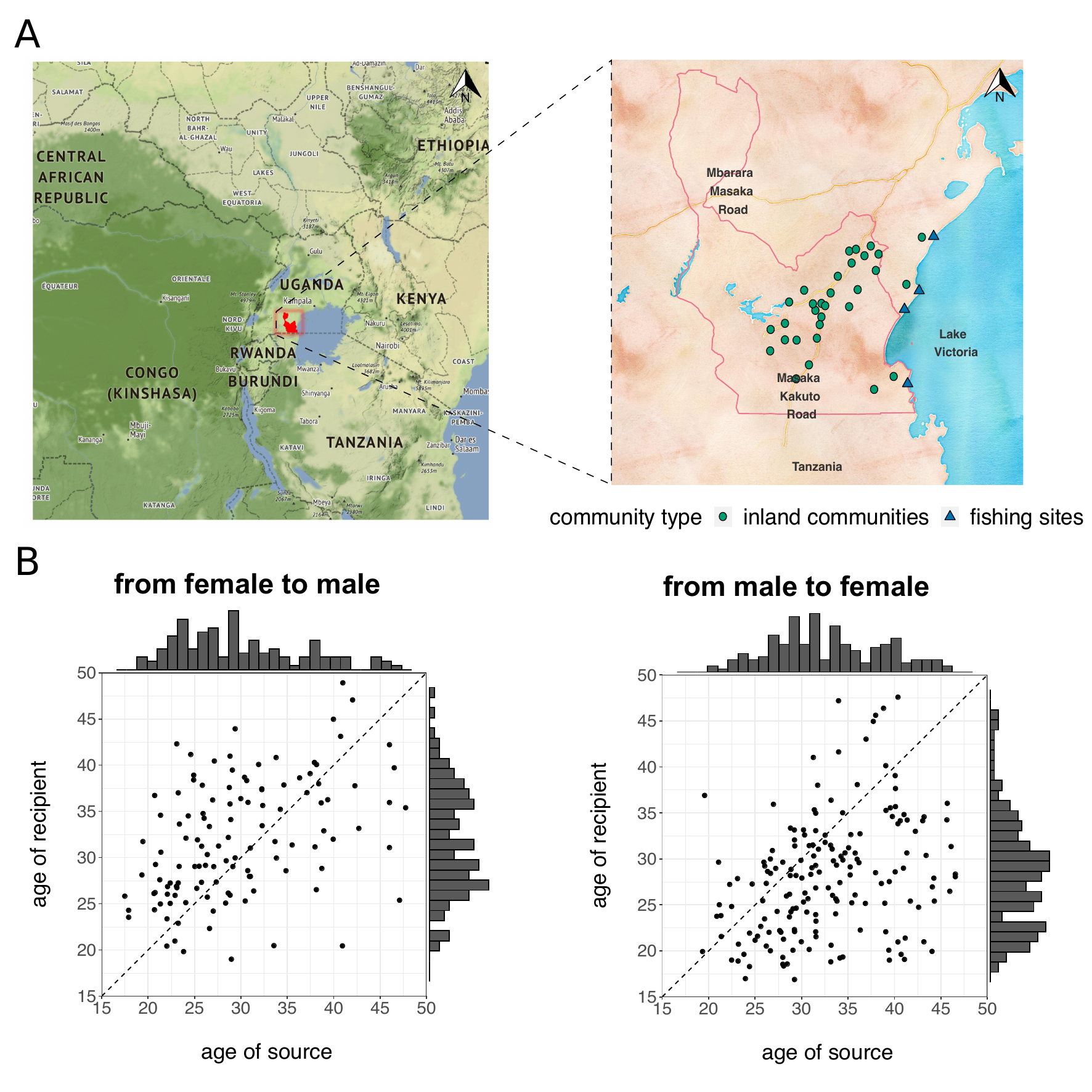}
    \caption{\textbf{Location of the Rakai Community Cohort Study, and data.} (\textbf{A}) Location of Rakai District (red) in south-eastern Uganda at the shores of Lake Victoria. HIV surveillance data were obtained from 2 survey rounds in 36 inland communities of the Rakai Community Cohort Study (green circles) and three survey rounds in the main 4 fishing communities within 3km of Lake Victoria (green triangles) between August 10, 2011 and January 30, 2015. (\textbf{B}) The study did a household census, and all individuals aged 15-49 years capable to provide informed consent and resident for at least 1 month with the intention to stay were invited to participate. Viral deep-sequencing was performed on plasma blood samples from HIV infected participants who reported no ART use. Deep-sequence phylogenetic analysis returned phylogenetic transmission scores between individuals, and 293 pairs had strong support of phylogenetic linkage and transmission direction. 173 pairs were male-to-female and 120 were female-to-male. The figures show the phylogenetically likely source-recipient pairs by age of the source and recipient at the midpoint of the study period.}
    \label{fig:data}
\end{figure}

\section{Methodology} \label{sec:methods}
%
%
%

\subsection{Notation and Definitions}
In this section we present the notation that is used to estimate transmission flows in a population $\mathcal{P}$ of size $N$, during a study period $\mathcal{T}=[t_1,t_2]$. We define by $i=1,\dotsc,N$ the identifier of infected individuals in $\mathcal{P}$ during $\mathcal{T}$. The transmission events during the study period can thus be described in a $N\times N$ binary matrix $\boldsymbol{Z}$, where $z_{ij}=1$ denotes transmission from person $i$ to person $j$, and $z_{ij}=0$ denotes no transmission. The transmission matrix is not symmetric, and diagonal entries are zero. 

We estimate transmission flows between population strata, and denote the strata by $a$ and the set of strata by $\mathcal{A}$, which is of dimension $A>0$. The number of transmission events in $\mathcal{T}$ from group $a$ to group $b$ are $z_{ab}= \sum_{i\in a, j\in b} z_{ij}$, and the primary object of interest is the $A\times A$ flow matrix $\boldsymbol{\pi}$ with entries $\pi_{ab}= z_{ab}/z^+$ where $z^+= \sum_{ab} z_{ab}$. The flow matrix
is in general not symmetric, and is subject to $\sum_{a,b \in \mathcal{A}} \pi_{ab} =1$. The matrix may contain structural zeros, for example in the case of HIV female-to-female transmission is extremely unlikely
. We denote the number of structurally non-zero entries by $L$, which satisfies $L\leq A^2$.

In general the flow matrix is time-dependent due to changes in population composition and varying transmission rates \citep{anderson1992infectious}. For instance in a compartment model of susceptible ($S$), infected ($I$) and treated ($T$) men and women of high ($h$) and low risk ($l$) of onward transmission, the ODE equations pertaining to the male ($m$) high risk population are
\begin{linenomath*}
\begin{equation}\label{eq:ode}
\begin{split}
\dot{S}_{mh} & = - \lambda(t) S_{mh}(t) + \mu - \mu S_{mh}(t) \\
\dot{I}_{mh} & = \lambda(t) I_{mh}(t) - \gamma(t) I_{mh}(t)  - \mu I_{mh}(t) \\
\dot{T}_{mh} & = \gamma(t) I_{mh}(t)  - \mu T_{mh}(t),
\end{split}
\end{equation}
\end{linenomath*}
where the force of infection is $\lambda(t)= \beta_{fh}(t)I_{fh}(t)/N_{fh}(t) + \beta_{fl}(t)I_{fl}(t)/N_{fl}(t)$, the birth/death rate $\mu$ is constant, and the viral suppression rate $\gamma$ and transmission rates $\beta_{fh}$, $\beta_{fl}$ are time-dependent. The actual, unobserved number of transmissions from high risk women to high risk men in $\mathcal{T}=[t_1,t_2]$ are
\begin{linenomath*}
\begin{equation}\label{eq:timedependenttransmissions}
z_{fh,mh}([t_1,t_2])= \int_{t_1}^{t_2} \beta_{fh}(t)I_{fh}(t)S_{mh}(t)/N_{fh}(t) dt,
\end{equation}
\end{linenomath*}
and the corresponding proportion of transmissions is
\begin{linenomath*}
\begin{equation}\label{eq:timedependentflows}
\pi_{fh,mh}([t_1,t_2]) = \frac{ z_{fh,mh}([t_1,t_2]) }{ Z([t_1,t_2]) },
\end{equation} 
\end{linenomath*}
where $Z$ is the sum of transmission events in $\mathcal{T}=[t_1,t_2]$. Here, we focus on estimating transmission flows in a given study period, and for ease of notation drop the dependence of our data and estimates on $\mathcal{T}=[t_1,t_2]$.

Pathogen deep sequence data are available from sampled, infected individuals. We denote the sampling status vector for all individuals in $\mathcal{P}$ by $\boldsymbol{s}=(s_i)_{i=1,\dotsc,N}$, where $s_i=1$ denotes that person $i$ is sampled, and $s_i=0$ that person $i$ is not sampled. The number of sampled individuals is $N^s$, which corresponds here to the $2652$ individuals for whom a viral deep sequence is available for analysis. We will characterise population sampling in terms of individual-level characteristics, such as age or location of residence, that are described with $p$ covariates, which we denote with the $N\times p$ matrix $\mathbf{X}$. The output of the phylogenetic deep sequence analysis can be summarised in an $N\times N$ direction score matrix $\mathbf{W}$ that describes the evidence for transmission from $i$ to $j$ with the weight $w_{ij}\in [0,1]$. The direction score matrix is not symmetric, diagonal entries are zero, and entries involving unsampled individuals are missing. To estimate the flow matrix $\boldsymbol{\pi}$, consider the observed flow counts   
\begin{linenomath*}
\begin{equation}\label{eq:flowcounts}
    n_{ab}= \sum_{i\in a, j\in b} \Indicator{s_i=1}\Indicator{s_j=1} \hat{z}_{ij} = \sum_{i\in a, j\in b} \Indicator{s_i=1}\Indicator{s_j=1} \Indicator{w_{ij}>\zeta},
\end{equation}
\end{linenomath*}
where $\zeta\in (0,1)$ is a threshold that can be used to select phylogenetically highly supported source-recipient pairs, and $\hat{z}_{ij}$ is taken as a perfect predictor of $z_{ij}$ among sampled individuals. The counts can be arranged into the $A\times A$ count matrix $\mathbf{n}$, and sum to $n^+=\sum_{a,b} n_{ab}$. In previous studies, $\zeta$ was set to $0.5$ or $0.6$, and $n^+$ was between $100$ to $500$ \citep{hall2019improved, ratmann2020quantifying}. 

The na\"{\i}ve flow estimator is defined by 
$
\hat{\pi}_{ab}=\frac{n_{ab}}{\sum_{c,d} n_{cd}}.
$
If we suppose that each population group $a$ is independently sampled at random with probability $\xi_a$, and the actual flows from group $a$ to group $b$ are $z_{ab}$, then 
$
\mathbb{E}(\hat{\pi}_{ab})= (z_{ab}\xi_a\xi_b)/
(\sum_{c,d} z_{cd}\xi_c\xi_d).
$
Consequently, the na\"{\i}ve flow estimator is only unbiased when the population groups were homogeneously sampled, i.e.~$\xi_a$ is the same for all $a$, which is rarely the case \citep{ratmann2020quantifying}. 

\subsection{Inferring flows from heterogeneously sampled data} \label{sec:model}
The data inputs for estimating transmission flows from pathogen deep-sequence data~\eqref{eq:flowcounts} are of the same form as for estimating origin-destination matrices with unobserved transport routes~\citep{hazelton2001inference}, migration flows~\citep{raymer2013integrated}, or contact intensities~\citep{van2017efficient}, prompting us to formulate the statistical model in general terms.  Considering the actual, unobserved number of flow events (i.e.,~transmissions) between all population groups, $\mathbf{z}=(z_{ab})_{a,b\in\mathcal{A}}$, the complete data likelihood that arises under mathematical models of the form \eqref{eq:ode} in a fixed study period is the multinomial
\begin{linenomath*}
\begin{equation}\label{eq:comp_likelihood}
 p(\mathbf{z}|z^+,\boldsymbol{\pi})\propto\prod_{a,b} (\pi_{ab})^{z_{ab}},
\end{equation}
\end{linenomath*}
where for ease of notation we denote by $\boldsymbol{\pi}$ the vector of non-zero elements of the flow matrices (\ref{eq:highlowflow}-\ref{eq:agesexflow}), and similarly for $\mathbf{n}$ and $\mathbf{z}$. Model \eqref{eq:comp_likelihood} ignores potential second-order correlations between flows, for example that a female infected by an older male may be more likely to transmit to men of older age. Since the total number of transmissions is in itself a random variable, we consider the related Poisson model
\begin{linenomath*}
\begin{equation}\label{eq:comp_likelihood2}
p(\mathbf{z}|\boldsymbol{\lambda})\propto\prod_{a,b} (\lambda_{ab})^{z_{ab}} \exp(-\lambda_{ab})= \bigg[ \prod_{a,b} (\pi_{ab})^{z_{ab}} \bigg]\bigg[ \eta^{z^+} \exp(-\eta)\bigg],
\end{equation}
\end{linenomath*}
where $\lambda_{ab}$ can be interpreted as the flow intensities from group $a$ to group $b$, $\eta=\sum_{c,d} \lambda_{cd}$, and $\pi_{ab}$ are recovered via $\pi_{ab}=\lambda_{ab}/\eta$.

The actual flows $\mathbf{z}$ are not observed. We assume that individuals are sampled at random within strata (SARWS). SAWRS implies that sampling is independent of being a source or not, and the likelihood of the observed counts conditional on the complete data is
$p(\mathbf{n}|\mathbf{z},\boldsymbol{\xi})=\prod_{a,b} \mbox{Binomial}(n_{ab} ; z_{ab},\xi_a \xi_b)$,
where $\xi_a$ is the sampling probability in group $a$. In this class of models the latent flow counts $z_{ab}$ can be conveniently integrated out, yielding for the observed flow counts the Poisson model
\begin{linenomath*}
\begin{equation}\label{eq:data_likelihood}
p(\mathbf{n}|\boldsymbol{\lambda},\boldsymbol{\xi})=
\prod_{a,b} (\lambda_{ab}\xi_a\xi_b)^{n_{ab}} \exp(-\lambda_{ab}\xi_a\xi_b).
\end{equation}
\end{linenomath*}
An important point in this construction is that we are free to choose the stratification $\mathcal{A}$ in order to accommodate the SARWS assumption. We will show that flow estimates on different, coarser population stratifications that are of primary interest are easily obtained through the aggregation property of the Poisson system \eqref{eq:comp_likelihood2}.

The sampling-adjusted maximum-likelihood estimates of $\lambda_{ab}$ and $\pi_{ab}$ under \eqref{eq:data_likelihood} can be derived under the SARWS assumption. The number of sampled individuals in $a$, $N^s_a= \sum_{i \in a} \Indicator{s_i=1}$, is a Binomial sample of the number of all individuals in $a$, $N_a$, which leads to
\begin{linenomath*}
\begin{equation}\label{eq:MLE}
\hat{\pi}_{ab}= \frac{n_{ab}}{\hat{\xi}_a \hat{\xi}_b} \Big/ \Big[ \sum_{c,d} \frac{n_{cd}}{\hat{\xi}_c \hat{\xi}_d} \Big],
\end{equation}
\end{linenomath*}
where $\hat{\xi}_a= N^s_a/N_a$ (Supplementary Material, section~\ref{supp:sec:mle}). 

\subsection{Inferring high-resolution flows with  regularising priors}\label{sec:priorConstruction}
Bayesian regularisation techniques play a central role in obtaining robust and suitably smoothed flow estimates. Considering population sampling, we exploit additional information on the sampling vector $\boldsymbol{s}$. We assume that flows are independent of sampling, allowing us to decompose the joint posterior distribution into
\begin{linenomath*}
\begin{equation}\label{eq:postdist1}
\begin{split}
& 
p(\boldsymbol{\lambda}, \boldsymbol{\xi}|\mathbf{n},\boldsymbol{s}, \mathbf{X}) 
\propto  
p(\mathbf{n}|\boldsymbol{\lambda},\boldsymbol{\xi},\boldsymbol{s},\mathbf{X}) p(\boldsymbol{\lambda}|\boldsymbol{\xi},\boldsymbol{s},\mathbf{X}) p(\boldsymbol{\xi}|\boldsymbol{s},\mathbf{X})
\\
&\quad\quad
= 
p(\mathbf{n}|\boldsymbol{\lambda},\boldsymbol{\xi}) p(\boldsymbol{\lambda}|\boldsymbol{\xi}) p(\boldsymbol{\xi}|\boldsymbol{s},\mathbf{X}). 
\end{split}
\end{equation}
\end{linenomath*}
A possible limitation of \eqref{eq:MLE} is that the counts $N_a$, $N^s_a$ can be small when the population is finely stratified. It is thus often advantageous to model individual-level sampling probabilities in terms of a linear combination of predictors. Using, for example, a logistic regression approach, we obtain 
\begin{linenomath*}
\begin{equation}\label{eq:logisticprior}
    p(\xi_a | \boldsymbol{s}, \mathbf{X})= \int \mbox{logit}^{-1}(\boldsymbol{x_a} \boldsymbol{\beta}) p(\boldsymbol{\beta}|\boldsymbol{s}, \mathbf{X})
    d \boldsymbol{\beta},
\end{equation}
\end{linenomath*}
where $\boldsymbol{x_a}$ is the row vector of population characteristics that specify group $a$, $\boldsymbol\beta$ are the regression coefficients, and $p(\boldsymbol{\beta}|\boldsymbol{s}, \mathbf{X})$ is the posterior density of the regression coefficients, estimated from the sampling status vector $\boldsymbol{s}$ of all individuals in the study population.

Considering the prior density on the transmission intensities $\boldsymbol{\lambda}$, in some applications the population strata are unordered such as in application~\eqref{eq:highlowflow}. In this case we propose using
\begin{linenomath*}
\begin{equation}\label{eq:lambdaGamma}
    \lambda_{ab}|\xi_a,\xi_b \sim \mbox{Gamma}(\alpha_{ab}, \beta), \quad \alpha_{ab}=0.8/L, \quad \beta=0.8/Z^p(\xi_a,\xi_b),
\end{equation}
\end{linenomath*}
where $L$ is the length of the flow vector $\boldsymbol{\pi}$, which is equivalent to the number of structurally non-zero entries in the flow matrices (\ref{eq:highlowflow}-\ref{eq:agesexflow}), and $Z^p$ is the number of expected transmission events, $Z^p= \sum_{a,b : n_{ab}\neq 0} \frac{n_{ab}}{\xi_a\xi_b} + \sum_{a,b : n_{ab}= 0} \frac{1- \xi_a\xi_b}{\xi_a\xi_b}$. This choice is motivated by the fact that \eqref{eq:lambdaGamma} induces on $\boldsymbol{\pi}$ an objective Dirichlet prior density with parameters $\alpha_{ab}=0.8/L$ \citep{berger2015overall}, such that the likelihood \eqref{eq:data_likelihood} dominates the prior \eqref{eq:lambdaGamma} regardless of the number of flows to estimate. However when the population groups can be ordered, such as the $K$ 1-year age bands in \eqref{eq:agesexflow}, the structure of the flow model \eqref{eq:postdist1} enables using regularising prior densities that penalise against large deviations in flow intensities between similar source and recipient populations. For \eqref{eq:agesexflow}, we opted for (stacked) two-dimensional Gaussian-process priors on the entries of $\boldsymbol{\lambda}$,
\begin{linenomath*}
\begin{equation}\label{eq:gp}
\begin{split}
    & \log \boldsymbol{\lambda} = \mu \mathbf{1}+ \nu\mathds{1}_{mf} + \boldsymbol{f},\\
    & \boldsymbol{f}=(\boldsymbol{f}^T_{mf}, \boldsymbol{f}^T_{fm})^T,\\  
    & \boldsymbol{f}_{mf}\sim \mathcal{GP}(0, k_{mf}),\quad \boldsymbol{f}_{fm}\sim \mathcal{GP}(0, k_{fm}),
    \\
    & k_{mf}\big( (a_1,b_1), (a_2,b_2) \big)= \sigma_{mf}^2 \exp\Big( - \Big[ \frac{(a_2-a_1)^2}{2\ell_{mf,a}^2} + \frac{(b_2-b_1)^2}{2\ell_{mf,b}^2} \Big] \Big)\\
    & k_{fm}\big( (a_1,b_1), (a_2,b_2) \big)= \sigma_{fm}^2 \exp\Big( - \Big[ \frac{(a_2-a_1)^2}{2\ell_{fm,a}^2} + \frac{(b_2-b_1)^2}{2\ell_{fm,b}^2} \Big] \Big)\\
    & \sigma_{mf}^2, \sigma_{fm}^2 \sim \text{Half-Normal}(0,10)
    \\
    & \ell_{d,i} \sim \text{Inv-Gamma}(\alpha_{d,i}, \beta_{d,i}),
\end{split}    
\end{equation}
\end{linenomath*}
where the first $K^2$ entries of $\boldsymbol{\lambda}$ correspond to flows in the male-female direction and the remaining $K^2$ entries correspond to flows in the female-male direction, $\mu$ is the baseline log transmission intensity, $\nu$ is a scalar on the elements of $\boldsymbol{\lambda}$ in the male-female direction, and $k_{mf}$, $k_{fm}$ are gender-specific squared exponential kernels with variance parameters $\sigma_{mf}^2, \sigma_{fm}^2$ and length scales $\ell_{mf,a}$, $\ell_{mf,b}$, $\ell_{fm,a}$, $\ell_{fm,b}$ \citep{rasmussen2003gaussian}. 

\subsection{Scalable numerical inference}\label{sec:numerics}
The semi-parametric Poisson model can be efficiently fitted with the dynamic Hamiltonian Monte Carlo sampler of the Stan probabilistic programming framework \citep{carpenter2017stan}. 
The implementation 
uses the Hilbert Space Gaussian process approximation (HSGP) to the GP prior \eqref{eq:gp} developed by \cite{solin2014hilbert}, in which the squared exponential covariance kernel $k$ in \eqref{eq:gp} is approximated through a series expansion of eigenvalues and eigenfunctions of the Laplacian differential operator on a compact domain $\Omega$ of the input space. In our two-dimensional case, we consider $\Omega= [-B_1,B_1] \times [-B_2,B_2]$, and the boundary points play an important role in the approximation, and need to be specified appropriately. 

Briefly, the HSGP approximation involves the spectral density associated with the stationary kernel of the GP prior. For the squared exponential kernel with two-dimensional inputs, it is given by
\begin{linenomath*}
\begin{equation}\label{eq:spectraldensitySE}
S_{\theta}(\boldsymbol{\omega)}= 2 \pi \sigma^2   \prod_{d=1}^{2} \ell_d  \exp \left(-\frac{1}{2} \sum_{i=1}^{2} \ell_d^2 \omega_d^2 \right),  
\end{equation}
\end{linenomath*}
where $\boldsymbol{\omega}=(\omega_1, \omega_2)$ denote the frequencies, and $\theta=(\sigma,\ell_1,\ell_2)$ are the kernel parameters \citep{rasmussen2003gaussian}. The approximation further involves the eigenvalues and eigenfunctions of the Laplacian differential operator. On the compact domain $\Omega$, the $j$th univariate eigenvalues and eigenfunctions in dimension $d=1,2$ can be computed \citep{solin2014hilbert}, and are
$\lambda_{dj}= \big( \frac{j\pi}{2B_d} \big)^2$, 
$\phi_{dj}(x_d)= \sqrt{1/B_d} \sin\big( \sqrt{\lambda_{dj}} (x_d+B_d)\big)$.
The HSGP approximation involves the first $m_1$ and $m_2$ such terms of both dimensions. There are $m=m_1\times m_2$ possible combinations of such terms, which we index through $\boldsymbol{K}\in \mathbb{N}^{2\times m}$. For example, if $m_1=3$ and $m_2=4$, then
\begin{linenomath*}
\begin{equation*}
\boldsymbol{K}=
\left(
\begin{array}{rrrrrrrrrrrr} 
1 & 1 & 1 & 1 & 2 & 2 & 2 & 2 & 3 & 3 & 3 & 3 \\
1 & 2 & 3 & 4 & 1 & 2 & 3 & 4 & 1 & 2 & 3 & 4
\end{array}\right).
\end{equation*}
\end{linenomath*}
For 2D inputs, the  $j= 1,\dotsc,m$ eigenvalues and eigenfunctions are the combinations of the univariate eigenvalues and eigenfunctions, 
$\boldsymbol{\lambda}_j= (\lambda_{\boldsymbol{K}_{1j}},\lambda_{\boldsymbol{K}_{2j}})$, and 
$\phi_j(a,b)= \phi_{1\boldsymbol{K}_{1j}}(a)\phi_{2\boldsymbol{K}_{2j}}(b)$.
This fully specifies the HSGP approximation to Gaussian processes with 2D inputs,
\begin{linenomath*}
\begin{equation}\label{seq:hgsp_model}
\begin{split}
& \boldsymbol{f}\sim\mathcal{HSGP}(0,k^{\text{HSGP}}) \\
& 
k^{\text{HSGP}}\big( (a,b), (a^\prime,b^\prime) \big)
=
\sum_{j=1}^m 
S_{\theta}\left(\sqrt{\boldsymbol{\lambda}_j}\right)
\phi_j(a,b) \phi_j(a^\prime,b^\prime),
\end{split}
\end{equation}
\end{linenomath*}
which depends on the choice of $m_1$, $m_2$ and $B_1$, $B_2$. 
In our work, we applied the approximation~\eqref{seq:hgsp_model} to each of the stacked Gaussian process prior components in~\eqref{eq:gp}. 
To our knowledge, this is one of the first applications of the HSGP approximation. Note that the structure of the data inputs and the form of the Poisson model is the same in many flow applications~\citep{raymer2013integrated,van2017efficient}, and so the HSGP approximation could make estimation of high-resolution flows also numerically scalable in these settings. 

Full details on the algorithm are reported in Supplementary Text S\ref{sec:algorithms} and a tutorial is provided in \url{https://github.com/BDI-pathogens/phyloscanner/blob/master/phyloflows/vignettes/08_practical_example.md}.

\begin{figure}[!t]
    \centering
    \includegraphics[width=0.95\textwidth]{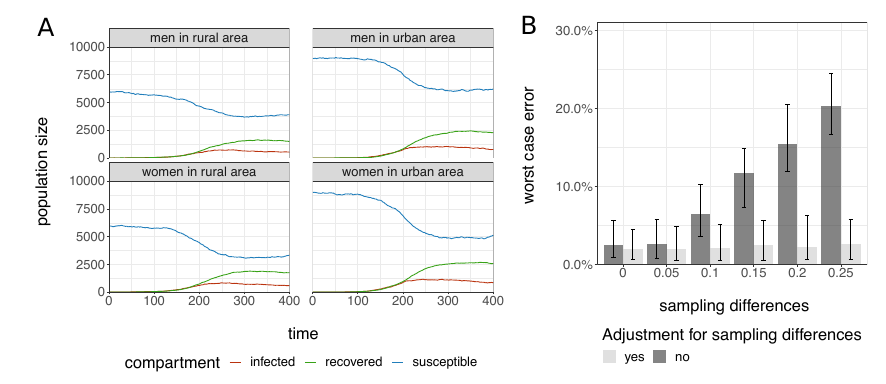}
    \caption{{\bf Simulation experiments to assess the accuracy of flow estimates under sampling bias.} ({\bf A}) ODE-based models were used to simulate epidemic trajectories of susceptible, infected and treated men and women across two population strata, and transmission flows between them. Panel A visualises one of the simulate trajectories. ({\bf B}) Transmission flows were estimated under the Poisson likelihood model~\eqref{eq:data_likelihood}, without adjustments for sampling differences (dark grey), and with adjustments for sampling differences (light grey). Accuracy was measured with the worst case error between posterior median estimates and the simulated true values, and shown are the average error and 95\% range in $100$ replicate simulations.}
    \label{fig:odesim}
\end{figure}

\section{Applications} \label{sec:application}

\subsection{Accuracy with and without adjustments for sampling heterogeneity}\label{sec:result:noadj}
Standard phylodynamic methods ignore sampling differences between population strata~\citep{volz2009phylodynamics, le2019hiv, scire2020improved}. We first assessed the impact of sampling heterogeneity on estimating transmission flows in simulation experiments, that are fully reported in Supplementary Material, section~\ref{sec:supp:simulation}. The first experiment is a minimal example involving flows between two population groups, which for simplicity we refer to as individuals in rural areas (group $a$) and individuals in large communities (group $b$). Transmission chains were simulated under the ODE model \eqref{eq:ode}, i.e. not from our simpler likelihood model~\eqref{eq:data_likelihood}, and the simulated flow matrix $\boldsymbol{\pi}^{0}_r$ was recorded in $r=1,\dotsc,100$ replicate simulations (Figure \ref{fig:odesim}A). Observations were drawn under heterogeneous sampling, with fixed sampling probabilities $\xi_a=0.6$ in group $a$, and decreasing sampling probabilities in groub $b$, $\xi_b=0.6, 0.55, \dotsc, 0.35$. First, we estimated flows from \eqref{eq:postdist1} assuming no sampling differences between population strata, and which we implemented by setting $p(\xi_a |\boldsymbol{s}, \mathbf{X})$ and $p(\xi_b |\boldsymbol{s}, \mathbf{X})$ to the Beta density with parameters 25.5, 25.5. Second, we estimated flows with information on sampling differences included, by setting  $p(\xi_a|\boldsymbol{s},\mathbf{X})$ to the Beta density with shape parameters $N^s_a+\alpha_\xi$, $N^a-N^s_a+\beta_\xi$, where $N^s_a$ are the number of sampled inviduals in group $a$, $N_a$ is the population denominator, and the hyperparameters $\alpha_\xi$, $\beta_\xi$ were both set to $0.5$ under Jeffrey's prior on the Binomial sampling probabilities. The density $p(\xi_b|\boldsymbol{s},\mathbf{X})$ was specified analogously. Figure \ref{fig:odesim}B compares the accuracy in the posterior median flow estimates $\hat{\pi}_{r,ab}$ in terms of the worst case error (WCE) $\varepsilon_r=\max_{a,b}  |\widehat{\pi}_{r,ab}-\pi^T_{r,ab}|$, and illustrates that the Poisson model~\eqref{eq:postdist1} can minimise the impact of sampling heterogeneity on flow estimates. More complex simulation experiments yielded similar results (Supplementary Material, section~\ref{sec:supp:simulation}).

\begin{figure}[!t]
    \centering
    \includegraphics[width=0.95\textwidth]{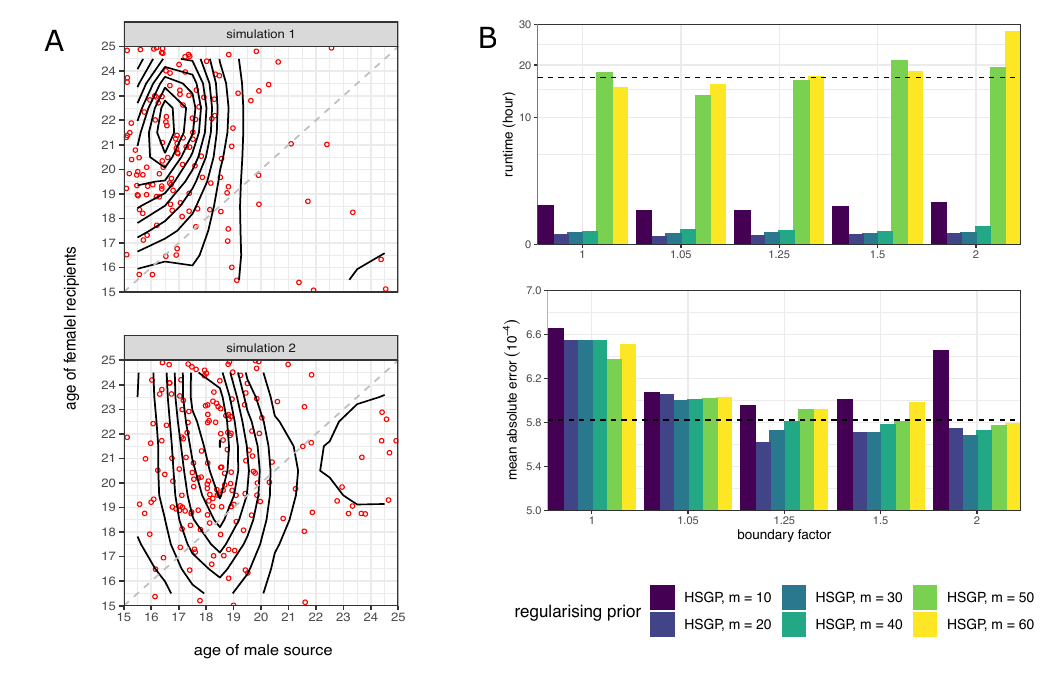}
    \caption{{\bf Simulation experiments to assess the accuracy of flow estimates under different smoothing priors.} ({\bf A}) Age- and gender-specific transmission pairs were simulated from the GP model \eqref{eq:gp}, and shown are the simulated male to female transmission pairs for two simulations (red), along with contours of the simulated ground-truth transmission flow density.
    ({\bf B}) Average runtimes of transmission flow inferences across $20$ simulated data sets using different HSGP and GP priors, and average mean absolute errors in inferred posterior median flow estimates. The dashed line corresponds to average runtimes and average mean absolute errors when using the GP prior, under which the simulations were generated.}
    \label{fig:gpsim}
\end{figure}

\subsection{Accuracy with different smoothing priors}\label{sec:performance_hsgp}
Next, we assessed on simulations the impact of different regularising prior densities on estimating flows across orderable population strata, which is typically not considered in standard phylodynamic inference approaches~\citep{volz2009phylodynamics, de2017transmission, le2019hiv, scire2020improved, bbosa2020phylogenetic}. We focused on age- and gender-specific transmission flows by $1$-year age inputs between $15$-$24$ years, resulting in $8 \times 10^2=800$ flow combinations, and simulated $300$ transmission pairs from the GP model~\eqref{eq:gp} using ground-truth parameters that were motivated by analyses of the Rakai data set (see Supplementary Material, section \ref{sec:supp:simulation}). Figure~\ref{fig:gpsim}A illustrates the simulated transmission pairs from men to women, and the corresponding underlying flows in $2$ of $20$ simulations generated. Transmission flows were then inferred using the HSGP approximation~\eqref{seq:hgsp_model} to~\eqref{eq:gp} in our semi-parametric Poisson flow model. We varied the number $m$ of basis functions from $100$ to $3600$ by setting $m_1=m_2=10,20,\dotsc,60$, and chose as HSGP domain $\Omega$ an expanded version of the input domain, $[15/B, 50B]\times [15/B, 50B]$, with boundary factor $B=1,1.05,1.25,1.5,2$. To have a benchmark for the performance of the HSGP approximations, inferences were also performed using the GP prior~\eqref{eq:gp}, from which the data were simulated. Priors for all parameters are described in the Supplementary Material, section \ref{sec:supp:simulation}. Figure~\ref{fig:gpsim}B shows that relative to using GP priors, average HMC runtimes across the $20$ simulated data sets improved by more than ten-fold with HSGP priors when the number of basis functions was less than $50$. Figure~\ref{fig:gpsim}B also summarises the average mean absolute error in posterior median flow estimates. Similar to~\cite{rituort2020hilbertstan}, our results indicate that the HSGP approximations were least accurate for small $B\leq 1.05$, and for larger boundary factors when  the number of basis functions was not increased simultaneously. On our simulations, HSGP approximations performed almost as well as GPs for the tuning parameters $(B=1.25, m=20-30)$, $(B=1.5, m=20-40)$ or $(B=2, m=30-40)$. As computational cost increases with $m$, we chose $B=1.25, m=30$.


%
%
%
\subsection{Application to population-based deep-sequence data from Rakai, Uganda} \label{sec:deepsamplinghet}
We illustrate application of the semi-parametric Poisson flow model \eqref{eq:postdist1} on a population-based sample of HIV deep sequences from the RCCS in south-eastern Uganda at the shores of Lake Victoria \citep{ratmanninfer, ratmann2020quantifying}. Between 2011/08/10 to 2015/01/30, two survey rounds were conducted in 36 inland communities, and three survey rounds in 4 fishing communities (Figure~\ref{fig:data}A). Preceeding each survey, a household census was conducted to identify individuals aged 15-49 years who lived in the communities for at least one month and with intention to stay, who were eligible to participate. 
In brief, there were 37645 census-eligible individuals, of whom 25882 (68.8\%) participated in the RCCS. Participation was higher among women than men, increased with age for both men and women, and was similar in fishing and inland communities (Supplementary Material, section~\ref{sec:supp:regression}). 11404 (96.9\%) of non-participants were absent for school or work. Infected individuals who did not report ART use were selected for sequencing, and deep-sequencing rates were higher among men than women, similar by age for men and women, and were higher in fishing communities (Supplementary Material, section~\ref{sec:supp:regression}).
There were $293$ heterosexual pairs with phylogenetic support for linkage and direction of transmission (source-recipient pairs) when using the threshold $\zeta=0.6$ in \eqref{eq:flowcounts}. 
The estimated infection times of the recipients were between 2009/10/01 and 2015/01/30, which defined the study period $\mathcal{T}$ during which we estimated transmission flows. Figure~\ref{fig:data}B illustrates the reconstructed source-recipient pairs by age of both individuals at the midpoint of the study period. 

\begin{figure}[!t]
    \centering
    \includegraphics[width=0.95\textwidth]{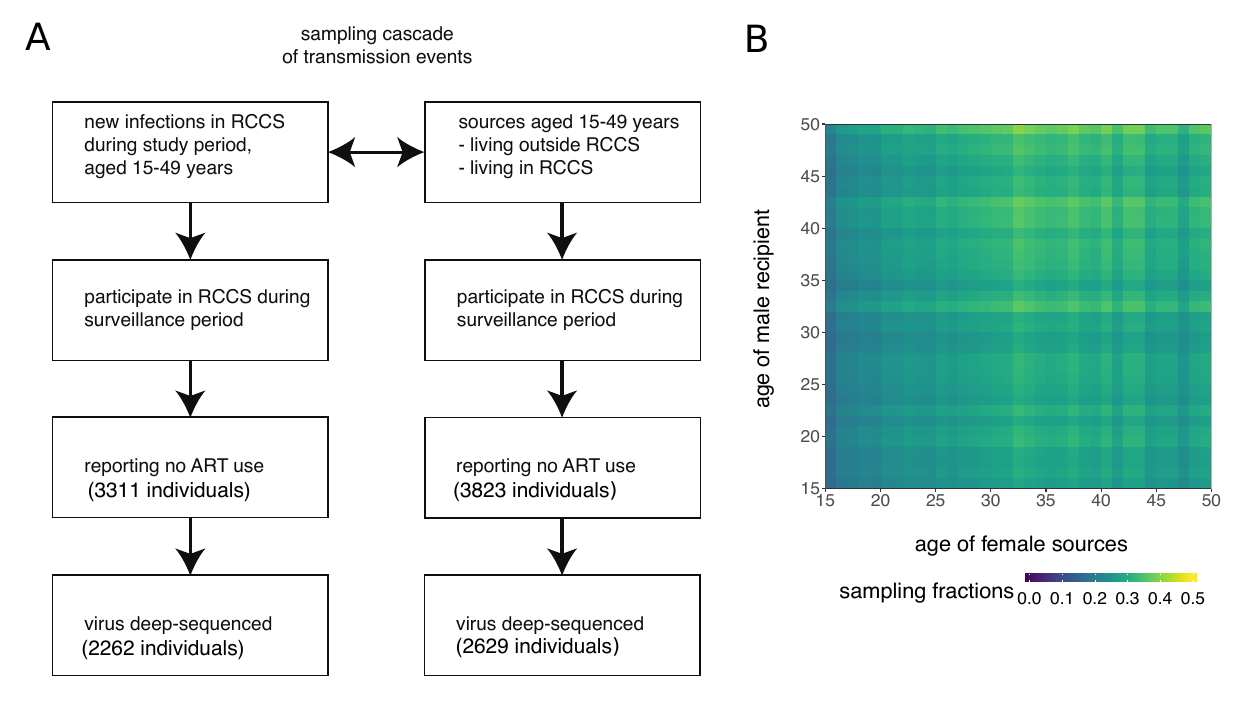}
    \caption{\textbf{Model of the sampling cascade of transmission events.} (\textbf{A}) The sampling cascade formalises the steps involved in sampling sources and recipients of transmission events. Recipients were defined as individuals aged 15-49 years who acquired infection in one of the RCCS communities during the study period. Sources were defined as individuals aged 15-49 years who transmitted to one of the recipients. Each arrow corresponds to a sampling step of the source and recipient populations, which we model using cohort data.(\textbf{B}) Estimated posterior median sampling probabilities of female to male transmission events in fishing communities by age of source and recipient. Conditional sampling probabilities were numerically estimated for each step of the sampling cascade using logistic Binomial regression, and then multiplied to give overall sampling probabilities.}
    \label{fig:samplingcascade}
\end{figure}

To interpret these observations, we defined as denominator transmission events to census-eligible individuals in RCCS communities who were infected during the study period $\mathcal{T}$, and formalised the individual steps in the sampling cascade of transmission events (Figure~\ref{fig:samplingcascade}A). The sources and recipients had to participate in at least one survey round between 2011/08/10 and 2015/01/30, report no ART use, and have virus sequenced successfully. In Rakai, each survey was preceeded by a household census, and with this denominator we numerically estimated age-, gender-, and location-specific conditional sampling probabilities at each step of the sampling cascade using Bayesian logistic-Binomial regression models, and then multiplied Monte Carlo draws from these distributions to numerically approximate the posterior distribution of the overall sampling probabilities $\xi_a$, $\xi_b$, $\forall a,b$ (see Supplementary Material, section~\ref{sec:supp:regression}). Figure~\ref{fig:samplingcascade}B illustrates the resulting, overall sampling probabilities of female-to-male transmissions in fishing communities. The estimated sampling probabilities indicate that the observed data over-represent transmissions between older individuals.

\begin{figure}[!t]
    \centering
    \includegraphics[width=0.8\textwidth]{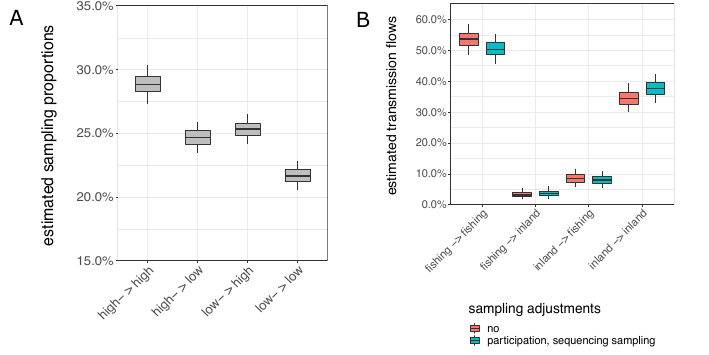}
    \caption{\textbf{Estimated sampling probabilities of transmission events, and impact on flow estimates.} (\textbf{A}) Boxplots of estimated pairwise sampling probabilities of transmission events between low-prevalence inland and high-prevalence fishing communities of the RCCS. The sampling probabilities were obtained by marginalising over age- and gender-specific differences, and indicate that transmission events between high-prevalence fishing communities were over-represented in the data set. (\textbf{B}) Transmission flow estimates between low-prevalence inland and high-prevalence fishing communities of the RCCS. Shown are posterior median estimates (horizontal line), interquartile ranges (box), and 95\% credible ranges when sampling heterogeneity by gender, age, and locations was ignored (red), and when sampling heterogeneity was accounted for as described in the main text.}
    \label{fig:flowchart}
\end{figure}

\subsection{Transmission flows between areas with high and low disease prevalence}\label{sec:highlowprevalence}
We then used the source-recipient data of Figure~\ref{fig:data}B to address problem~\eqref{eq:highlowflow} and estimate transmission HIV flows within and between high- and low-prevalence RCCS communities. The high-prevalence communities comprised the four fishing communities, and the low-prevalence communities comprised the remaining 36 inland communities. Detailed analyses have been reported in \cite{ratmann2020quantifying}; here we focus on illustrating how known sampling heterogeneities can be accounted for, and how they affect inferences of transmission flow.

The participation, ART use, and sequence sampling probabilities differed by gender, age, and location, and we stratified the population accordingly to meet the SARWS assumption that underlies the Poisson flow model (Section~\ref{sec:model}). Specifically, we stratified populations by gender, 1-year age bands (between 15 and 49 years), and resident location (low or high prevalence), which resulted in 140 sampling groups. Following our sampling cascade model (Figure~\ref{fig:samplingcascade}), we then sought to estimate transmission flows between the $2\times 35^2$ age- and gender-specific transmission flows for each of the $4$ combinations of geographic source and recipient locations, through the joint posterior distribution~\eqref{eq:postdist1}. We further accounted for geographic in-migration, resulting in $14,700$ flow variables. On this high-resolution flow space, we were able to directly apply the estimated, structured sampling probabilities that are illustrated in 
Figure~\ref{fig:samplingcascade}B. This shows that accounting for the observed heterogeneities in how the census population was sampled resulted in a more complex inferential problem than~(\ref{eq:highlowflow}-\ref{eq:agesexflow}) suggest. 

\begin{figure}[!t]
    \centering
    \includegraphics[width=0.95\textwidth]{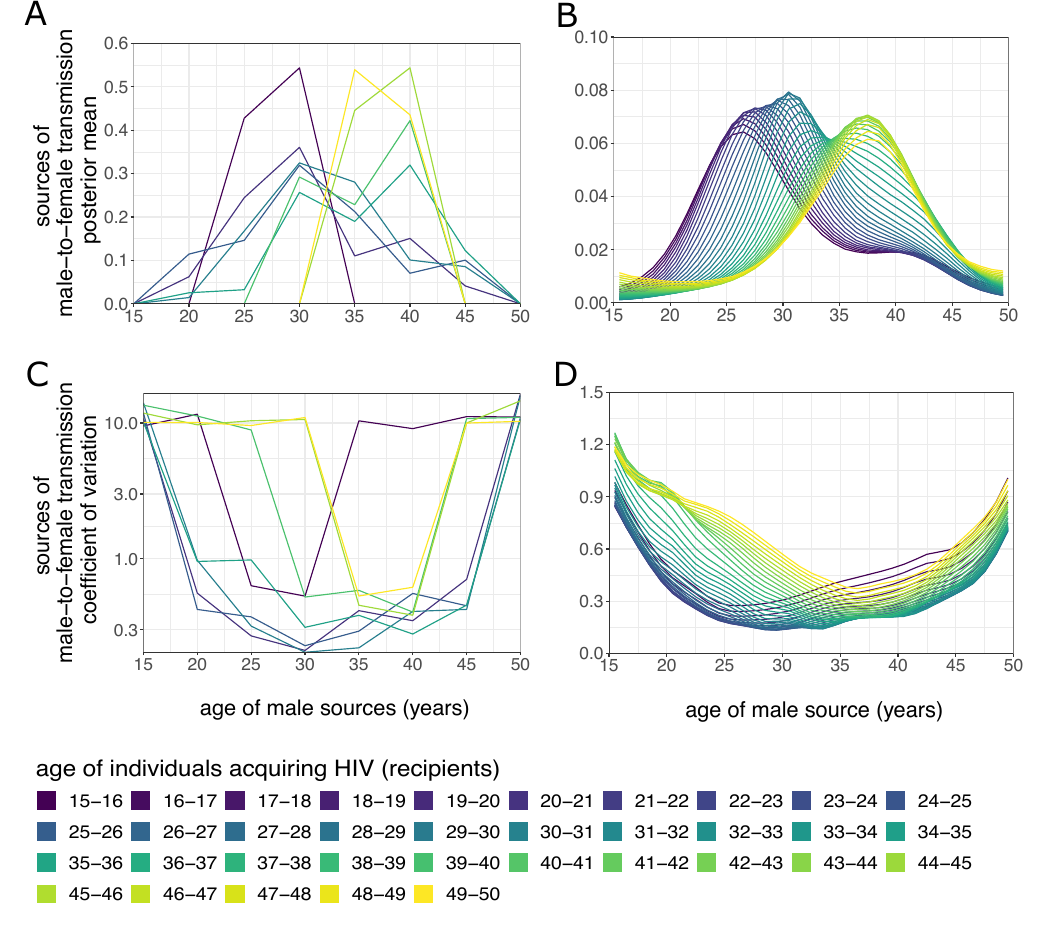}
    \caption{\textbf{Impact of using reqularising prior densities on estimation of age- and gender-specific sources of infection.} We compared estimates of the sources of HIV infection in women without regularisation, using the Gamma prior densities~\eqref{eq:lambdaGamma}, to estimates obtained under the regularising HSGP prior on the log transmission intensities~\eqref{seq:hgsp_model}. Each colour corresponds to infection recipients of a particular age. \textbf{(A)} Each line shows the posterior median estimates of the probability of infections attributed to the age of individuals of the opposite gender, i.e. the sources of infection, obtained without regularisation when using $5$-year age strata. \textbf{(B)} Same as A, using the HSGP prior with boundary factor $B=1.25$ and $m=30$ basis functions when using $1$-year age strata. \textbf{(C-D)} Corresponding estimates of the posterior coefficient of variation in the source estimates.}
    \label{fig:age_analysis}
\end{figure}

To regularise inferences, we used the HSGP approximation~\eqref{seq:hgsp_model} to the stacked Gaussian process prior~\eqref{eq:gp}. We further sought to allow for differences in transmission dynamics across locations, and for this reason specified independent HSGP priors on the parts of the flow space that correspond to transmissions to low prevalence areas, and on the parts of the flow space that correspond to transmissions to high prevalence areas. Numerical inference of the joint posterior density \eqref{eq:postdist1} took $89$ hours on four $2.4$ Ghz processors with Stan version 2.19. There were no convergence, mixing, or divergence warnings, as long as informative prior densities on the length scale hyperparameters were chosen, which we set by matching their 99\% credible ranges to the empirical 99\% quantiles in Figure~\ref{fig:data}. Thus flow inferences remained computationally manageable even in the high-resolution space considered here.



Figure~\ref{fig:flowchart}B shows the marginal posterior estimates of the aggregated flow vector $\boldsymbol{\pi}=(\pi_{hh}, \pi_{hl}, \pi_{lh}, \pi_{ll})$, Equation~\eqref{eq:highlowflow}, when sampling heterogeneity was ignored by setting all $\xi_a$ to the average sampling probability (red), and when gender, age, and location-specific participation and sequence sampling probabilities of sources and recipients were accounted for as described above (turquoise). The average sampling difference between individuals in fishing and inland communities was 7.16\%, suggesting based on our results in Figure~\ref{fig:odesim}B that after accounting for sampling differences, the sampling-adjusted estimates could differ by up to 5\% from the unadjusted estimates. Figure~\ref{fig:flowchart}B shows that our results are in line with this expectation.
The estimated flow ratio (inland$\rightarrow$fishing / fishing$\rightarrow$inland) was 2.18 (1.06-4.71) when sampling heterogeneity was accounted for, and 2.58 (1.23-5.86) when sampling heterogeneity was not accounted for. We thus see that sampling heterogeneity can have an impact on flow estimates, and that the finding that high-prevalence fishing communities were net sinks, and not sources, of local infection flows is robust to sampling heterogeneity. 

\FloatBarrier

\subsection{Transmission flows between age groups} \label{sec:age}
We next turned to problem \eqref{eq:agesexflow} and estimated transmission flows by age and gender from the source-recipient data shown in Figure~\ref{fig:data}. Here, we focus on illustrating how the HSGP prior in the Poisson model allows us to borrow information across data points, and thereby go beyond existing phylodynamic methods~\citep{de2017transmission, le2019hiv, bbosa2020phylogenetic} and make flow inferences by 1-year age bands.

To do this, we compared estimates of the source vectors $\boldsymbol{\delta}$, defined in~\eqref{eq:sources}, when we used the independent Gamma prior density~\eqref{eq:lambdaGamma} (no regularisation) to those when we used the HSGP prior density~\eqref{seq:hgsp_model} (with regularisation). We focused the comparison on the age- and gender-specific sources of infection regardless of location, i.e. the source vector that corresponds to Equation~\eqref{eq:agesexflow}, which was obtained by aggregating over the high- and low-prevalence locations of the source and recipient population groups. Figures~\ref{fig:age_analysis}A-B show the posterior median source estimates for each recipient group respectively without regularisation when using $5$-year age bands and with regularisation using $1$-year age bands, and Figures~\ref{fig:age_analysis}C-D show the corresponding posterior coefficients of variation. 
The estimated coefficients of variation were similar with and without regularisation, and well below $1$ with regularisation, except from sources associated with little contribution to onward transmission and for very young or very old recipients.
These findings suggest that the $1$-year flow estimates are statistically meaningful, and at high resolution provide better insights.
More detailed analyses are reported in Supplementary Text~S5. First, we document the obvious, that  it is not possible to estimate $2450$ flow variables from $293$ data points without regularisation. Second, we show that, as age bands are widened, estimates increasingly depend on the particular start and end points of the chosen age strata, and so we caution against inferences by $5$-year age bands or wider. 
%
%
%
%
%
%


\begin{figure}[!t]
    \centering
    \includegraphics[width=0.95\textwidth]{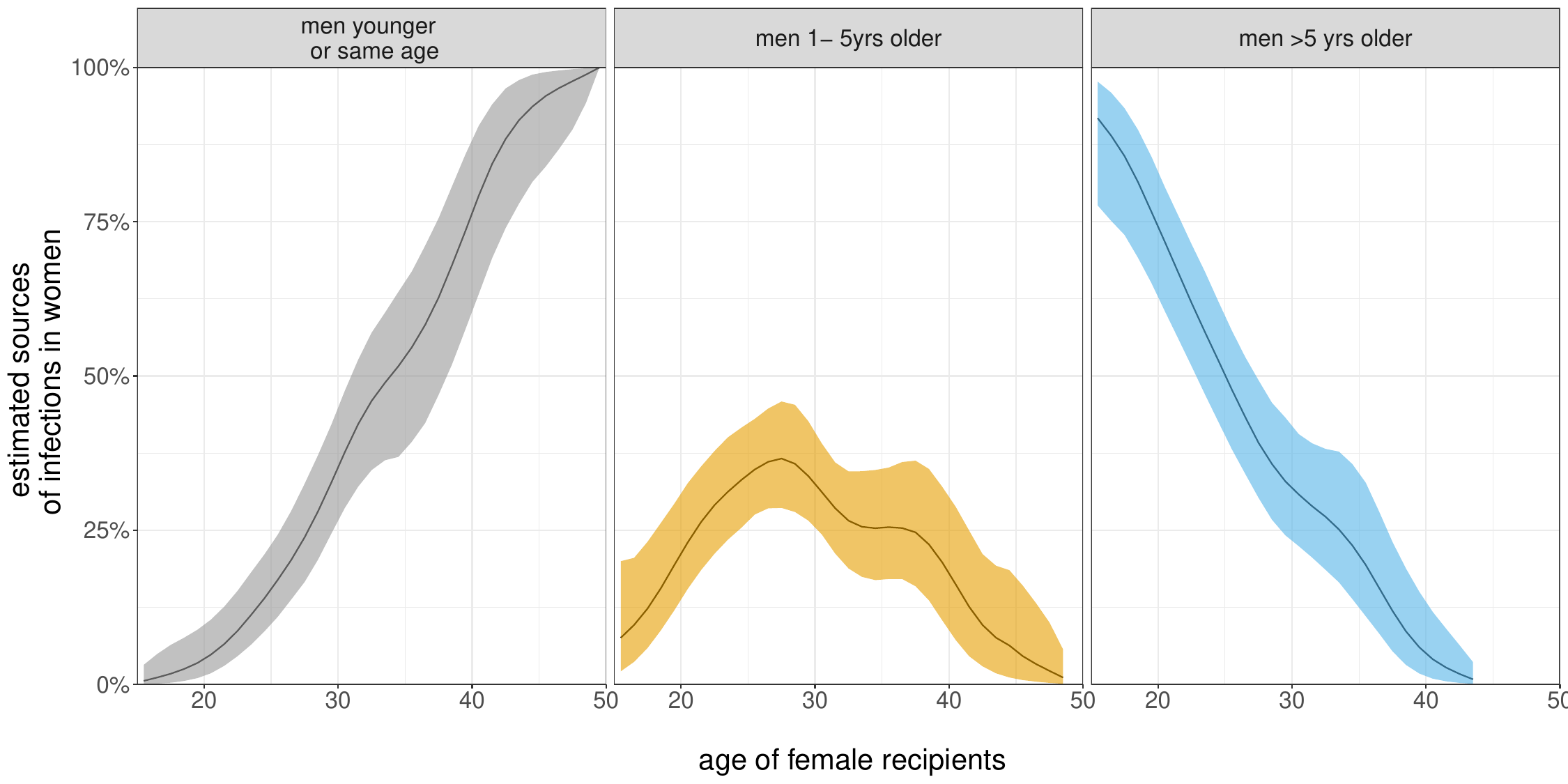}
    \caption{{\bf Estimated sources of infection in women in Rakai, Uganda, 2009-2015.} We estimated the sources of HIV infection in women of increasing age. Sources were defined as men that were younger or the same age as the women (grey), 1-5 years older (orange), and over 5 years older (blue), and sum to 100\% for each age of infected women on the x-axis. Posterior median source estimates are shown along with 95\% marginal credibility intervals.}    \label{fig:source_age_of_young_women}
\end{figure}

\subsection{Sources of transmission to women aged <25 years}\label{sec:youngwomen}
Across sub-Saharan Africa, HIV prevalence rises rapidly among young women aged <25 years 
\citep{unaids2018},
which has prompted efforts to prevent infection among adolescent girls and young women, most notably the DREAMS partnership \citep{sauldetermined}. Infections among young women aged <25 years are commonly attributed to older men, with a recent phylogenetic study from South Africa finding that of 60 identified transmission pairs involving women aged <25 years, 42 (70.0\%) had a probable male partner aged >25 years \citep{de2017transmission}. Our larger data set with directional phylogenetic information allowed us to revisit these estimates.

Our data contained 96 source-recipient pairs involving women aged <25 years, of whom 59 (61.5\%) originated from men and 37 (38.5\%) from women. However, under the regularising HSGP prior density~\eqref{seq:hgsp_model}, our gender- and age-specific flow estimates also borrow information from the other source-recipient pairs that involved older women. We report in Figure \ref{fig:source_age_of_young_women} the estimated sources of HIV infection in women of increasing age. The facets show the contribution of each source, men younger or the same age, men up to $5$ years older, and men more than $5$ years older, and sum to $100$\% for each age of infected women on the x-axis. At age 15, an estimated 91.8\% (77.6\% - 97.7\%) of women were infected by men more than 5 years older, while at age 20, this was  71.8\% (60.6\% - 80.7\%), and at age 25 this was  47.8\% (38.2\% - 57.4\%). These estimates document the overwhelming impact that men more than $5$ years older have on driving infection in very young women in our observation period $2010$-$2015$, and they show that the contribution of these men on infection in women declines rapidly with the age of the women. We provide exact estimates in Table~S9.

\section{Discussion}\label{sec:discussion}
In this study we introduce a class of semi-parametric Bayesian Poisson models for estimating high-resolution flows between population strata, and apply the model to estimate the sources of HIV infections from pathogen deep-sequence data. The modelling framework is flexible, and enables addressing a range of epidemiological questions on pathogen spread between geographic areas, by $1$-year age bands, gender, or indeed other discretely-valued sociodemographic characteristics. We templated the model with and without Hilbert space approximations for scalable inference of high-resolution flows in generic Stan model files, and hope that given the canonical structure of the semi-parametric Poisson flow model, these will be helpful in other movement, origin-destination or flow applications as well~\citep{tebaldi1998bayesian, hazelton2001inference, raymer2013integrated, lindstrom2013bayesian, faye2015chains, van2017efficient, sun2021transmission}.


Existing phylodynamic estimation approaches are tailored for pathogen consensus sequences \citep{lemey2009bayesian, vaughan2014efficient,  volz2009phylodynamics, scire2020improved}. The approach described here is tailored for pathogen deep-sequence data, in that an observed, time-homogeneous flow matrix is required as input, which can be derived through aggregation from individual  source-recipient relationships in deep-sequence phylogenies. 
The main advantages are first, that population-level spread can be directly estimated from individual source-recipient relationships, and modelled in terms of associated individual covariates. In comparison, standard phylogeographic models estimate transition rates from the shape of viral phylogenies captured in times to lineage coalescence \citep{lemey2009bayesian, stadler2013uncovering, scire2020improved}, which are often harder to interpret. Second, relatively little computational effort is needed to fit the Poisson flow model (\ref{eq:data_likelihood}-\ref{eq:postdist1}) to deep-sequence data, because it falls within the class of Bayesian hierarchical models for binary data, for which efficient fitting and regularisation techniques exist \citep{carpenter2017stan, rasmussen2003gaussian,solin2014hilbert}. This makes it computationally feasible to investigate complex aspects of disease spread such as population-level HIV transmission by 1-year age bands. 
Third, differences in how the phylogenetic data were sampled for each stratum can be explicitly accounted for in the model, which is particularly important for characterising HIV transmission, which tends to concentrate in marginalised, vulnerable, and hard to reach populations~\citep{unaids2019}. We found relatively small differences in the estimated sources of infection in the study communities by location and age when inferences were performed with and without sampling adjustments, in line with the relatively limited differences in sampling inclusion probabilities in the RCCS communities and the expected impact on source attribution on simulated data (Figure~\ref{fig:odesim}B). However in other use cases, the impact of sampling heterogeneity on source attribution can be substantially larger. For example, the RCCS included in surveyed communities an estimated 75.7\% of the lakeside population within 3km of the shoreline of Lake Victoria along the Rakai region, and an estimated 16.2\% of the inland population of the Rakai region~\citep{ratmann2020quantifying}. We can use the proposed framework to extrapolate our inferences from the RCCS communities to the underlying population, and given the larger sampling differences we estimated that 88.7\% (84.5–91.9\%) of transmissions in the Rakai region occurred in the inland population~\citep{ratmann2020quantifying}. Specifying and characterising the denominator population is thus crucial for interpreting phylogenetic source attribution estimates, and we believe the proposed Bayesian semi-parametric flow model provides a useful tool in this endeavour.


The method we propose has limitations. First, the method requires pathogen deep-sequence data instead of consensus sequences, which at present are uncommon, though increasingly generated in routine clinical care \citep{houlihan2018use}.
Second, current deep-sequencing protocols typically generate short sequence fragments, usually of $200$ to $300$ base pairs in length after trimming adaptors and low quality ends, and merging paired end fragments. This implies that pathogens need to evolve at a fast rate, because otherwise reconstructed deep-sequence phylogenies do not contain the pattern of ancestral subgraphs that is characteristic of pathogen spread in one direction. Such high evolutionary rates are typical for viral pathogens that infect and evolve in humans over long periods of time, such as HIV or hepatitis C. We expect that the methods developed here will become applicable to a broad range of viral and bacterial infectious diseases as existing deep-sequencing methods that generate substantially longer pathogen sequence fragments become cheaper \citep{rhoads2015pacbio}. 
Third, our inferences are based on source-recipient pairs with strong evidence for the direction of transmission, which is a subset of all the data available, and we cannot exclude that this selection step introduces bias into flow estimates.
Fourth, the model was not designed to estimate time changes in transmission flows. While in principle it is possible to add time as a covariate to the linear predictor of the log transmission intensities~\eqref{eq:gp}, the resulting flow estimates will in general not be consistent with the constraints imposed by standard assumptions on disease spread, as in Equations~(\ref{eq:ode}-\ref{eq:timedependentflows}), 
and other techniques such as the structured coalescent may be better suited 
\citep{volz2009phylodynamics}.

Reducing HIV incidence among adolescent and young women is a key priority for public health programs across sub-Saharan Africa to achieve epidemic control milestones \citep{unaids2018}. The DREAMS intervention aims to promote determined, resilient, empowered, AIDS-free, mentored, and safe adolescent girls and young women, and includes educational programs that aim to address the socio-behavioral factors that underlie vulnerability and infection risk \citep{sauldetermined}. 
Our analysis of a large cross-sectionally sampled HIV deep-sequence data set from Rakai, Uganda, supports previous analyses \citep{de2017transmission,Probert2019,bbosa2020phylogenetic} and indicates that 68.9\% (60.2\%-76.9\%) of adolescent and young women aged <25 years acquired HIV in age-disparate relationships with men at least 5 years older. The estimated proportion of infections attributable to age-disparate relationships was approximately 90\% among adolescent girls, and decreased to approximately 50\% among women aged 25 years. 
Taken together, the data from this study and other phylogenetic studies from Uganda, Zambia, and South Africa suggest that rapid increases in HIV prevalence among adolescent and young women may be driven by the same source populations across sub-Saharan Africa, and support DREAMS interventions that include clear prevention messages about age-disparate sexual relationships.

%
%

\section*{Acknowledgements}
This study was supported by the Bill \& Melinda Gates Foundation (OPP1175094, OPP1084362), the National Institute of Allergy and Infectious Diseases (R01AI110324, U01AI100031, U01AI075115, R01AI102939, K01AI125086-01), National Institute of Mental Health (R01MH107275), the National Institute of Child Health and Development (RO1HD070769, R01HD050180), the Division of Intramural Research of the National Institute for Allergy and Infectious Diseases, the World Bank, the Doris Duke Charitable Foundation, the Johns Hopkins University Center for AIDS Research (P30AI094189), and the Presidents Emergency Plan for AIDS Relief through the Centers for Disease Control and Prevention (NU2GGH000817). We acknowledge data management support provided in part by the Office of Cyberinfrastructure and Computational Biology at the National Institute for Allergy and Infectious Diseases, computational support through the Imperial College Research Computing Service, doi: \url{10.14469/hpc/2232}. We thank the  participants of the Rakai Community Cohort Study and the many staff and investigators who made this study possible, as well as the PANGEA-HIV steering committee, the RCCS leadership, and two anonymous reviewers for their helpful comments on this manuscript.

\section*{List of Supplementary Material}
\noindent
\textbf{S1.} Maximum likelihood flow estimates under heterogeneous sampling \\
\textbf{S2.} Numerical inference algorithms \\
\textbf{S3.} Simulation experiments \\
\textbf{S4.} Modelling and estimation of the sampling cascade \\
\textbf{S5.} Analyses using different age bands \\
\textbf{S6.} Supplementary Figures and Tables \\
Stan files were provided as text documents. \\
prior\_gamma.txt \\
prior\_gp.txt \\
prior\_gp\_approx.txt

\end{bibunit}

\setlength{\parskip}{0pt}
\linespread{0}


\setcounter{equation}{0}
\setcounter{section}{0}
\setcounter{figure}{0}
\setcounter{table}{0}
\renewcommand{\theequation}{S\arabic{equation}}
\renewcommand{\thesection}{S\arabic{section}}
\renewcommand{\thefigure}{S\arabic{figure}}
\renewcommand{\thetable}{S\arabic{table}}

\begin{bibunit}[rss]

\section{Maximum likelihood flow  estimates under heterogeneous sampling}\label{supp:sec:mle}
We here describe derivation of the maximum likelihood estimates of the Poisson model \eqref{eq:data_likelihood},
\begin{linenomath*}
\begin{equation*}
p(\mathbf{n}|\boldsymbol{\lambda},\boldsymbol{\xi})\propto
\prod_{a,b} (\lambda_{ab}\xi_a\xi_b)^{n_{ab}} \exp(-\lambda_{ab}\xi_a\xi_b),
\end{equation*}
\end{linenomath*}
when the number of sampled individuals in $a$, $N^s_a= \sum_{i \in a} \Indicator{s_i=1}$, is a Binomial sample of the number of all individuals in $a$, $N_a$. We denote the vector of sampled individuals across population strata by $\boldsymbol{N}^s=(N^s_a)_{a\in\mathcal{A}}$, and similarly the vector of individuals in each population group by $\boldsymbol{N}=(N_a)_{a\in\mathcal{A}}$. Then the product likelihood is 
\begin{linenomath*}
\begin{equation}\label{seq:data_likelihood_binomial}
p(\mathbf{n},\boldsymbol{N},\boldsymbol{N}^s|\boldsymbol{\lambda},\boldsymbol{\xi})
=
\bigg(
\prod_{a,b}
\mbox{Poisson}(n_{ab}; \lambda_{ab}\xi_a \xi_b)
\bigg)
\bigg(
\prod_a
\mbox{Binomial}(N^s_a;N^a,\xi_a)
\bigg).
\end{equation}
\end{linenomath*}
Taking the derivative of the log-likelihood \eqref{seq:data_likelihood_binomial} to zero gives
\begin{linenomath*}
\begin{equation}
    \begin{split}
& \hat{\pi}_{ab}\hat{\eta} = \frac{n_{ab}}{\hat{\xi}_a \hat{\xi}_b}\\
&  \hat{\xi}_a =\frac{N^s_a}{N_a} 
    \end{split}
\end{equation}
\end{linenomath*}
As $\sum_{ab} \hat{\pi}_{ab} = 1$,
\begin{linenomath*}
\begin{equation}
    \hat{\eta} = \sum_{a,b}\frac{n_{ab}}{\hat{\xi}_a \hat{\xi}_b},
\end{equation}
\end{linenomath*}
and we obtain
\begin{linenomath*}
\begin{equation}
    \hat{\pi}_{ab}=\frac{n_{ab}}{\hat{\xi}_a \hat{\xi}_b}\big/\sum_{a,b}\frac{n_{ab}}{\hat{\xi}_a \hat{\xi}_b}.
\end{equation}
\end{linenomath*}
\FloatBarrier

\section{Numerical inference algorithms}\label{sec:algorithms}
We here describe numerical algorithms for estimating transmission flows with the Poisson flow model~\eqref{eq:data_likelihood}. 

It is possible to implement the flow model~\eqref{eq:data_likelihood} in the Stan computing language \citep{carpenter2017stan}. We hope this feature enhances robust numerical inference in a well-tested software environment, facilitates model sharing and model extensions, and gives end-users the option to apply alternative inference algorithms. For the work presented in this paper, we have used the dynamic Hamiltonian Monte Carlo algorithm of the Stan probabilistic programming framework in Stan version 2.21. Section~\ref{s:stan_independet_prior} describes the Stan model specification for inference of the joint posterior distribution~\eqref{eq:postdist1} when population strata are unordered and the prior density on the transmission intensities $\boldsymbol{\lambda}$ is independent Gamma~\eqref{eq:lambdaGamma}. Section~\ref{s:stan_gp_prior} describes the Stan model specification for inference when population strata can be ordered and the prior densities for the transmission intensities are correlated through a Gaussian process prior \eqref{eq:gp}. For simplicity, we describe models and algorithms for the case that sampling probabilities are the same among source and recipient cases. Extensions to different sampling probabilities are straightforward. Stan is not necessary for numerical inference. To illustrate this, we describe in Section~\ref{sec:mcmc} a custom MCMC within Gibbs sampler for inference of the joint posterior distribution~\eqref{eq:postdist1} when population strata are unordered and the prior density on the transmission intensities $\boldsymbol{\lambda}$ is independent Gamma~\eqref{eq:lambdaGamma}.

\subsection{Stan implementation using independent prior on transmission intensities}\label{s:stan_independet_prior}
When population strata are unordered, we focus on inference of the posterior distribution
\begin{linenomath*}
\begin{equation}\label{seq:postdist1}
\begin{split}
& 
p(\boldsymbol{\lambda}, \boldsymbol{\xi}|\mathbf{n},\boldsymbol{s}, \mathbf{X}) 
\propto  
p(\mathbf{n}|\boldsymbol{\lambda},\boldsymbol{\xi},\boldsymbol{s},\mathbf{X}) p(\boldsymbol{\lambda}|\boldsymbol{\xi},\boldsymbol{s},\mathbf{X}) p(\boldsymbol{\xi}|\boldsymbol{s},\mathbf{X})
\\
&\quad\quad
= 
p(\mathbf{n}|\boldsymbol{\lambda},\boldsymbol{\xi}) p(\boldsymbol{\lambda}|\boldsymbol{\xi}) p(\boldsymbol{\xi}|\boldsymbol{s},\mathbf{X})\\
&\quad\quad
= 
\bigg( 
\prod_{a,b}\mbox{Poisson}(n_{ab};\lambda_{ab}\xi_a\xi_b) 
\bigg)
\bigg( 
\prod_{a,b}\mbox{Gamma}(\lambda_{ab}; \alpha_{ab}, \beta)
\bigg)
p(\boldsymbol{\xi}|\boldsymbol{s},\mathbf{X})\\
\end{split}
\end{equation}
\end{linenomath*}
where 
\begin{longtable}{ll}
    $\boldsymbol{\lambda}$& vector of transmission intensities from group $a$ to group $b$, with length $L$,\\ 
    $\boldsymbol{\xi}$& vector of sampling proportions for each population group $a$, $a\in\mathcal{A}$,\\
    $\boldsymbol{n}$& vector of observed transmission counts, with length $L$,\\
    $\boldsymbol{s}$& sampling status vector, with length $N$,\\
    $\mathbf{X}$& matrix of sampling characteristics, with dimension $N \times p$,
\end{longtable}
\noindent
and $\alpha_{ab}$, $\beta$ are as in \eqref{eq:lambdaGamma}, and the flow vector $\boldsymbol{\pi}$ is obtained from the transformation $\pi_{ab}=\lambda_{ab}/( \sum_{c,d}\lambda_{cd} )$. 

To specify the posterior density $p(\boldsymbol{\xi}|\boldsymbol{s},\mathbf{X})$ in Stan, we further approximated its components $p(\xi_a |\boldsymbol{s},\mathbf{X})$ through a suitable closed-form density. In our applications, we opted for Beta densities with shape and rate parameters estimated via maximum likelihood. The resulting Stan model file was provided in the file named prior\_gamma.txt.

Use of the Stan model is illustrated in the online example script \url{https://github.com/BDI-pathogens/phyloscanner/blob/master/phyloflows/vignettes/07_age_analysis.md}.

\subsection{Stan implementation using Gaussian process prior on transmission intensities}\label{s:stan_gp_prior}

When population strata can be ordered as in the application to age-specific transmission flows, we focus on inference of the posterior distribution \eqref{seq:postdist1} with the independent Gamma priors replaced by a two-dimensional Hilbert space Gaussian process priors on the non-zero components of $\boldsymbol{\lambda}$, \eqref{eq:gp}. To implement this in Stan, Equation \eqref{seq:hgsp_model} can be more compactly written as
\begin{linenomath*}
\begin{equation}\label{seq:hgsp_kernel_compact}
k^{\text{HSGP}}\big( (a,b), (a^\prime,b^\prime) \big)
= \boldsymbol{\phi}(a,b)^T\Delta\: \boldsymbol{\phi}(a^\prime,b^\prime),
\end{equation}
\end{linenomath*}
where $\boldsymbol{\phi}(a,b)=(\phi_j(a,b))_{j=1}^m \in \mathbb{R}^m$ is the column vector of eigenfunctions and $\Delta \in \mathbb{R}^{m\times m}$ is the diagonal matrix whose $(j,j)$th entry is $S_\theta(\sqrt{\boldsymbol{\lambda_j}})$. Consequently the HSGP Gram matrix for $n$ observations and associated inputs $\{(a_i,b_i)\}_{i=1}^n \in \Omega$ is
\begin{linenomath*}
\begin{equation}\label{seq:hgsp_gram}
K^{\text{HSGP}}
= \boldsymbol{\Phi}\Delta \boldsymbol{\Phi}^T,
\end{equation}
\end{linenomath*}
where
\begin{linenomath*}
\begin{equation*}
\boldsymbol{\Phi} =
\begin{pmatrix}
\phi_1(a_1,b_1) &  \cdots & \phi_m(a_1,b_1) \\
\vdots & \ddots & \vdots \\
\phi_1(a_n,b_n) &  \cdots & \phi_m(a_n,b_n) \\
\end{pmatrix}
\in \mathbb{R}^{n\times m}.
\end{equation*}
\end{linenomath*}
The two-dimensional zero-mean HSGP model is fully defined in terms of the kernel \eqref{seq:hgsp_model}, as the stochastic process for which any finite set of two-dimensional inputs in $\Omega$ follows a multivariate normal distribution with variance-covariance matrix specified by \eqref{seq:hgsp_model}, and this is vectorised as \eqref{seq:hgsp_gram}.
Using Cholesky decomposition, the HSGP model \eqref{seq:hgsp_model} can be conveniently calculated from $m$ standard normal random variables through
\begin{linenomath*}
\begin{equation}\label{seq:hgsp_f}
f(a,b)
=
\sum_{j=1}^m \sqrt{S_\theta(\sqrt{\boldsymbol{\lambda}_j}))} \phi_j(a,b) \beta_j
\end{equation}
\end{linenomath*}
where $\beta_j\sim \mathcal{N}(0,1)$ for $j=1,\dotsc,m$. The resulting Stan model file is prior\_gpapprox.txt. 
Use of the Stan model is illustrated in the online example script \url{https://github.com/BDI-pathogens/phyloscanner/blob/master/phyloflows/vignettes/07_age_analysis.md}. We explored optimal choices of the tuning parameters of the HSGP approximation on simulations, please see Section~\ref{sec:performance_hsgp} and \ref{sec:hsgp_approx} for details. The Stan file using the GP prior without approximation is in prior\_gp.txt as a comparison to approximated GP priors.

\subsection{Markov Chain Monte Carlo within Gibbs algorithm}\label{sec:mcmc}
This section describes implementation details of a custom Metropolis-within-Gibbs algorithm to estimate the posterior distribution \eqref{seq:postdist1},
\begin{linenomath*}
\begin{equation*}
p(\boldsymbol{\lambda}, \boldsymbol{\xi}|\mathbf{n},\boldsymbol{s}, \mathbf{X}) 
\propto  
\bigg( 
\prod_{a,b}\mbox{Poisson}(n_{ab});\lambda_{ab}\xi_a\xi_b) 
\bigg)
\bigg( 
\prod_{a,b}\mbox{Gamma}(\lambda_{ab}; \alpha_{ab}, \beta)
\bigg)
p(\boldsymbol{\xi}|\boldsymbol{s},\mathbf{X}).
\end{equation*}
\end{linenomath*}
The algorithm is also implemented in the \texttt{phyloflows} R package,  version 1.2.0.

\subsubsection{Overall structure of algorithm} \label{sec:supp:mcmc:structure}
The algorithm exploits the factorisation of the posterior density \eqref{seq:postdist1} into the full conditionals
\begin{linenomath*}
\begin{equation*}
\begin{split}
        &p(\xi_a| \boldsymbol{\xi}_{-a},\boldsymbol{\lambda}, \boldsymbol{s}, \mathbf{n}, \mathbf{X}) \text{ for all }a,\\
        &p(\lambda_{ab} | \boldsymbol{\xi}, \boldsymbol{\lambda}_{-ab},
        \boldsymbol{s},\mathbf{n}, \mathbf{X}) \text{ for all }a,b,
\end{split}
\end{equation*}
\end{linenomath*}
where $\xi_a$ is the sampling probability of the $a$th population group, $\lambda_{ab}$ denotes the average transmission counts from population group $a$ to population group $b$, $\boldsymbol{\xi}_{-a}=\boldsymbol{\xi}\setminus\xi_a$ and $\boldsymbol{\lambda}_{-ab}=\boldsymbol{\lambda}\setminus\lambda_{ab}$. The density $p(\xi_a| \boldsymbol{\xi}_{-a},\boldsymbol{\lambda}, \boldsymbol{s}, \mathbf{n}, \mathbf{X})$ is not available in closed form, and we performed Metropolis-within-Gibbs updates for each $\xi_a$. The density $p(\lambda_{ab} | \boldsymbol{\xi}, \boldsymbol{\lambda}_{-ab},\boldsymbol{s},\mathbf{n}, \mathbf{X})$ is of Gamma form, and so a Gibbs step can be used to update $\lambda_{ab}$. The resulting MCMC algorithm iterates through Metropolis-within-Gibbs updates for each $\xi_a$, followed by a series of Gibbs updates for each $\lambda_{ab}$.

\subsubsection{Metropolis-Hastings within Gibbs steps} 
The algorithm starts by updating in turn the sampling proportions $\xi_a$ of each population group $a\in\mathcal{A}$. The full conditional distribution of $\xi_a$ is
\begin{linenomath*}
\begin{equation}\label{eq:fcxi}
\begin{split}
& 
p(\xi_a|\boldsymbol{\xi}_{-a},\boldsymbol{\lambda},\mathbf{n},\boldsymbol{s},\mathbf{X}) 
\\
&
\quad\quad\propto 
\Pi_{c=a \mbox{or} d=a} p(n_{cd}|\lambda_{cd},\xi_c\xi_d)p(\xi_a|\boldsymbol{s},\mathbf{X}) \\ 
&
\quad\quad\propto 
\Pi_{c=a \mbox{or} d=a}  \mbox{Poi}(n_{cd};\lambda_{cd}\xi_c \xi_d) 
\mbox{Gamma}(\lambda_{cd};\xi_c \xi_d) 
p(\xi_a|\boldsymbol{s},\mathbf{X}),
\end{split}
\end{equation}
\end{linenomath*}
which does not have closed form,  and so a Metropolis-Hastings update is used. We sought to avoid tuning parameters as much as possible, and for this reason proposed moves from the prior,
\begin{linenomath*}
\begin{equation}\label{eq:pd1}
q(\xi_a^{\prime}|\xi_a) =p(\xi_a^{\prime}|\boldsymbol{s},\mathbf{X}).
\end{equation}
\end{linenomath*}
The resulting Metropolis Hasting ratio is 
\begin{linenomath*}
\begin{equation}\label{eq:pd3}
\begin{split}
r &= 
\frac{
p(\xi_a^\prime|\boldsymbol{\xi}_{-a},\boldsymbol{\lambda},\mathbf{n},\boldsymbol{s},\mathbf{X})
}
{
p(\xi_a|\boldsymbol{\xi}_{-a},\boldsymbol{\lambda},\mathbf{n},\boldsymbol{s},\mathbf{X})
}
\times
\frac{
q(\xi_a)
}
{
q(\xi_a^{\prime})
}
\\[2mm] 
&= 
\frac{
\Pi_{b} p(n_{ab}|\lambda_{ab},\xi_a^\prime \xi_b) 
\Pi_{b\neq a}
p(n_{ba}|\lambda_{ba},\xi_b \xi_a^\prime)
\Pi_{b} p(\lambda_{ab} | \boldsymbol{\xi}^\prime)
\Pi_{b\neq a} p(\lambda_{ba}| \boldsymbol{\xi}^\prime)
p(\xi_a^\prime|\boldsymbol{s},\mathbf{X})
}
{
\Pi_{b} p(n_{ab}|\lambda_{ab},\xi_a \xi_b) \Pi_{b\neq a} p(n_{ba}|\lambda_{ba},\xi_b \xi_a) 
\Pi_{b} p(\lambda_{ab} | \boldsymbol{\xi})
\Pi_{b\neq a} p(\lambda_{ba}| \boldsymbol{\xi})
p(\xi_a|\boldsymbol{s},\mathbf{X})
} \\
&\quad\quad\quad
\times
\frac{
p(\xi_a|\boldsymbol{s},\mathbf{X})
}
{
p(\xi_a^{\prime}|\boldsymbol{s},\mathbf{X})
}
\\[2mm] 
&=
\frac{
\Pi_{b} p(n_{ab}|\lambda_{ab},\xi_a^\prime \xi_b) \Pi_{b\neq a} p(n_{ba}|\lambda_{ba},\xi_b \xi_a^\prime) \Pi_{b} p(\lambda_{ab} | \boldsymbol{\xi}^\prime)
\Pi_{b\neq a} p(\lambda_{ba}| \boldsymbol{\xi}^\prime)
}
{
\Pi_{b} p(n_{ab}|\lambda_{ab},\xi_a \xi_b) \Pi_{b\neq a} p(n_{ba}|\lambda_{ba},\xi_b \xi_a)
\Pi_{b} p(\lambda_{ab} | \boldsymbol{\xi})
\Pi_{b\neq a} p(\lambda_{ba}| \boldsymbol{\xi})
}, 
\end{split}
\end{equation}
\end{linenomath*}
where for brevity we denoted by $\boldsymbol{\xi}^\prime$ the vector of sampling probabilities such that $\xi_c^\prime=\xi_c$ for all indices $c \neq a$ and $\xi^\prime_a$ inserted at index $a$. Proposed moves are accepted with probability $\min(1,r)$.

In our applications, different sampling cascades apply to source and recipient population groups, which we denote respectively by $\xi_a^s$ and $\xi_a^r$. Equations (\ref{eq:fcxi}-\ref{eq:pd3}) generalise straightforwardly to this setting, and the algorithms updates the sampling probabilities $\boldsymbol{\xi}^s$ and $\boldsymbol{\xi}^r$ in turn.

\subsubsection{Gibbs step}
The conditional distribution 
$p(\lambda_{ab} | \boldsymbol{\xi}, \boldsymbol{\lambda}_{-ab}, \boldsymbol{s},\mathbf{n}, \mathbf{X})$ is 
\begin{linenomath*}
\begin{equation}
\begin{split}
& p(\lambda_{ab} | \boldsymbol{\xi}, \boldsymbol{\lambda}_{-ab}, \boldsymbol{s},\mathbf{n}, \mathbf{X}) \propto p(\lambda_{ab}) p(n_{ab}|\lambda_{ab},\xi_a,\xi_b) \\
& \propto\exp(-\lambda_{cd}\xi_c\xi_d) (\lambda_{cd}\xi_c\xi_d)^{n_{cd}} \lambda_{cd}^{\alpha_{cd}-1}\exp(-\beta\lambda_{cd}) \\
& \propto \exp(-\lambda_{cd}(\xi_c\xi_d+\beta)) \lambda_{cd}^{n_{cd}+\alpha_{cd}-1}\\
& \sim \mbox{Gamma}(n_{cd}+\alpha_{cd},\xi_c\xi_d+\beta).
\end{split}
\end{equation}
\end{linenomath*}
Each $\lambda_{ab}$ is updated in turn by sampling from the right hand side.
\FloatBarrier

\section{Simulation experiments} \label{sec:supp:simulation}
We here describe several simulation experiments that we implemented to validate the hierarchical Poisson model~\eqref{eq:data_likelihood} to estimate transmission flows in the context of sampling heterogeneity. Our overall strategy was to simulate transmission counts, re-estimate transmission flows. and compare the estimated flows with the true simulated flows.

\subsection{ODE-based simulations experiments}\label{sec:supp:2by2}
The first experiment was a minimal example to assess (1) the impact of sampling differences on flow estimates and (2) the basic performance of our approach. The main results were reported in Figure~\ref{fig:odesim}B. We here provide further detail on these simulation experiments.

We considered transmission flows between two population groups $A=(a,b)$, which for ease of illustration we refer to as individuals living in rural areas or small communities (group $a$), and individuals living in large communities (group $b$). The population was further structured by men and women, yielding in total $4$ population strata. We then simulated epidemics based on the compartmental model~\eqref{eq:ode} among susceptible, infected, and treated individuals in the four population strata. The parameters of the model were specified such that 40\% of were in group $a$ and 60\% of were in group; half of individuals in each group were men and women respectively; and HIV prevalence was approximately 60\% at equilibrium. Stochastic simulations in a population of 30000 individuals were performed using MASTER \citep{vaughan2013stochastic}, and run for 400 time units. The simulations are specified in the xml file \url{https://github.com/BDI-pathogens/phyloscanner/blob/master/phyloflows/inst/misc/Master.xml}. Table \ref{tab:parameters} lists the model parameters that we used in the simulations, where in terms of notation we replaced the population subscripts $l$ with $a$ and $h$ with $b$ compared to~\eqref{eq:ode}.

\begin{table}
     \caption{Parameter values used for the ODE-based simulations \label{tab:parameters}} 
  \begin{tabular}{lll}
    \toprule
    {\bf Parameter} & {\bf Symbol} & {\bf Value} \\[1mm]
    \midrule
    transmission rates & & \\[1mm]
    female to male & & \\[1mm]
    (further divided by pop size) & & \\[1mm]
    \quad a->a & $\beta_{fm,aa}$ & 0.0713\\[1mm]
    \quad b->a & $\beta_{fm,ba}$ & 0.0071\\[1mm]
    \quad a->b & $\beta_{fm,ab}$ & 0.0122\\[1mm]
    \quad b->b & $\beta_{fm,bb}$ & 0.0713\\[1mm]
    \midrule
    transmission rates & & \\[1mm]
    male to female & & \\[1mm]
    (further divided by pop size) & & \\[1mm]
    \quad a->a & $\beta_{mf,aa}$ & 0.1019\\[1mm]
    \quad b->a & $\beta_{mf,ba}$ & 0.0173\\[1mm]
    \quad a->b & $\beta_{mf,ab}$ & 0.0224\\[1mm]
    \quad b->b & $\beta_{mf,bb}$ & 0.1019\\[1mm]
    \midrule
    viral suppression rate & $\gamma$ & 0.0444\\[1mm]
    \midrule
    birth/death rate & $\mu$ & 0.01667\\[1mm]
    \bottomrule
   \end{tabular}
\end{table}

To assess the impact of sampling heterogeneity on flow estimates, we kept the sampling probability in population group $a$ at 60\%, and varied the sampling probability in group $b$ from 60\% to 35\% in order to assess the impact of sampling differences of 0\%, 5\%, 10\%, 15\%, 20\%, 25\%. The sampling status of each individual in $a$ was simulated under a Bernoulli draw with probability $\xi_a$, and respectively for each individual in $b$ with probability $\xi_b$, and observed transmissions $n^r_{ab}$ from group $a$ to group $b$ were calculated as in~\eqref{seq:actualflowcounts}. 

From the simulated data, we first re-estimated the transmission flows while ignoring sampling heterogeneity. We estimated the posterior distribution~\eqref{eq:postdist1} with the phyloflows MCMC-within-Gibbs sampler of section~\ref{sec:mcmc}, with $p(\xi_a|\boldsymbol{s},\mathbf{X})$ and $p(\xi_b|\boldsymbol{s},\mathbf{X})$ set to Beta($25.5, 25.5$). Thus, the specified sampling densities ignored sampling heterogeneity. Code is available at \url{https://github.com/BDI-pathogens/phyloscanner/blob/master/phyloflows/inst/misc/ode_estimation.R}. For each simulation, we then calculated the worst case error. The main results are shown in Figure~\ref{fig:odesim}B, and indicate that worst case error increased substantially with sampling heterogeneity.


\begin{figure}[!t]
    \centering
    \includegraphics[width=0.75\textwidth]{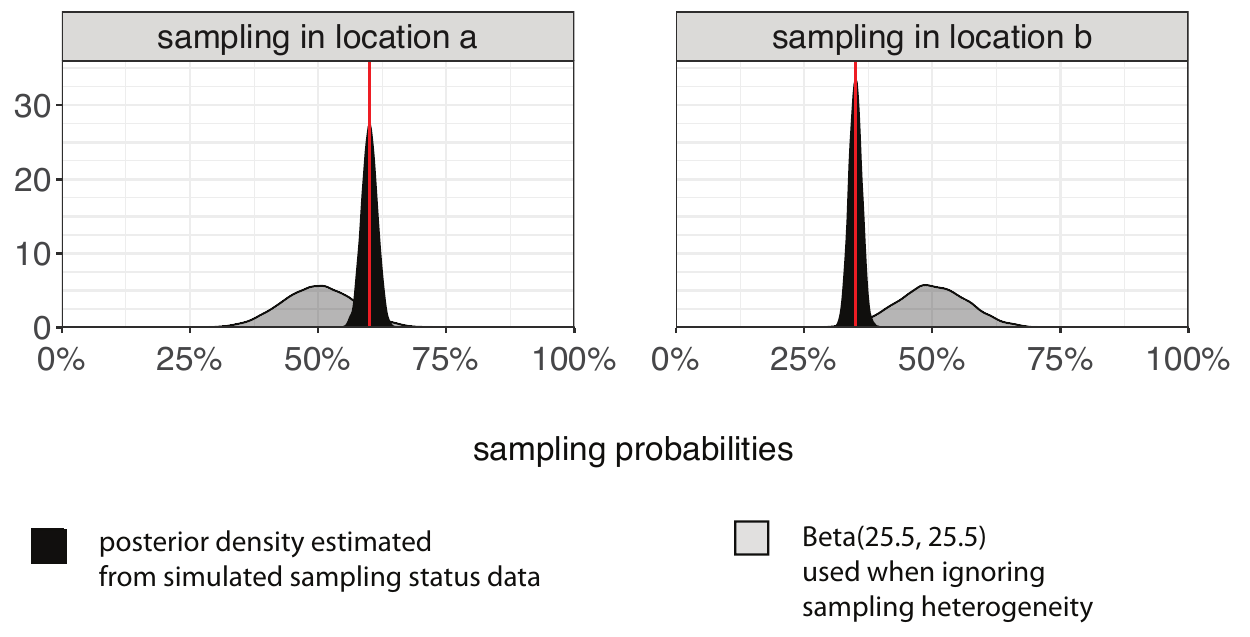}
    \caption{\textbf{Re-estimated sampling probabilities in the ODE-based simulations.} We simulated transmission flows between men and women in two locations $a$ and $b$ under the ODE model~\eqref{eq:ode} in Section~\ref{sec:supp:2by2}. 
    From the simulated sampling status data, posterior estimates of the sampling proportions in $a$ and $b$ were obtained with Equation~\eqref{supp:eq_beta_prior}, and then used in flow inference. We show posterior densities of the sampling probabilities in $a$ and $b$ (grey), compared to a $Beta(25.5,25.5)$ prior density.
    Vertical lines indicate the true sampling fractions, $\xi_a=0.6$ and $\xi_b=0.35$.}
    \label{fig:simulationprior}
\end{figure}

Second, we re-estimated the transmission flows while accounting for sampling heterogeneity. The posterior distribution of sampling probabilities $p(\xi_a|\boldsymbol{s},\mathbf{X})$ was estimated from the number of sampled individuals in the populations $N^s_a$, assuming we know the total population size $N_a$, through
\begin{linenomath*}
\begin{equation}\label{supp:eq_beta_prior}
    p(\xi_a|\boldsymbol{s},\mathbf{X})=\mbox{Beta}(\xi_a, N^s_a+\alpha_\xi, N_a-N^s_a+\beta_\xi),
\end{equation}
\end{linenomath*}
where we set $\alpha_\xi=0.5$, $\beta_\xi=0.5$, and similarly for $\xi_a$. Figure~\ref{fig:simulationprior} illustrates the estimated posterior sampling distributions, in comparison to the true values. We then estimated the posterior distribution~\eqref{eq:postdist1} with the phyloflows MCMC-within-Gibbs sampler of section ~\ref{sec:mcmc}, using the posterior distribution of sampling probabilities~\eqref{supp:eq_beta_prior}.  Figure~\ref{fig:odesim}B reports the worst case error of posterior flow estimates. 



\FloatBarrier

\subsection{Sensitivity to overall sample size of observed transmission flows}\label{sec:supp:sensitivity}
We further assessed the accuracy of flow inferences as a function of overall sample size of the observed transmission flows, $n^+=\sum_{a,b} n_{ab}$. We re-considered the ODE-based simulation experiments of Section~\ref{sec:supp:2by2}, with a 25\% sampling difference between $a$ and $b$ locations. The overall population size parameter was set to $11200$, $18000$, $52000$ such that the overall sample size was $n^+\approx 100$, $300$, $600$. 

\begin{figure}[!t]
    \centering
    \includegraphics[width=0.75\textwidth]{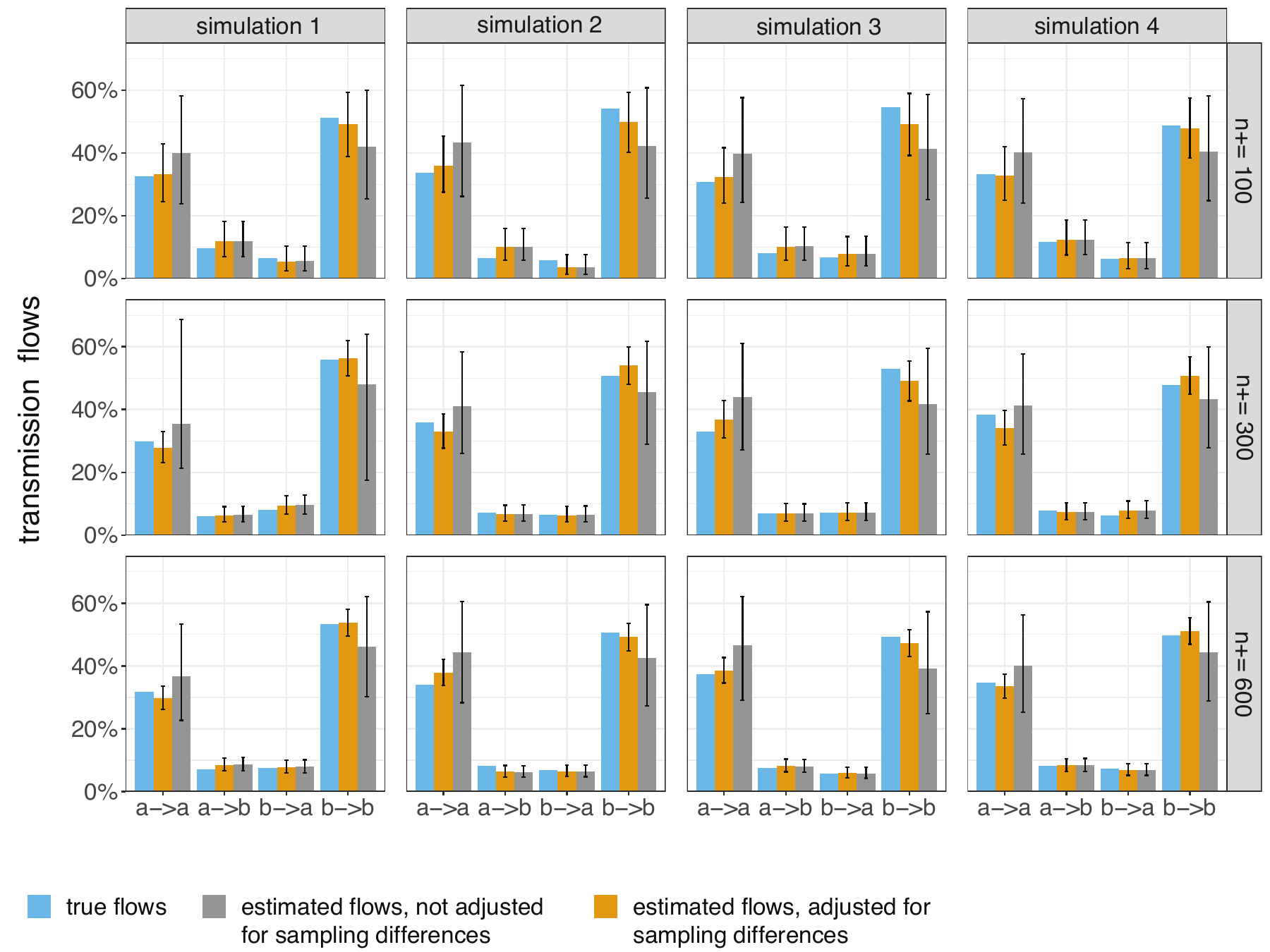}
    \caption{{\bf True and re-estimated flows in four ODE-based simulations, varying in sample size.} We simulated transmission flows between men and women in two locations $a$ and $b$ under the ODE model~\eqref{eq:ode} in Section~\ref{sec:supp:sensitivity}. Columns show results for four simulated data sets, and rows show results for increasing samples size. The true transmission flows are show in blue. Posterior flow estimates (median and 95\% credibility intervals) are shown when sampling heterogeneity was ignored (grey), and when sampling heterogeneity was accunted for as in~\eqref{supp:eq_beta_prior} (orange).}
    \label{fig:posterior_ODE_sensitivity}
\end{figure}

\begin{figure}[!t]
    \centering
    \includegraphics[scale=0.4]{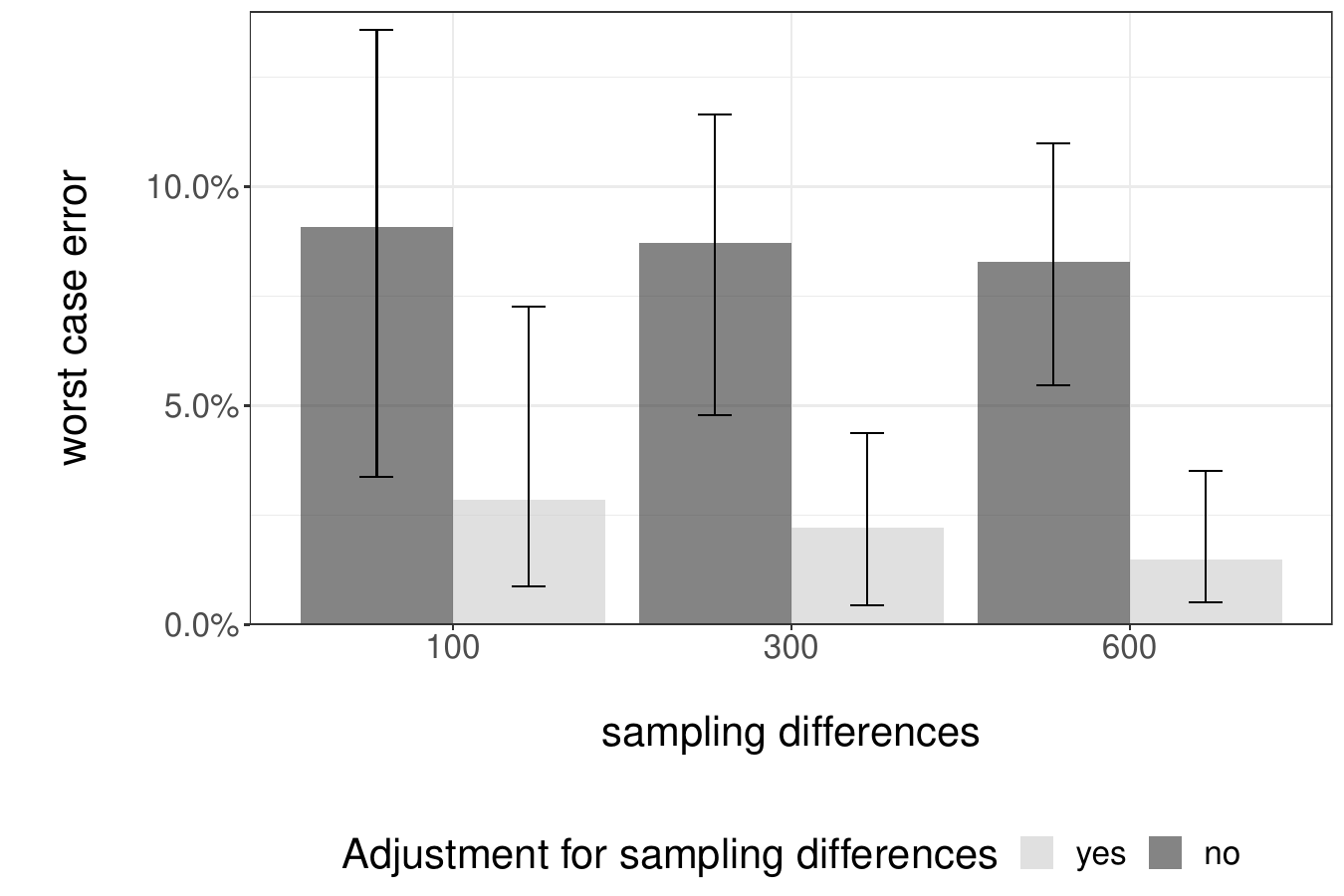}
    \caption{\textbf{Worst case error in estimated transmission flows with increasing sample size.} We simulated transmission flows between men and women in two locations $a$ and $b$ under the ODE model~\eqref{eq:ode} as described in Section~\ref{sec:supp:2by2}. Transmission flows were re-estimated with the hierarchical Poisson model~\eqref{eq:data_likelihood} on simulated data sets that varied in sample size $n^+=100$, $300$, $600$ (x-axis). In the first set of inference runs (dark grey), sampling differences across locations $a$, $b$ were not adjusted for, and in the second set of inferences (light grey), sampling differences were accounted for based on counts of sampled and infected individuals. The worst case error between the true transmission flows and median posterior estimates was calculated in each scenario, and the median (bar) and 2.5\% and 97.5\% quantiles (error bars) are shown. Errors decreased with increasing sample size, although errors remained very large when sampling differences were not accounted for.}
    \label{fig:wce_ode_samplesize}
\end{figure}

Figure \ref{fig:posterior_ODE_sensitivity} (grey bars) shows posterior estimates of the  transmission flows on $4$ randomly selected simulated data sets, for the case that 60\% of the population in $a$ and 35\% of the population in $b$ were sampled, and sampling heterogeneity was ignored. Figure \ref{fig:posterior_ODE_sensitivity} (orange bars) shows posterior estimates of the same transmission flows on the same $4$ randomly selected data sets, when sampling heterogeneity was accounted for as in~\eqref{supp:eq_beta_prior}. There was considerable variability in the flow data sets when sample size was low ($n^+=100$). However, information on population sampling substantially improved flow estimates regardless of sample size. Figure~\ref{fig:wce_ode_samplesize} summarises the worst case error in these simulations. When sampling heterogeneity was not accounted for, variability in flow estimates decreased with increasing sampling size, but error magnitude did not decrease. When sampling heterogeneity was accounted for, both the variability in flow estimates and error magnitude decreased with increasing sampling size.

\begin{figure}[!t]
    \centering
    \includegraphics[width=0.6\textwidth]{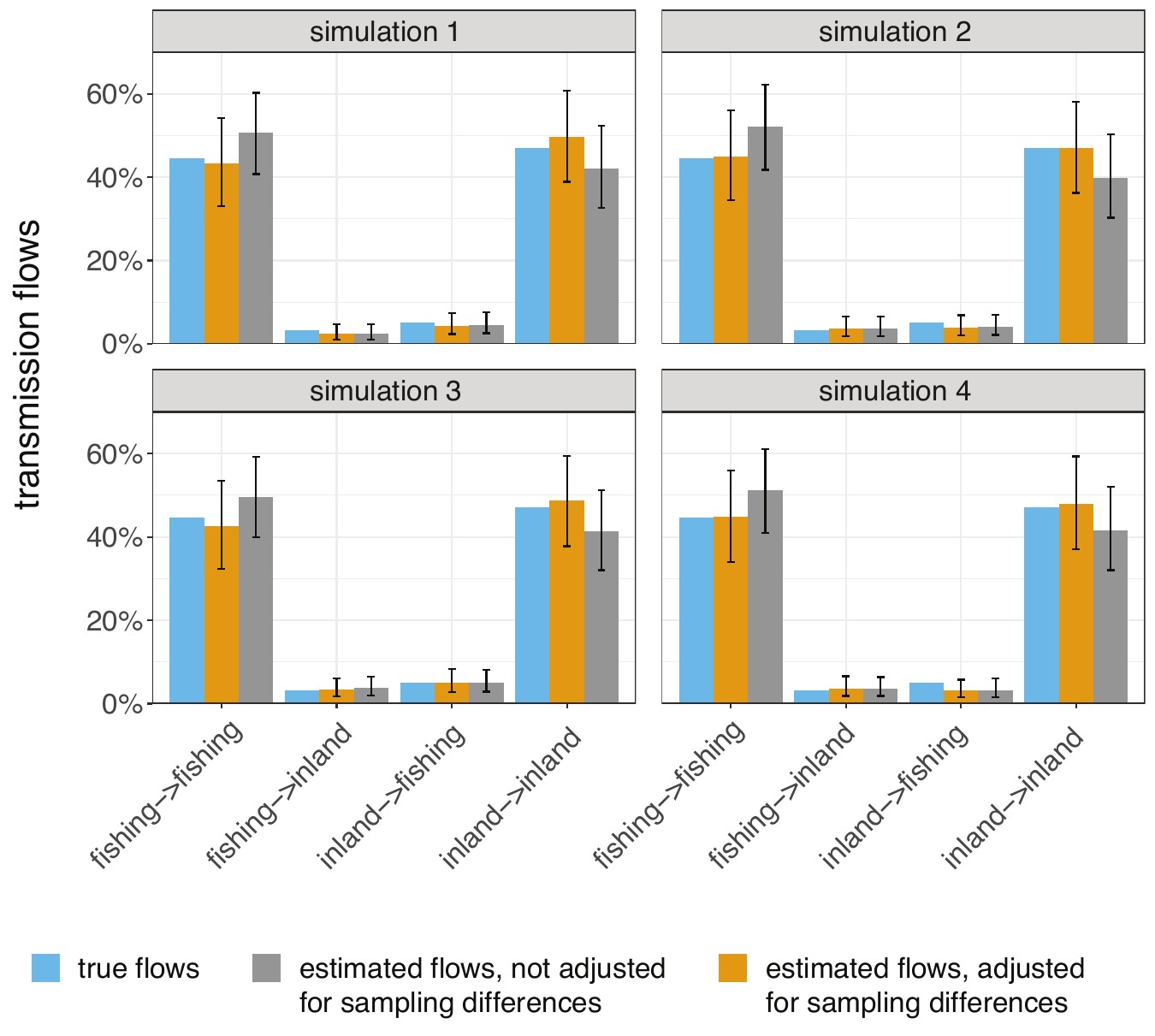}
    \caption{{\bf True and re-estimated flows in more complex simulations between 24 population groups.} We simulated transmission flows by gender, 3 age bands, and location under model parameters similar to those obtained for the Rakai analysis; see Section~\ref{sec:supp:rakai_type_simulations}. Panels show results for four simulated data sets. Posterior flow estimates (median and 95\% credibility intervals) are shown when sampling heterogeneity was ignored (grey), and when sampling heterogeneity was accounted for as in~\eqref{supp:eq_beta_prior} (orange).}
    \label{fig:rakai_simulation_flow}
\end{figure}

\begin{figure}[!t]
    \centering
    \includegraphics[width=0.6\textwidth]{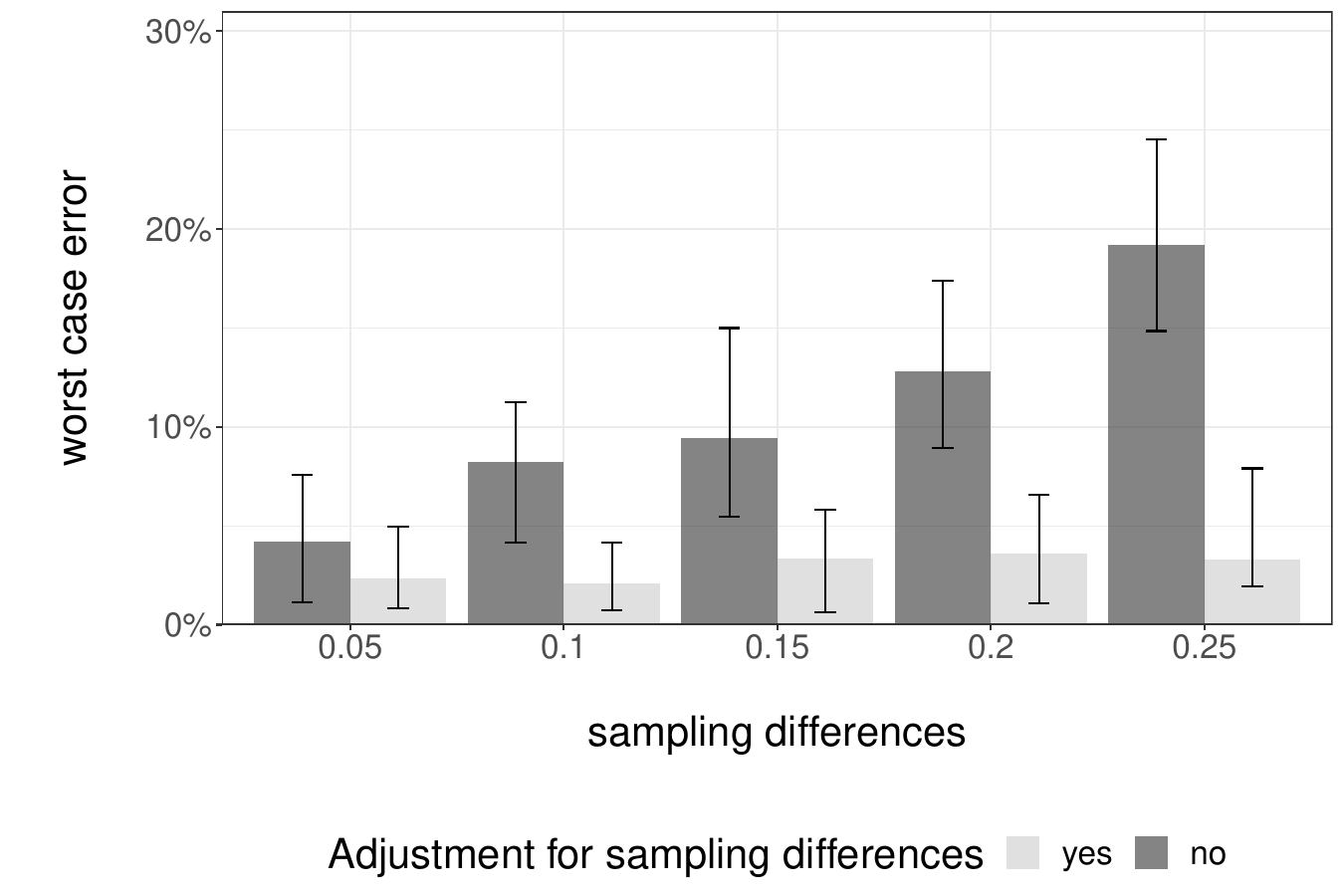}
    \caption{\textbf{Worst case error in estimated transmission flows in more complex simulations of transmission flow between 24 population groups.}
    We simulated transmission flows between men and women in 24 population groups as described in Section~\ref{sec:supp:rakai_type_simulations}. Simulation scenarios varied in average sampling differences between simulated inland and fishing communities (x-axis). Inferences were performed while ignoring for sampling heterogeneity (dark grey), and while accounting for sampling heterogeneity (light grey). In each simulation, worst case error was reported, after aggregating flow estimates to the 4 flow combination within and between inland and fishing combinations for comparison with the ODE-based simulations, and its median and 2.5\% and 97.5\% quantiles are shown (y-axis). Overall, trends in worst case error were similar those found in the more simple ODE-based simulation experiments.
    }
    \label{fig:wce:rakai}
\end{figure}

\subsection{More complex simulation experiments}\label{sec:supp:rakai_type_simulations}
The second experiment mimicked the more complex population structure and sampling heterogeneity as observed in the Rakai case study and described in Section~\ref{sec:supp:regression}. 

We considered simulated populations stratified into 24 population sub-groups, by gender (male, female), location (inland,fishing), in-migration status (in-migrant, resident), age (15-24, 25-34, 35-50 years), and simulated 576 transmission flows between the male-female and female-male sub-group combinations. For simplicity, we did not generate epidemic trajectories and instead simulated transmission counts $n_{ab}$ from pre-specified transmission flows $\boldsymbol{\pi}^0$, and pre-specified sampling probabilities $\xi_a$. For a fixed target sample size $n^+$, the total number of actual transmissions $z^+$ were simulated from $z^+ \sim \mbox{Poisson}(n^+/\overline{\xi})$, where $\overline{\xi}= \frac{1}{24} \sum_a \xi_a$ was the average sampling probability in the population. Then, the actual transmission counts between population groups were simulated from $\mathbf{z} \sim \mbox{Multinomial}(Z, \boldsymbol{\pi_0)}$. The sampling status of each individual in $a$ was simulated under a Bernoulli draw with probability $\xi_a$, and respectively in population $b$ with probability $\xi_b$, and observed transmissions $n_{ab}$ from group $a$ to group $b$ were calculated as in~\eqref{seq:actualflowcounts}. 100 such simulated data sets were generated, for a target sample size $n^+=300$. 

The difference between sampling proportions in inland and fishing communities was about 10\%. We transformed the sampling proportions with a sine function to obtain a greater range of average sampling differences between inland and fishing communities from 5\% to 25\%. Additional simulations were then generated under these sampling scenarios.

From the simulated data, we re-estimated the 576-dimensional transmission flows while ignoring sampling heterogeneity, and then while accounting for sampling heterogeneity as described in Section~\ref{sec:supp:2by2}. To facilitate comparison, we aggregated the true and re-estimated flows into the 4-dimensional vector of flows within and between inland and fishing communities, and calculated worst case errors between them. Figure \ref{fig:rakai_simulation_flow} shows posterior estimates of the transmission flows on 4 randomly selected simulated data sets, for the baseline case that the difference in sampling probabilities between inland and fishing communities was 10\%. Figure~\ref{fig:wce:rakai} shows that in these more complex simulation scenarios, trends in worst case error were overall similar compared to the more simple ODE-based simulation experiments in Section~\ref{sec:supp:2by2}. When ignoring sampling heterogeneity, worst case error increased considerably with average sampling differences between inland and fishing communities. When accounting for sampling heterogeneity, worst case error remained largely unaffected by increasing average sampling differences, and was overall slightly higher compared to that in the ODE-based simulation experiments.

\FloatBarrier

\subsection{Accuracy of HSGP approximation}\label{sec:hsgp_approx}
We finally assessed the accuracy of HSGP approximation in the context for heterogenous sampling. The main results are summarised in Section \ref{sec:performance_hsgp}. This supplementary material provides more details of this simulation experiment and visualises the performance of the final model specification. We considered transmission flows between one-year-increment age groups between 15 and 24 by gender and locations and simulated transmission intensities and counts via a slight extension of Equation \eqref{eq:gp},
\begin{linenomath*}
\begin{equation}\label{eq:ext_gp}
\begin{split}
    & \boldsymbol{n} \sim \text{Poisson}(\boldsymbol{\lambda} \boldsymbol{\xi}^T \boldsymbol{\xi}^R)\\
    & \log \boldsymbol{\lambda} = \mu_{mf,hh}\mathds{1}_{mf,hh} + \mu_{mf,hl}\mathds{1}_{mf,hl} + \mu_{mf,lh}\mathds{1}_{mf,lh} + \mu_{mf,ll}\mathds{1}_{mf,ll} \\
    & \quad + \mu_{fm,hh}\mathds{1}_{fm,hh} + \mu_{fm,hl}\mathds{1}_{fm,hl} + \mu_{fm,lh}\mathds{1}_{fm,lh} + \mu_{fm,ll}\mathds{1}_{fm,ll} + \boldsymbol{f},\\
    & \boldsymbol{f}=(\boldsymbol{f}^T_{mf}, \boldsymbol{f}^T_{fm})^T,\\  
    & \boldsymbol{f}_{mf}\sim \mathcal{GP}(0, k_{mf}),\quad \boldsymbol{f}_{fm}\sim \mathcal{GP}(0, k_{fm}),
    \\
    & k_{mf}\big( (a_1,b_1), (a_2,b_2) \big)= \sigma_{mf}^2 \exp\Big( - \Big[ \frac{(a_2-a_1)^2}{2\ell_{mf,a}^2} + \frac{(b_2-b_1)^2}{2\ell_{mf,b}^2} \Big] \Big)\\
    & k_{fm}\big( (a_1,b_1), (a_2,b_2) \big)= \sigma_{fm}^2 \exp\Big( - \Big[ \frac{(a_2-a_1)^2}{2\ell_{fm,a}^2} + \frac{(b_2-b_1)^2}{2\ell_{fm,b}^2} \Big] \Big),
\end{split}    
\end{equation}
\end{linenomath*}
\noindent
under the hyper-parameter values in Table~\ref{tab:parameters:hsgp}.
\begin{table}
     \caption{Parameter values used to simulate transmission flows in populations stratified by 1-year age bands under the extension of GP model~\eqref{eq:gp}. \label{tab:parameters:hsgp}} 
  \begin{tabular}{lll}
    \toprule
    {\bf Parameter} & {\bf Symbol} & {\bf Value} \\[1mm]
    \midrule
    intercept & & \\[1mm]
    female to male & & \\[1mm]
    \quad h->h & $\mu_{fm,hh}$ & -1\\[1mm]
    \quad h->l & $\mu_{fm,hl}$ & -10\\[1mm]
    \quad l->h & $\mu_{fm,lh}$ & -9\\[1mm]
    \quad l->l & $\mu_{fm,ll}$ & -2.5\\[1mm]
    \midrule
    intercept & & \\[1mm]
    male to female & & \\[1mm]
    \quad h->h & $\mu_{mf,hh}$ & -0.5\\[1mm]
    \quad h->l & $\mu_{mf,hl}$ & -9\\[1mm]
    \quad l->h & $\mu_{mf,lh}$ & -9\\[1mm]
    \quad l->l & $\mu_{mf,ll}$ & -1\\[1mm]
    \midrule
    lengthscale &  & \\[1mm]
    female to male & $\boldsymbol{\ell_{fm}} $& (4.1,2.3)\\[1mm]
    male to female & $\boldsymbol{\ell_{mf}} $& (2.3,4.6)\\[1mm]
    \midrule
     marginal standard deviation & &\\[1mm]
    female to male & $\sigma_{fm} $& 1.8 \\[1mm]
    male to female & $\sigma_{mf} $& 1.5\\[1mm]
    \bottomrule
   \end{tabular}
\end{table}

This extension allows to capture average transmission intensities by locations and genders and varying age-dependent transmission dynamics from male to female and from female to male. Meanwhile, samples from models for sampling probability (Supplementary Material \ref{sec:supp:regression}) were reused and their means were taken as ground-truth sampling intensities. 20 replicates of transmission intensities and counts were generated.

Flows and hyperparameters were re-estimated using both Gaussian Process model and Hilbert Space Gaussian Process model based on observed counts in the simulation, under priors
\begin{linenomath*}
\begin{equation*}
\begin{split}
    & \sigma_{mf}^2, \sigma_{fm}^2 \sim \text{Half-Normal}(0,10)
    \\
    & \ell_{d,i} \sim \text{Inv-Gamma}(\alpha_{d,i}, \beta_{d,i}) \\
    & \mu_{mf,hh}, \mu_{mf,hl}, \mu_{mf,lh}, \mu_{mf,ll},\mu_{fm,hh}, \mu_{fm,hl}, \mu_{fm,lh}, \mu_{fm,ll} \sim \text{Normal}(0,10).
\end{split}   
\end{equation*}
\end{linenomath*}

\begin{figure}[!t]
    \centering
    \includegraphics[width=0.85\textwidth]{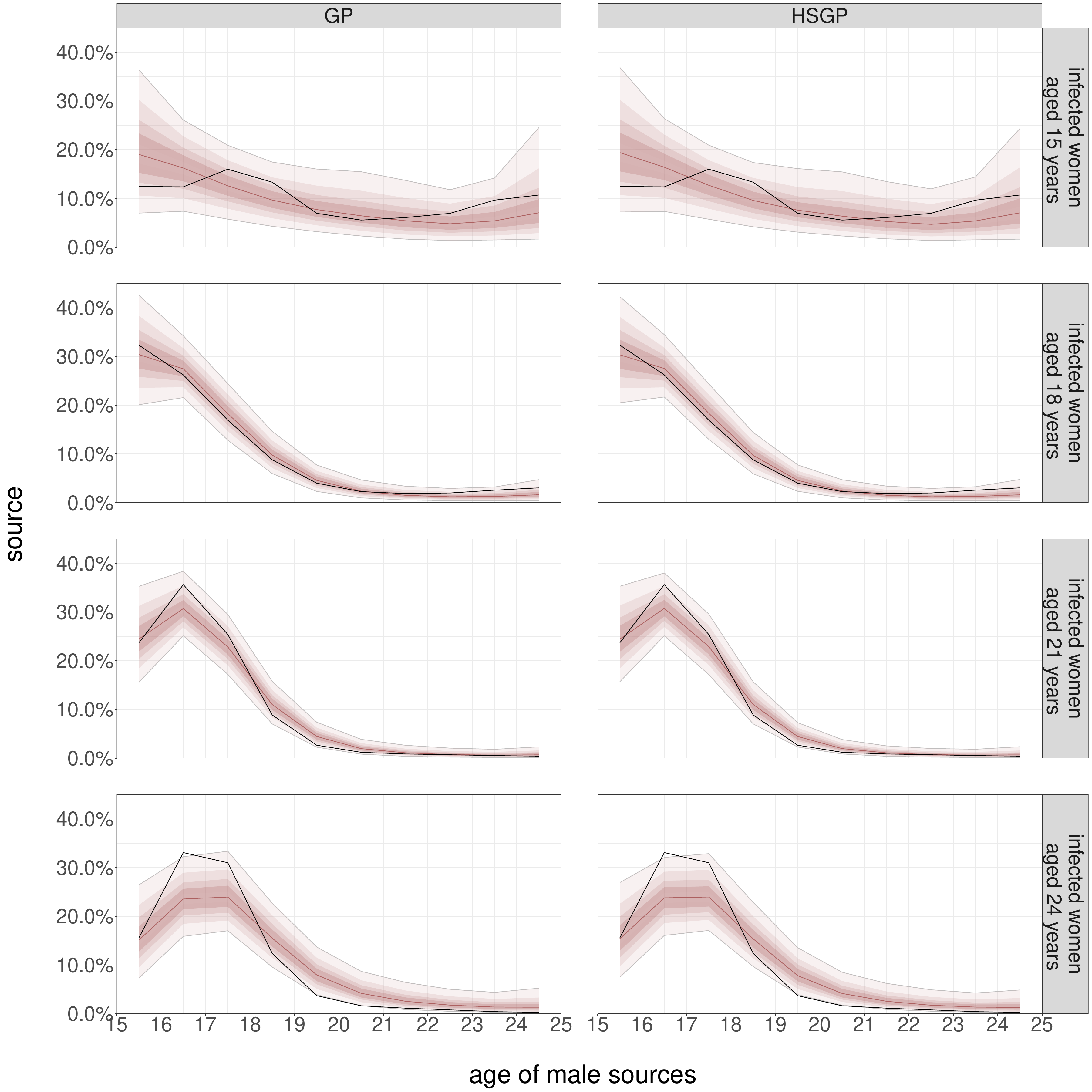}
    \caption{{\bf Estimated sources of infection in women under the final HSGP approximation}. We simulated age-specific transmission flows under the kernel parameters shown in Table~\ref{tab:parameters:hsgp}, and re-estimated transmission flows along with all other parameters under HSGP approximations to~\eqref{eq:gp}. The subfigures on the left illustrate the estimated sources of infection in women under the GP prior from which the data were simulated. The subfigures on the right illustrate the estimated sources under the final HSGP approximation. Shades illustrate the estimated posterior credibility intervals, showing 40\% credibility intervals in dark red to 80\% credibility intervals in light red. Posterior medians are shown as a red line. True values are shown in black. We found no systematic differences in source estimates when using the final HSGP approximation over the GP prior.}
    \label{fig:hsgp_sources}
\end{figure}

\begin{figure}[!t]
    \centering
    \includegraphics[width=0.85\textwidth]{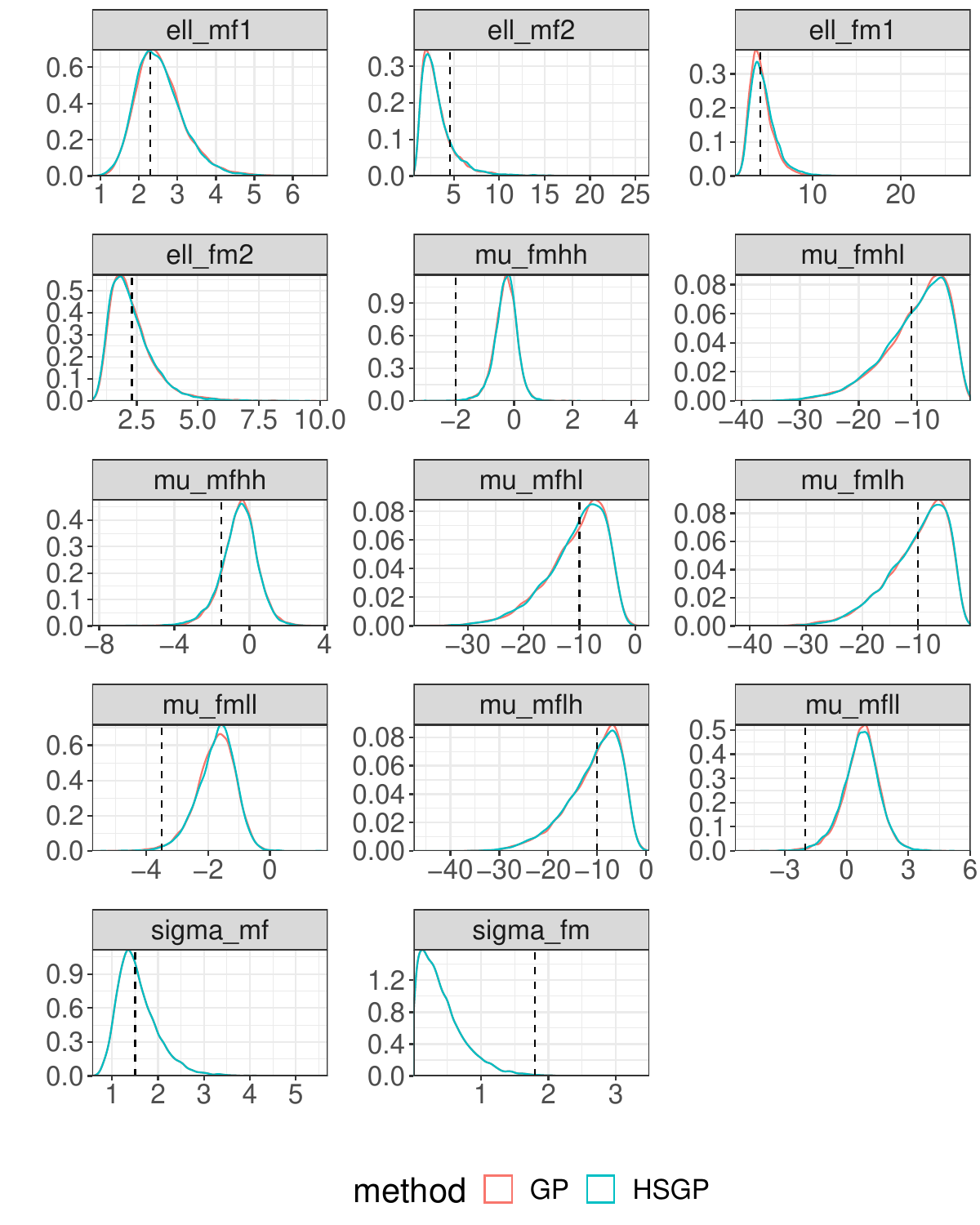}
    \caption{{\bf Estimated covariance kernel parameters under the final HSGP approximation}. We simulated age-specific transmission flows under the kernel parameters shown in Table~\ref{tab:parameters:hsgp}, and re-estimated transmission flows along with all other parameters under HSGP approximations to~\eqref{eq:gp}. The subfigures illustrate the marginal posterior densities of the kernel parameters for under the GP model (in red) and the HSGP approximation (in blue) based on $B=1.25, m=30$ (right panels).}
    \label{fig:finalhgsp_hyperparameters}
\end{figure}


\noindent
Here, we illustrate further the performance of the HSGP approximation under the final tuning parameters, $B=1.25, m=30$. Figure~\ref{fig:hsgp_sources} compares the marginal posterior distribution of the sources of infections in women when using the chosen HSGP approximation, compared to using the GP prior. We found no particular differences in the estimated sources of infection when using the  HSGP approximation over the GP prior. Figure~\ref{fig:finalhgsp_hyperparameters} compare the marginal posterior densities of the kernel parameters between the chosen HSGP approximation and the GP prior. Again, we found no systematic differences in the estimated kernel parameters when using the chosen HSGP approximation compared to the GP prior.

\FloatBarrier

\section{Modelling and estimation of the sampling cascade}\label{sec:supp:regression}
We here describe the statistical models used to characterise each step of the sampling cascade of transmission events, shown in Figure~\ref{fig:samplingcascade}A. 

To recall our setting, we defined in the main text transmission events as HIV infection events to individuals who lived in RCCS communities, were aged 15-49 years, and were infected in the study period $\mathcal{T}=\{2009/10/1, 2015/01/30\}$. The main study objective was to estimate transmission flows and related quantities~(\ref{eq:highlowflow}-\ref{eq:sourcesrecipientsratios}) among different population groups, expressed in proportions relative to this denominator of transmission events, denoted by $\mathcal{Z}$. Each transmission event in $\mathcal{Z}$ can be indexed by its source $i$ and recipient $j$ in the population of individuals that were infected by the end of $\mathcal{T}$, and we denote transmission from $i$ to $j$ with the indicator variable $z_{ij}=1$, and no transmission with $z_{ij}=0$. Thus, recipients are individuals in $\mathcal{P}$ that were infected during the study period $\mathcal{T}$, i.e. in our case 2009/10/1-2015/01/30. Sources are infected individuals who transmitted to one of the recipients. In our applications, population groups were defined by location (inland and fishing communities), gender, and 1-year age bands from age 15 to age 49, yielding $A=140$ population strata. The actual transmission events from group $a$ to group $b$ are 
\begin{linenomath*}
\begin{equation}\label{seq:actualflowcounts}
    z_{ab}= \sum_{i\in a, j\in b} z_{ij},
\end{equation}
\end{linenomath*}
and observing any such event depends on the sampling status of the source and recipient,
\begin{linenomath*}
\begin{equation}\label{seq:obsflowcounts}
    n_{ab}= \sum_{i\in a, j\in b} \Indicator{s^S_i=1}\Indicator{s^R_j=1} z_{ij}.
\end{equation}
\end{linenomath*}
To make inference on transmission flows, our strategy is to explicitly account for sampling heterogeneity in order to reduce bias in the flow estimates. We propose estimating sampling proportions and then propagate these estimates into inference of transmission flows. Overall, we assume that sources and recipients are independently sampled, and conditionally at random within each population sub-group $a$ with probabilities $\boldsymbol{\xi}^S= (\xi^S_a)_{a\in\mathcal{A}}$ and $\boldsymbol{\xi}^R= (\xi^R_a)_{a\in\mathcal{A}}$. For sake of brevity, we derived the flow model in equations  (\ref{eq:comp_likelihood}-\ref{eq:postdist1}) using the same sampling probabilities for sources and recipients. In our applications we consider its extension to distinct sampling probabilities for sources and recipients, with equations generalising straightforwardly. 

We modelled the overall sampling probabilities $\xi^S_a$, $\xi^R_a$ as the product of conditional sampling probabilities 
along the sampling cascade of Figure~\ref{fig:samplingcascade}A, see the forthcoming Equations \eqref{seq:sampling_of_sources} and \eqref{seq:sampling_of_recipients}.

The sampling cascade involved that sources and recipients participate in the cohort, report no ART use, and that virus was successfully deep-sequenced. Section~\ref{sec:part_prob_sources} describes estimation of participation probabilities for source cases, and Section~\ref{sec:reg:trseq} describes estimation of sequence sampling probabilities conditional on participation among source cases. Section~\ref{sec:part_prob_recipients} describes estimation of participation probabilities for recipient cases, and Section~\ref{sec:seq_prob_recipients} describes estimation of sequence sampling probabilities conditional on participation among recipient cases. Section~\ref{sec:pw_sampling_probs} illustrates the resulting sampling probabilities of transmission events.

In each step of the cascade, we compared the prediction accuracy of candidate models through the ten-fold cross validation. To do it, the data $\mathbf{y} \in \mathbb{R}^N$ is divided into 10 sets, the $k$th test set is denoted as $\mathbf{y}_k$ and the rest of observations forms the training set $\mathbf{y}_{-k}$. Here we defined accuracy measures.
\begin{itemize}
    \item Hold-out data in 95\% PPCrI (Posterior Predictive Credibility Interval): we simulated data (sampled individuals) from the fitted model on $\mathbf{y}_{-k}$, found the 95\% credibility intervals of the simulated data, and counted the percentage of observed data $\mathbf{y}_{k}$ included in the credibility interval.
    \item MAE (Mean Absolute Error): we simulated data (sampled individuals) from the fitted model on $\mathbf{y}_{-k}$, found the median estimates, and calculated the absolute differences of the observations $\mathbf{y}_{k}$ and the median estimates.
    \item ELPD (expected log point-wise predictive density for a new dataset): this evaluates the log predictive density at each data point in the $k$th fold,  
    \begin{equation}
    \text{ELPD}_k = \sum_{i=1}^{N_k} \text{ELPD}_{ki} = \sum_{i=1}^{N_k}  \log(\frac{1}{S} \sum_{s=1}^S p(\mathbf{y}_{ki}|\theta^{k,s})),
    \end{equation}
where $N_k$ is the number of data in the $k$th fold, $\mathbf{y}_{ki}$ denotes the $i$th observation in the $k$th test set, and $\theta^{k,s}$ denote the $s$th samples drawn from $p(\theta|\mathbf{y}_{k})$.
\end{itemize}

\subsection{Modelling and estimating participation probabilities of source cases} \label{sec:part_prob_sources}

We are interested in participation probabilities among infected individuals who transmitted virus. In general, HIV infections in surveyed locations may have their source outside these locations, with previous work suggesting that approximately 30\% of infections in RCCS communities originate from outside the cohort \citep{grabowski2014role}. Sources from outside the RCCS communities by definition did not participate in any survey round, and in the absence of any data we focused on sources from RCCS communities. This is a general limitation of our flow inferences.

As part of demographic surveillance, the RCCS conducts a population census immediately prior to each survey round in the cohort communities, which provides an  enumeration of the underlying age-eligible population of 15-49 years. We used these data to estimate participation probabilities $\xi^p_a$ over the survey period in each of the underlying population sub-groups $a$, and thereby approximated the participation probabilities of sources cases. This involved three assumptions. First, we assumed that individuals participated regardless of infection status, and second regardless of being a source or not, and third that participation probabilities were the same in the survey period (2011/08/10-2015/01/30) and in the slightly longer study period (2009/10/01-2015/01/30), during which transmission flows were estimated. 

\begin{figure}[!t]
    \centering
    \includegraphics[width=0.95\textwidth]{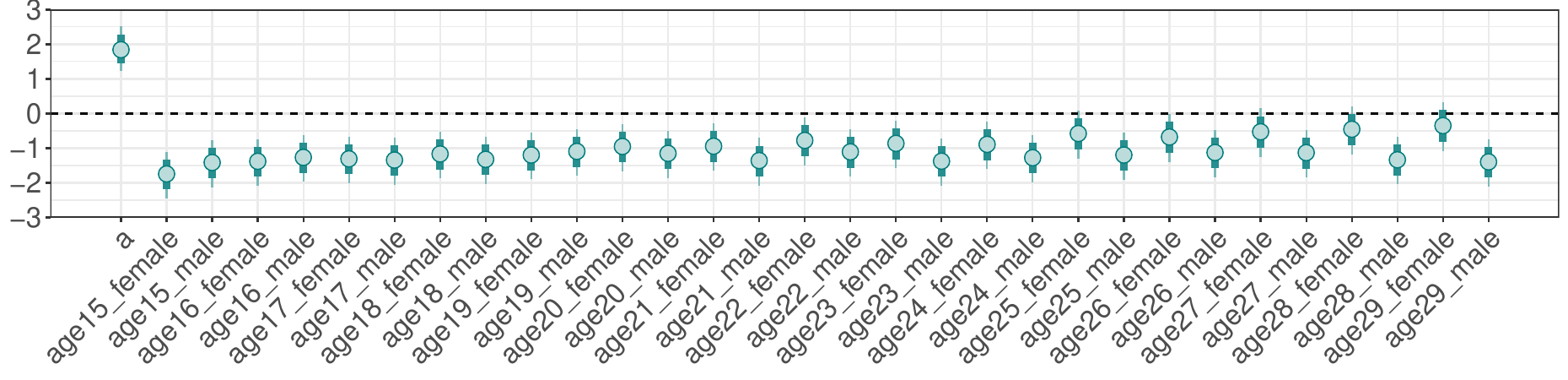} \\
     \includegraphics[width=0.95\textwidth]{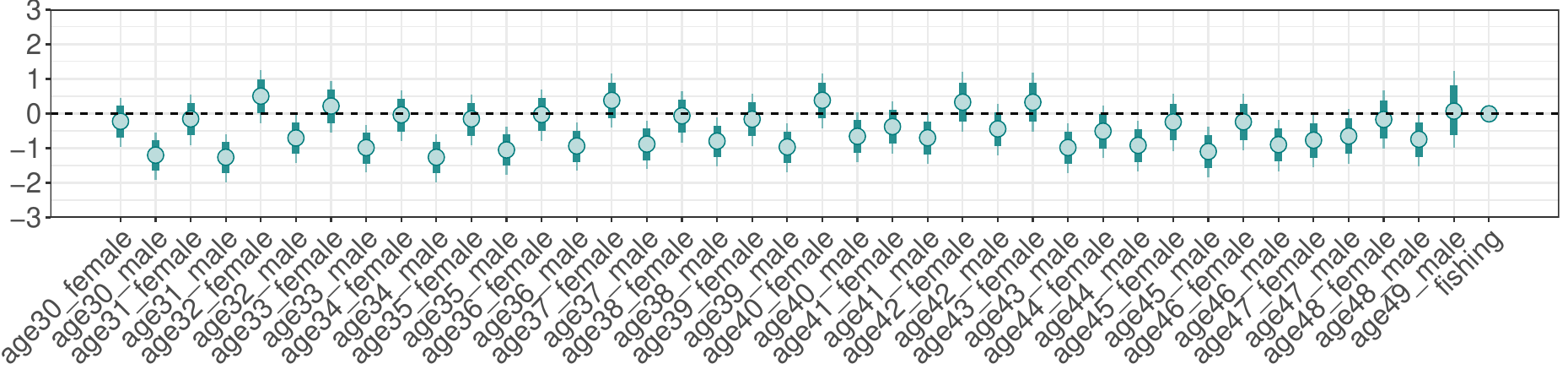}
    \caption{\textbf{Marginal posterior densities of regression coefficients of model P4 to estimate population participation probabilities.} Circle: median; bar: interquartile range; line: 95\% credibility interval. The first variable $a$ refers to the intercept.}
    \label{fig:participation_posci}
\end{figure}

\begin{figure}[!t]
    \centering
    \includegraphics[width=0.85\textwidth]{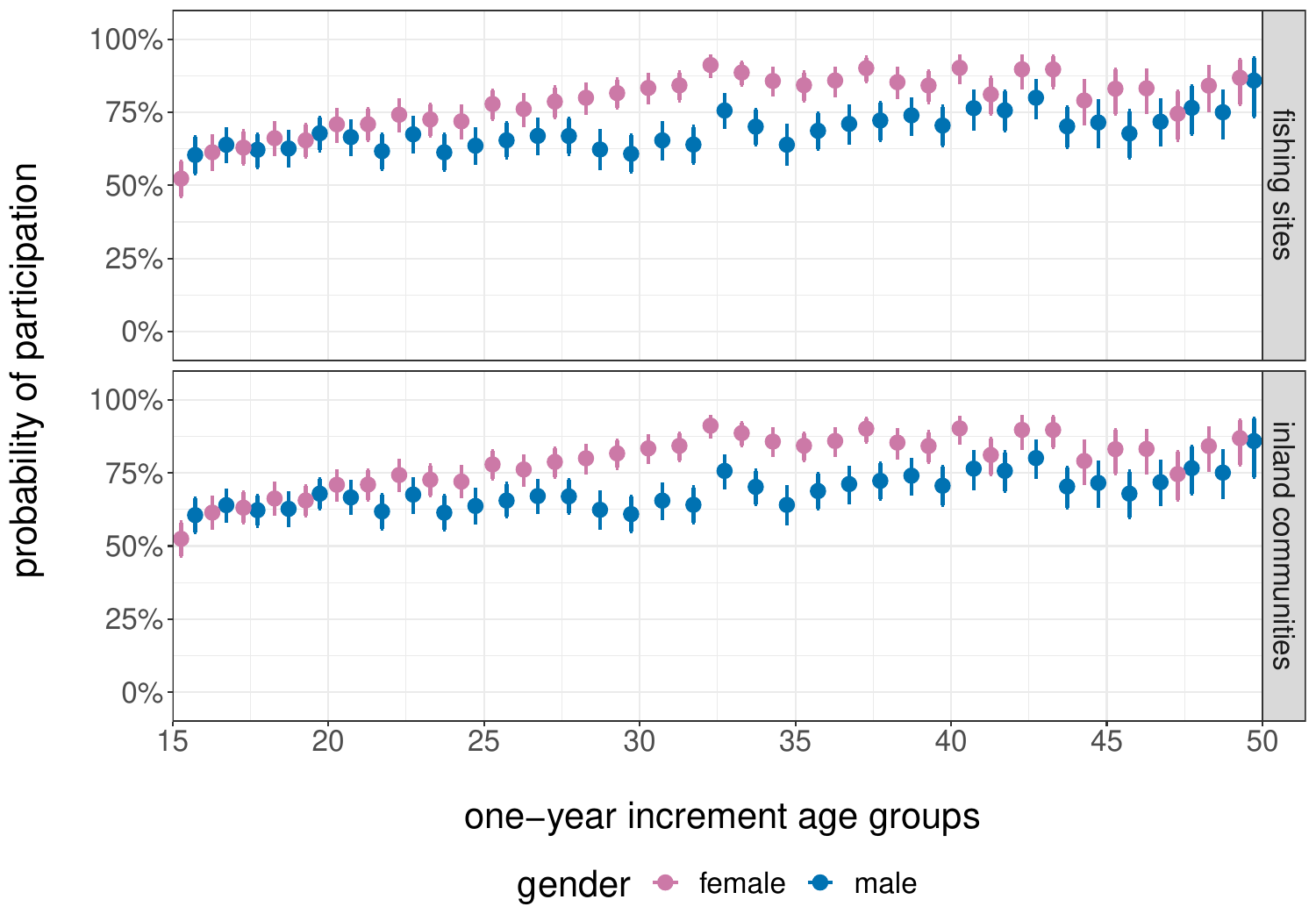}
    \caption{\textbf{Marginal posterior densities of population participation probabilities under model P4.} Circle: median; line: 95\% credibility interval.}
    \label{fig:participation_rate}
\end{figure}

We used a Beta-Binomial regression framework to estimate the posterior distribution of population participation probabilities $\boldsymbol{\xi}^p=(\xi^p_a)_{a\in\mathcal{A}}$. Denote by $N_a^e$ the number of participation-eligible individuals in population group $a$, by $N_a^p$ the number of participants in population group $a$, by $\mathbf{X}_a \in \{0,1\}^p$ a row vector of $p$ predictor variables associated with population $a$, including location status, gender, age band, and possible interaction terms, and by $\mathbf{X} \in \{0,1\}^{A\times p}$ the design matrix that stacks the row vectors $\mathbf{X}_a$ for each population group $a$. The general form of the regression models considered was 
\begin{linenomath*}
\begin{equation}\label{eq:partregression}
    \begin{split}
    & N_a^p \sim \mbox{Beta-Binomial}(N_a^e,\xi_a^p,\gamma), \:\forall a \\
    & \mbox{logit}(\boldsymbol{\xi^p}) = \beta_0\boldsymbol{1} + \boldsymbol{\beta X} \\
    & \beta_0 \sim \mathcal{N}(0,100) \\
    & \boldsymbol{\beta} \sim \mathcal{N}(0,10\boldsymbol{I})\\
    & \gamma \sim \mbox{Exp}(1)
    \end{split}
\end{equation}
\end{linenomath*}
where the Beta-Binomial is parameterised in terms of the number of Bernoulli trials, the mean of the marginal probabilities of successes and the dispersion parameter, $\beta_0\in\mathbb{R}$ is the baseline participation probability, $\boldsymbol{\beta}\in\mathbb{R}^p$ is the vector of regression coefficients, and $\boldsymbol{I}$ is the $p\times p$ identity matrix. Four models were considered and assessed using 10-fold cross-validation.
\begin{itemize}
    \item Model P1. Model~\eqref{eq:partregression} with indicator variables on location, gender, and 1-year age bands as contrasts, leading to $p=36$ predictors. The corresponding regression coefficients were given independent normal prior densities, and the dispersion coefficient was set to $\gamma=0$ to reflect no overdispersion.
\item Model P2. As model P1 but with $\gamma$ estimated to allow for overdispersion.
\item Model P3. Model~\eqref{eq:partregression} with indicator variables on location, and gender and 1-year age band interactions as contrasts, leading to $p=70$ predictors. The corresponding regression coefficients were given independent normal prior densities, and the dispersion coefficient was set to $\gamma=0$ to reflect no overdispersion. 
\item Model P4. As model P3 but with $\gamma$ estimated to allow for overdispersion.
\end{itemize}
The models were fitted in Stan version 2.21, convergence diagnostics \citep{vehtari2019rank} were checked, and effective sample sizes were $\gg 10000$. Table~\ref{table:participation_postcheck} shows the proportion of hold-out data in 95\% posterior predictive credibility intervals (PPCrI), the mean absolute error (MAE), and the expected log predictive density in 10-fold cross-validation (ELPD) for each model. Based on these statistics, model P4 was chosen to model participation probabilities across population strata. Figure \ref{fig:participation_posci} summarises the marginal posterior densities of the regression coefficients under model P4. Figure~\ref{fig:participation_rate} shows the marginal posterior participation probabilities by location, gender, and age-bands.

\begin{table}
\caption{\label{table:participation_postcheck} Prediction accuracy of models P1-P4 on hold-out  population participation data in 10-fold cross-validation.}
\begin{threeparttable}
\begin{tabular}{l l l l}
\hline\hline
{\bf Model} &  {\bf Hold-out data} & {\bf MAE}  & {\bf ELPD} \\[0.5ex] 
& {\bf in 95\% PPCrI} & & \\[0.5ex] 
& median (range)  & median (range) &   median (std dev)\\
\hline
P1  &   93.1\%  (91.5\% - 95.1\%)    &  1.79  (1.50 - 1.92)  &   -3705.27 ( 26.78 ) \\
P2  &  98.0\%  (97.2\% - 99.2\%)   &1.76  (1.49 - 1.98) &  -4168.58 ( 37.05 )  \\
P3  &  93.1\%  (92.4\% -  95.3\%)   &  1.67  (1.58 - 1.90) & -3651.39 ( 27.61 )\\
P4  & 98.4\%  (96.5\% - 98.8\%)   &  1.70  (1.53 - 1.90)   & -4072.42 ( 37.21 ) \\
\hline
\end{tabular}
\begin{tablenotes}
   \item[*] PPCrI is posterior predictive credibility interval, MAE is abbreviated from mean absolute error, ELPD is the abbreviation of expected log predictive density
  \end{tablenotes}
  \end{threeparttable}
\end{table}



\FloatBarrier

\subsection{Modelling and estimating sequencing probabilities of source cases}\label{sec:reg:trseq}
We are now interested in the probability that (participating) source cases report no ART use, and the probability that source cases and reported no ART use also had virus successfully deep-sequenced. We combined both steps and sought to estimate the probability that source cases had virus successfully deep-sequenced. Considerable demographic and clinical data were collected among RCCS participants, which allowed us to relax the assumptions of Section~\ref{sec:part_prob_sources} when modelling and estimating sequencing probabilities of source cases. In general, many infected individuals do not transmit the virus onwards, and individuals who transmitted virus in the study period $\mathcal{T}$ must have had unsuppressed HIV at some point in $\mathcal{T}$. A routinely collected proxy variable of viral suppression is self-reported ART use \citep{grabowski2018validity}. On this basis, we excluded from consideration of potential source cases infected individuals who consistently reported ART use, defined as reporting ART use at each survey visit, and approximated the probability of obtaining a deep-sequence sample from source cases who participated in the cohort by the probability of obtaining a deep-sequence sample from infected participants who reported no ART use on at least one visit during the survey period. In this approximation, we further assumed that individuals who consistently reported ART use during the survey period (2011/08/10-2015/01/30) would also have consistently reported ART use during the study period (2009/10/01-2015/01/30). 

\begin{table}
\caption{Prediction accuracy of models S1-S8 on hold-out deep-sequencing data in 10-fold cross-validation. \label{table:sequence_source_postcheck}}
\begin{threeparttable}
\begin{tabular}{l l l l}
\hline\hline
{\bf Model} &  {\bf Hold-out data} & {\bf MAE}  & {\bf ELPD} \\[0.5ex] 
& {\bf in 95\% PPCrI} & & \\[0.5ex] 
& median (range)  & median (range) &   median (std dev)\\[1mm]
\hline
S1 & 99.2\%  (98.6\% - 100.0\%)   &  0.60  (0.55 - 0.65) & -1082.79 ( 12.99 )\\[1mm]
S2 &  100\%  (99.2\% -  100\%)  &   0.58  (0.54 - 0.65) & -1130.18 ( 15.91 ) \\[1mm]
S3 &  99.2\%  (98.4\% - 100.0\%)   & 0.60  (0.53 - 0.70)  & -1057.71 ( 13.1 ) \\[1mm]
S4 &  99.2\%  (99.2\% - 100.0\%)  &  0.61  (0.56 - 0.70) & -1097.95 ( 15.9 ) \\[1mm]
S5 & 99.2\%  (99.2\% - 100\%) &   0.58  (0.56 - 0.61)   & -1106.56 ( 12.79 )\\[1mm]
S6 &   100\%  (99.2\% - 100\%) &  0.57  (0.55 - 0.62)& -1159.65 ( 15.77 )\\[1mm]
S7 & 99.2\%  (98.4\% - 100\%) &   0.58  (0.56 - 0.61)   &  -1118.38 ( 12.58 )\\[1mm]
S8 &   100\%  (99.2\% - 100\%) &   0.58  (0.55 - 0.61)  &  -1174.11 ( 15.6 ) \\[1mm]
\hline
\end{tabular}
\begin{tablenotes}
   \item[*] PPCrI is posterior predictive credibility interval, MAE is abbreviated from mean absolute error, ELPD is the abbreviation of expected log predictive density
 \end{tablenotes}
 \end{threeparttable}
\end{table}

\begin{figure}[!t]
    \centering
    \includegraphics[width=0.95\textwidth]{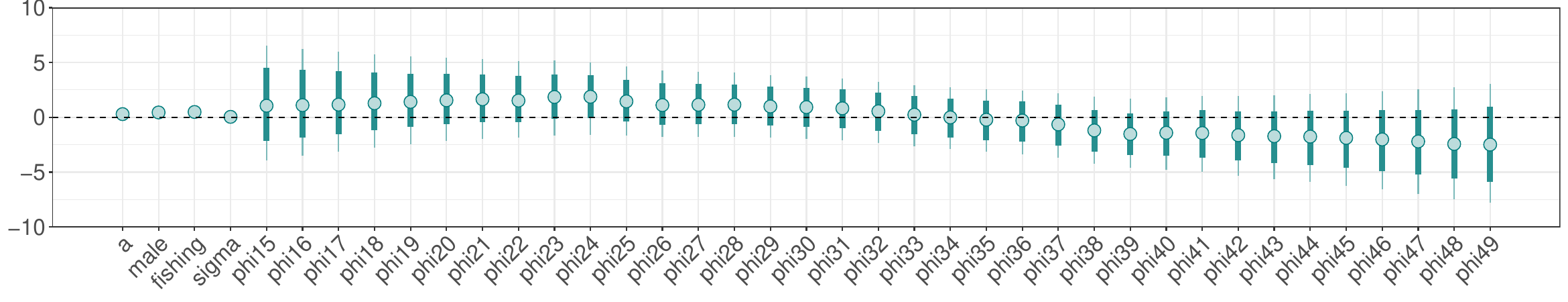}
    \caption{\textbf{Marginal posterior densities of regression coefficients of model S5 to estimate conditional deep-sequencing probabilities.} Circle: median; bar: interquartile range; line: 95\% credibility interval. The first variable $a$ refers to the intercept.}
    \label{fig:sequencing_source_posci}
\end{figure}

\begin{figure}[!t]
    \centering
    \includegraphics[width=0.95\textwidth]{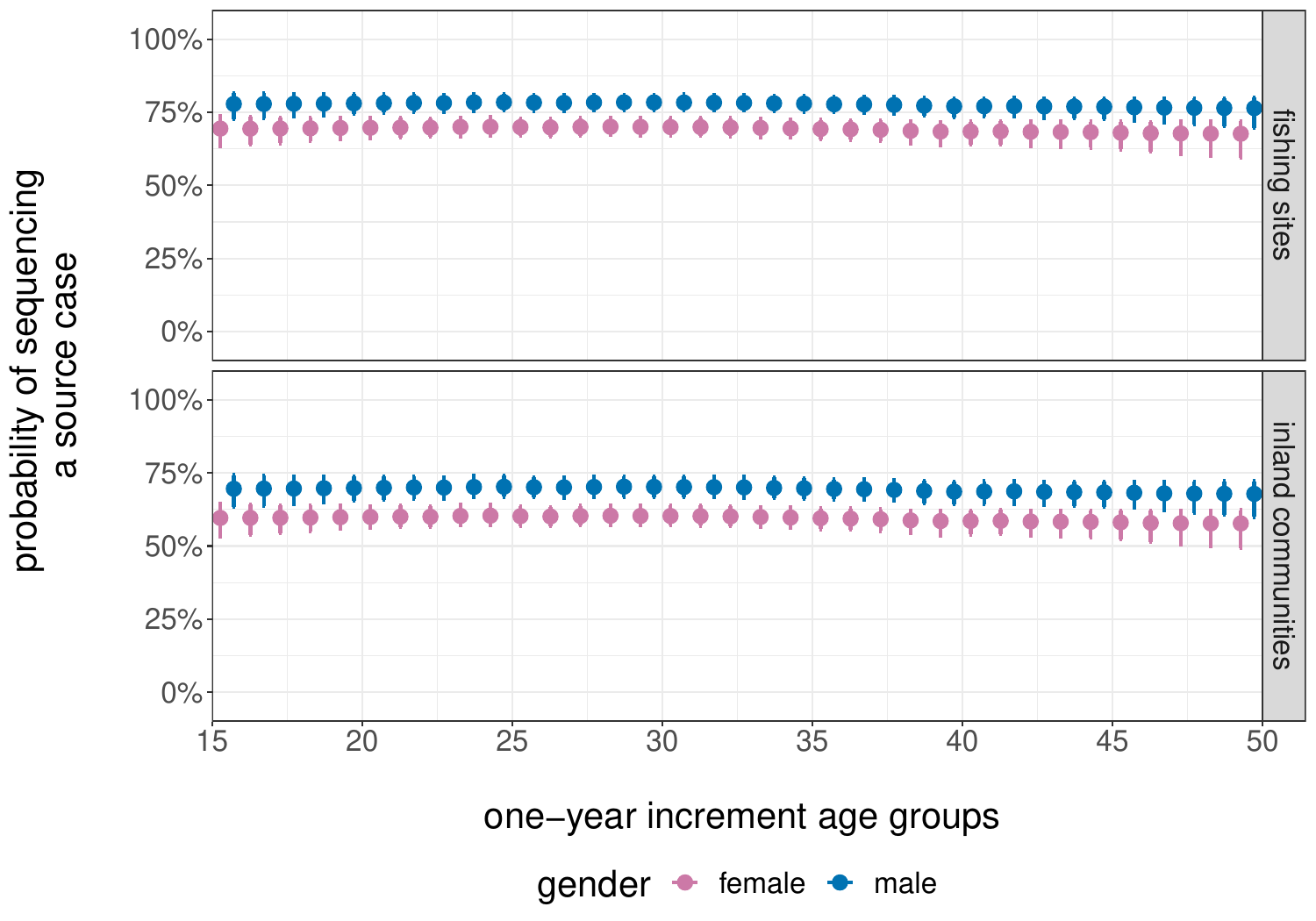}
    \caption{\textbf{Marginal posterior densities of source sequencing probabilities under model S5.} Circle: median; line: 95\% credibility interval.}
    \label{fig:sequence_source_rate}
\end{figure}

We again used a Beta-Binomial regression framework to estimate the posterior distribution of deep-sequencing probabilities $\boldsymbol{\xi}^s=(\xi^s_a)_{a\in\mathcal{A}}$, where  $\xi^s_a$ denotes the probability of obtaining a deep-sequence sample from potential source cases, defined as infected participants who reported no ART use on at least one visit during the survey period in population sub-group $a$. However due to smaller numbers we also considered regularised regression models with correlations imposed among the regression coefficients. We follow the notation of \eqref{eq:partregression} and further denote by $N^{\text{na\"{\i}ve}}_a$ the number of infected participants in $a$ who did not report consistent ART use, and by $N^{s}_a$ the number of infected participants in $a$ who did not report consistent ART use and had virus deep-sequenced successfully. We considered models of the form
\begin{linenomath*}
\begin{equation}\label{eq:seq_source_regression}
    \begin{split}
    & N_a^{s} \sim \mbox{Beta-Binomial}(N^{\text{na\"{\i}ve}}_a,\xi^s_a,\gamma), \: \forall a \\
    & \mbox{logit}(\boldsymbol{\xi^{s}}) = \beta_0 \boldsymbol{1} + \boldsymbol{\beta X} \\
    & \beta_0 \sim \mathcal{N}(0,100) \\
    & \boldsymbol{\beta} \sim \mathcal{N}(0,10\boldsymbol{I})\\
    & \gamma \sim \mbox{Exp}(1),
    \end{split}
\end{equation}
\end{linenomath*}
and similar models with 1st-order intrinsic conditional auto-regressive (ICAR) prior distributions across adjacent age bands \citep{rue2005gaussian}. 
Eight models were considered and assessed using 10-fold cross-validation.
\begin{itemize}
    \item Model S1. Model~\eqref{eq:seq_source_regression} with indicator variables on location (1), gender (1), and 1-year age bands (34) as contrasts, leading to $p=36$ predictors. The corresponding regression coefficients were given independent normal prior densities, and the dispersion coefficient was set to $\gamma=0$ to reflect no overdispersion.
\item Model S2. As model S1 but with $\gamma$ estimated to allow for overdispersion.
\item Model S3. Model~\eqref{eq:seq_source_regression} with indicator variables on location (1), and gender and 1-year age band interactions (69) as contrasts, leading to $p=70$ predictors. The corresponding regression coefficients were given in independent normal prior densities, and the dispersion coefficient was set to $\gamma=0$ to reflect no overdispersion. 
\item Model S4. As model S3 but with $\gamma$ estimated to allow for overdispersion.
\item Models S5-S8. As models S1-S4 respectively, but using an ICAR prior density on adjacent age bands.
\end{itemize}
The models were fitted in Stan version 2.21, convergence diagnostics \citep{vehtari2019rank} were checked, and effective sample sizes were again $\gg 10000$. In Table~\ref{table:sequence_source_postcheck} we report as before the proportion of hold-out data in 95\% PPCrI, the MAE, and the ELPD for each model in 10-fold cross-validations. Based on these figures, model S5 was chosen to model deep-sequencing probabilities (conditional on participation) across population strata. Figure \ref{fig:participation_posci} shows the marginal posterior densities of the regression coefficients under model S5. Figure~\ref{fig:sequence_source_rate} summarises the marginal posterior deep-sequencing probabilities by location, gender, and age-bands.

To summarise, the sampling probability of source cases was approximated by
\begin{linenomath*}
\begin{equation}\label{seq:sampling_of_sources}
    \xi^S_a = \xi^p_a * \xi^s_a
\end{equation}
\end{linenomath*}
for all population sub-groups $a$.

\FloatBarrier

\subsection{Modelling and estimating participation probabilities of recipient cases}\label{sec:part_prob_recipients}
We are now interested in participation probabilities among recipients, i.e. individuals who acquired infection in the study period $\mathcal{T}$, 2009/10/01-2015/01/30. We proceeded as in Section~\ref{sec:part_prob_sources} and used the RCCS census data to approximate the participation probabilities among individuals who acquired HIV in $\mathcal{T}$ by participation probabilities over the survey period in each of the underlying population sub-groups. This involved two assumptions. First, we assumed that individuals participated regardless of infection status, and second that participation probabilities were the same in the survey period (2011/08/10-2015/01/30) and in the slightly longer study period (2009/10/01-2015/01/30), during which transmission flows were estimated. Following Section~\ref{sec:part_prob_sources}, we adjusted for heterogeneity in participation probabilities among newly infected individuals using the posterior estimates under model P4, reported in Figure~\ref{fig:participation_rate}.

\subsection{Modelling and estimating sequencing probabilities of recipient cases}\label{sec:seq_prob_recipients}
We are finally interested in the probability that newly infected participants report no ART use, and the probability that newly infected participants reporting no ART use had virus successfully deep-sequenced. As in Section~\ref{sec:reg:trseq}, we combined both steps, and sought to estimate the probability that newly infected participants had virus successfully deep-sequenced.

Based on available data collected among RCCS participants, we could again relax the assumptions of Section~\ref{sec:part_prob_recipients}. Our approach here differed in two aspects from the approach we employed to estimate sequencing probabilities among source cases in Section~\ref{sec:reg:trseq}. First, we predicted who of the infected participants acquired infection during the study period $\mathcal{T}$, as opposed to before the start of $\mathcal{T}$. We used longitudinal individual-level serostatus data and phylogenetic estimates of the infection time for this step \citep{grabowski2017hiv, golubchik2017}. Second, we estimated the probability that infected participants who were classified to have acquired HIV in $\mathcal{T}$ (regardless of ART use, in contrast to Section~\ref{sec:reg:trseq}) also had virus successfully deep-sequenced. 
\begin{table}
\caption{Prediction accuracy of models N1-N3 on hold-out infection time data in 10-fold cross-validation. \label{seq:table:newinfection}}
\begin{threeparttable}
\begin{tabular}{l l l l}
\hline\hline
{\bf Model} &  {\bf Hold-out data} & {\bf MAE}  & {\bf ELPD} \\[0.5ex] 
& {\bf in 95\% PPCrI} & & \\[0.5ex] 
& median (range)  & median (range) &   median (std dev)\\ [0.5ex] 
\hline
N1  & 87.3\%  (85.0\% - 88.9\%)  & 0.1458 (0.1364 - 0.1830)  & -537 (10.1)\\[1mm]
N2  &  97.2\%  (95.6\% - 98.5\%)   & 0.0284 (0.0152 - 0.0467) & -103 (3.61)\\[1mm]
N3 & 97.3\%  (95.6\% - 98.5\%) & 0.0303 (0.0152 - 0.0438)  & -101 (3.56)\\
[1ex]
\hline
\end{tabular}
\begin{tablenotes}
   \item[*] PPCrI is posterior predictive credibility interval, MAE is abbreviated from mean absolute error, ELPD is the abbreviation of expected log predictive density
 \end{tablenotes}
 \end{threeparttable}
\end{table}

We now describe step 1.  Of 5032 infected participants, 883 (17.5\%) had a positive HIV test before the start of $\mathcal{T}$,  65 (1.3\%) had a phylogenetically estimated mean time of infection before $\mathcal{T}$, 1698 (33.8\%) had a phylogenetically estimated time of infection in $\mathcal{T}$. Information on last negative test was not available for this analysis. The remaining 2,386 (47.4\%) had no data or phylogenetic estimates on the likely time of infection. Based on the 2646 individuals with data on infection times, we trained Bernoulli regression models to estimate predictor variables associated with HIV acquisition in $\mathcal{T}$ ($Y_i=1$) versus before $\mathcal{T}$ ($Y_i=0$). The predictive accuracy of several models was compared using 10-fold cross-validation. Using the best model, we predicted the timing of HIV acquisition among the 2,386 infected participants with no such data. We then retained for further analysis the individuals with evidence of HIV acquisition during $\mathcal{T}$, plus those individuals for whom HIV acquisition during $\mathcal{T}$ was predicted. 

\begin{figure}[!t]
    \centering
    \includegraphics[width=0.95\textwidth]{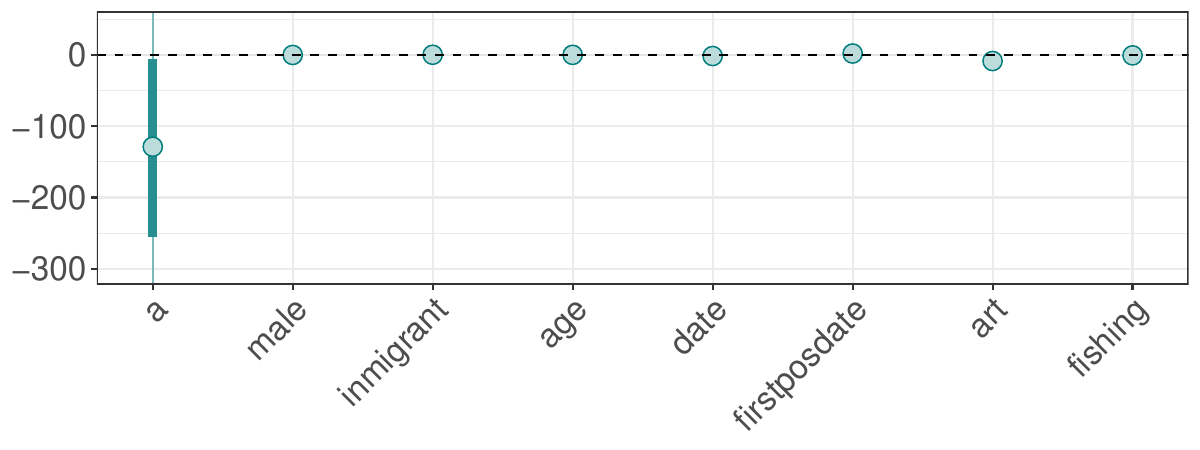}
     \caption{\textbf{Marginal posterior densities of regression coefficients of model N3 to estimate conditional deep-sequencing probabilities.} Circle: median; bar: interquartile range; line: 95\% credibility interval. The first variable $a$ refers to the intercept.}
    \label{fig:sequencing_newinfpred_posci}
\end{figure}

The Bernoulli regression models had the form
\begin{linenomath*}
 \begin{equation}\label{eq:regression_newly_infected}
    \begin{split}
    & Y_i \sim \mbox{Bernoulli}(\eta_i), \: \forall a \\
    & \mbox{logit}(\boldsymbol{\eta}) = \beta_0\boldsymbol{1} + \boldsymbol{\beta X} \\
    & \beta_0 \sim \mathcal{N}(0,100) \\
    & \boldsymbol{\beta} \sim \mathcal{N}(0,10\boldsymbol{I})
    \end{split}
\end{equation}
\end{linenomath*}
where $\eta_i$ is the probability that the $i$th infected participant acquired HIV in $\mathcal{T}$, and the design matrix was based on the covariates gender (indicator variable), first survey visit date (real valued), age of individual at first visit date (real valued), in-migration into the current RCCS community of residence in the two years prior to the first survey visit (indicator variable), date of first HIV positive test (real valued), residence in fishing or inland communities at time of first positive test (indicator variable), and ART use at first visit date when infected (indicator). Three models were considered.
\begin{itemize}
    \item Model N1. Model~\eqref{eq:regression_newly_infected} with the aforementioned covariates except ART status and time of first positive test, leading to $p=5$ predictors. 
\item Model N2. Model~\eqref{eq:regression_newly_infected} with the aforementioned covariates except ART status, leading to $p=6$ predictors. 
\item Model N3. Model~\eqref{eq:regression_newly_infected} with the aforementioned covariates, leading to $p=7$ predictors.
\end{itemize}
The models were fitted in Stan version 2.21, convergence diagnostics \citep{vehtari2019rank} were checked, and effective sample sizes were $\gg 5000$. Table~\ref{seq:table:newinfection} reports as before the proportion of hold data in 95\% PPCrI, the MAE, and the ELPD for each model in 10-fold cross-validations. Model N3 was chosen to model HIV acquisition during the survey period across population strata. Figure \ref{fig:sequencing_newinfpred_posci} shows the marginal posterior densities of the regression coefficients under model N3. Participants without evidence of infection times were classified as having acquired HIV infection in $\mathcal{T}$ if the median marginal posterior estimate of $\eta_i$ was above $0.37$, and were otherwise classified as having acquired HIV infection before $\mathcal{T}$. The threshold was selected by maximising the F1 score for binary outcomes. Table~\ref{seq:table:new_infection_classifications} summarises the number of infected participants classified as having acquired infection before or during the study period 2009/10/01-2015/01/30.

\begin{table}
\caption{\label{seq:table:new_infection_classifications} Demographic characteristics of infected participants by estimated infection time.} 
\centering
\begin{tabulary}{\textwidth}{>{\bfseries}llllll} 
\toprule  
 \textbf{Community} & \textbf{Gender} & \textbf{Age} & \textbf{Infection} & \textbf{Infection} & \textbf{Proportion of}\\[1mm]
 & & & \textbf{before $\mathcal{T}$} & \textbf{in $\mathcal{T}$} & \textbf{infection in $\mathcal{T}$}\\[1mm]
\midrule 
\multirow{14}{.15\textwidth}{Fishing communities} 
  & \multirow{8}{.1\textwidth}{Male} 
  & 15-19 years &   0  (0.0\%)&  15 (1.6\%) & 100\% \\[1mm]
  & & 20-24 years &  6 (3.9\%) &  104 (11.2\%) & 94.5\% \\[1mm]
  & & 25-29 years &  16 (10.5\%)&  256 (27.5\%)& 94.1\% \\[1mm]
  & & 30-34 years &  49 (32.0\%)&  253 (27.2\%)& 83.8\%\\[1mm]
  & & 35-39 years  & 39 (25.5\%)&  162 (17.4\%)& 80.6\%\\[1mm] 
  & & 40-44 years  & 25 (16.3\%)&  116 (12.5\%)& 82.3\%\\[1mm]
  & & 45-50 years  &  18 (11.8\%)&  25 (2.7\%)& 58.\% \\[1mm]
  & & subtotal  & 153 &  931 & 85.9\%\\[1mm]
  \midrule
  & \multirow{8}{.1\textwidth}{Female} 
  & 15-19 years &  7 (2.5\%)&  63 (5.9\%)& 90\% \\[1mm]
  & & 20-24 years & 27 (9.5\%)& 225 (21.2\%) & 89.3\% \\[1mm]
  & & 25-29 years &  50 (17.5\%) & 314 (29.6\%)& 86.3\% \\[1mm]
  & & 30-34 years &  85 (29.8\%)& 236 (22.2\%)& 73.5\%\\[1mm]
  & & 35-39 years  & 71 (24.9\%) & 145 (13.7\%)& 67.1\% \\[1mm] 
  & & 40-44 years  &  35 (12.3\%)& 55 (5.2\%)& 61.1\% \\[1mm]
  & & 45-50 years  & 10 (3.5\%) & 23 (2.2\%) & 69.7\%\\[1mm]
  & & subtotal  & 285 & 1061& 78.8\% \\[1mm]
  
  \midrule
  \midrule
  \multirow{14}{.15\textwidth}{Inland communities} 
  & \multirow{7}{.1\textwidth}{Male} 
    & 15-19 years &  9 (2.2\% &  14 (3\%)& 60.9\%\\[1mm]
  & & 20-24 years &  14 (3.5\%)&  72 (15.5\%)& 83.7\% \\[1mm]
  & & 25-29 years & 34 (8.5\%) &  129 (27.8\%)& 79.1\% \\[1mm]
  & & 30-34 years & 83 (20.8\%) &  99 (21.3\%)& 54.4\% \\[1mm]
  & & 35-39 years  &103 (25.8\%)  & 85 (18.3\%)& 45.2\% \\[1mm] 
  & & 40-44 years  &  81 (20.2\%)&  44 (9.5\%)& 35.2\% \\[1mm]
  & & 45-50 years  &  76 (19\%)&  21 (4.5\%)& 21.6\%\\[1mm]
  & & subtotal  &  400 & 464& 53.7\% \\[1mm]
  \midrule
  & \multirow{7}{.1\textwidth}{Female} 
  & 15-19 years &  19 (2.2\%) &  78 (9.1\%)& 80.4\% \\[1mm]
  & & 20-24 years & 62 (7.2\%) &  220 (25.7\%)& 78.0\%\\[1mm]
  & & 25-29 years & 143 (16.6\%) &  228 (26.7\%)& 61.5\%\\[1mm]
  & & 30-34 years & 203 (23.6\%) &  174 (20.4\%)&  46.2\%\\[1mm]
  & & 35-39 years  & 208 (24.2\%) & 37 (4.3\%) & 15.1\%\\[1mm] 
  & & 40-44 years  & 137 (15.9\%)  &  91 (10.6\%)& 39.9\%\\[1mm]
  & & 45-50 years  & 88 (10.2\%) &  27 (3.2\%)& 23.5\%\\[1mm]
  & & subtotal  & 860 & 855 & 49.85\%\\[1mm]
  \midrule
  & & total & 1698 & 3311 & 66.1\%\\[1mm]  
  \bottomrule
\end{tabulary}
\end{table}

\begin{table}
\caption{\label{table:sequence_recipient_postcheck} Prediction accuracy of models S$^\prime$1-S$^\prime$8 on hold-out deep-sequencing data among patients infected in $\mathcal{T}$ in 10-fold cross-validation.}
\begin{threeparttable}
\begin{tabular}{l l l l}
\hline\hline
{\bf Model} &  {\bf Hold-out data} & {\bf MAE}  & {\bf ELPD} \\[0.5ex] 
& {\bf in 95\% PPCrI} & & \\[0.5ex] 
& median (range)  & median (range) &   median (std dev)\\ [0.5ex] 
\hline
S$^\prime$1  &   98.5\%  (56\% - 100\%) &  0.65  (0.52 - 1.07) & -817.49 ( 14.23 )  \\[1mm]
S$^\prime$2  &   100\%  (98.6\% - 100\%)  &  0.62  (0.48 - 0.72) & -951.26 ( 15.04 )  \\[1mm]
S$^\prime$3 &     100.0\%  (98.1\% - 100.0\%)   & 0.66  (0.50 - 0.73)    & -883 ( 12.3 ) \\[1mm]
S$^\prime$4  &   100.0\% (98.3\% - 100.0\%)  & 0.67  (0.49 - 0.77)  & -917.12 ( 15.12 ) \\[1mm]
S$^\prime$5 &  99.5\%  (99\% - 100\%)  & 0.60  (0.51 - 0.68)  & -930.96 ( 11.93 ) \\[1mm]
S$^\prime$6  & 
100\%  (99\% - 100\%) &    0.61  (0.51 - 0.67) & -980.39 ( 14.95 ) \\[1mm]
S$^\prime$7  &  99.5\%  (97.5\% - 100\%)  &   0.62  (0.49 - 0.68)    & -940.31 ( 11.77 )\\[1mm]
S$^\prime$8  &   100.0\%  (98.5\% - 100.0\%) &   0.62 (0.47 - 0.66)   & -992.35 ( 14.84 ) \\
[1ex]
\hline
\end{tabular}
\begin{tablenotes}
   \item[*] PPCrI is posterior predictive credibility interval, MAE is abbreviated from mean absolute error, ELPD is the abbreviation of expected log predictive density
 \end{tablenotes}
 \end{threeparttable}
\end{table}

\begin{figure}[!t]
    \centering
    \includegraphics[width=0.95\textwidth]{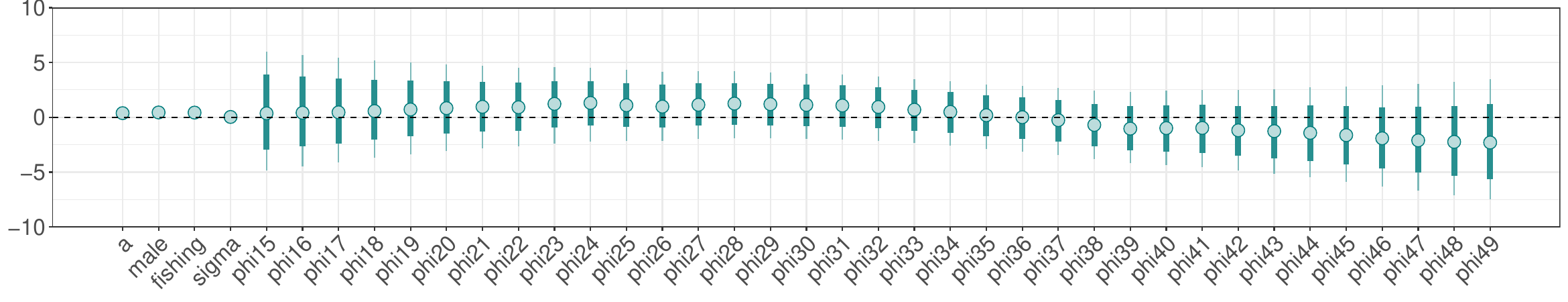}
    \caption{\textbf{Marginal posterior densities of regression coefficients of model S$^\prime$5 to estimate conditional deep-sequencing probabilities among infected participants who were classified to have acquired HIV during the study period.} Circle: median; bar: interquartile range; line 95\% credibility interval. The first variable $a$ refers to the intercept.}
    \label{fig:sequencing_recipient_posci}
\end{figure}

\begin{figure}[!t]
    \centering
    \includegraphics[width=0.75\textwidth]{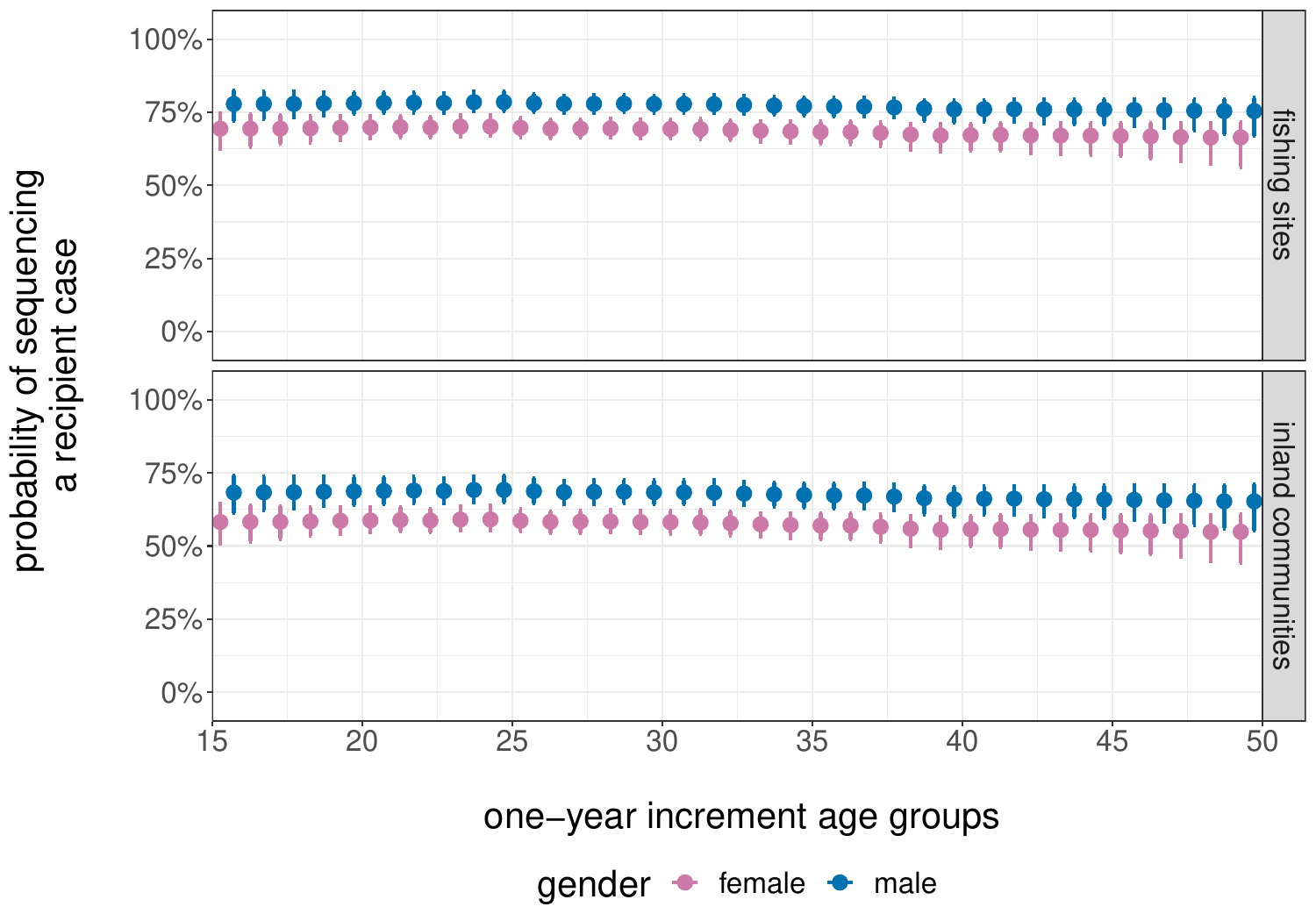}
    \caption{\textbf{Marginal posterior densities of recipient sequencing probabilities under model S$^\prime$5.} Circle: median; line: 95\% credibility interval.}
    \label{fig:sequence_recipient_rate}
\end{figure}

\begin{figure}[!t]
    \centering
    \includegraphics[width=0.85\textwidth]{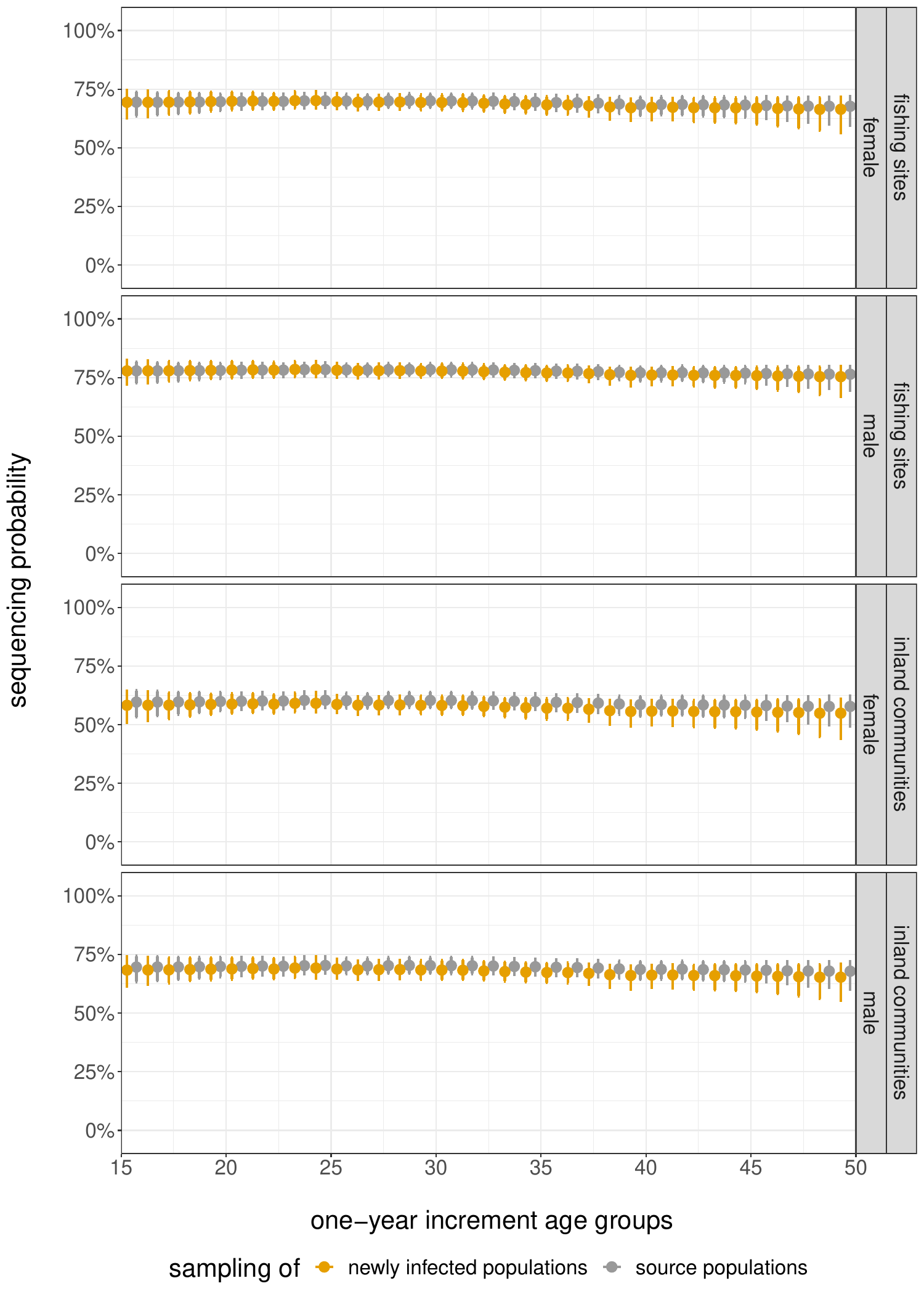}
    \caption{\textbf{Comparison of marginal posterior densities of regression coefficients between model S5 and model S$^\prime$5.} Circle: median; bar: interquartile range; line 95\% credibility interval.}
    \label{fig:sequencing_source_recipient_posci}
\end{figure}

We now describe step 2. To estimate the deep-sequencing probabilities among newly infected participants, we excluded from further consideration the 1698 participants that were classified to have acquired HIV before the start of the study period $\mathcal{T}$, and based estimates on the 3311 participants that were classified to have acquired HIV in $\mathcal{T}$. As in Section~\ref{sec:reg:trseq} we used a Beta-Binomial regression framework to estimate the posterior distribution of deep-sequencing probabilities $\boldsymbol{\xi}^{s^\prime}=(\xi^{s^\prime}_a)_{a\in\mathcal{A}}$, where  $\xi^{s^\prime}_a$ denotes the probability of obtaining a deep-sequence sample from (classified) newly infected participants in population sub-group $a$. We follow the notation of Section~\ref{sec:reg:trseq} and further denote by $N^{\text{study}}_a$ the number of infected participants in $a$ who were classified to have acquired HIV during the study period $\mathcal{T}$, and by $N^{s^\prime}_a$ the number of infected participants in $a$ who were classified to have acquired infection during the study period and had virus deep-sequenced successfully. We considered models of the form
\begin{linenomath*}
\begin{equation}\label{eq:seq_recipient_regression}
    \begin{split}
    & N_a^{s^\prime} \sim \mbox{Beta-Binomial}(N^{\text{study}}_a,\xi^{s^\prime}_a,\gamma), \: \forall a \\
    & \mbox{logit}(\boldsymbol{\xi}^{s^\prime}) = \beta_0\boldsymbol{1} + \boldsymbol{\beta X} \\
    & \beta_0 \sim \mathcal{N}(0,100) \\
    & \boldsymbol{\beta} \sim \mathcal{N}(0,10\boldsymbol{I})\\
    & \gamma \sim \mbox{Exp}(1),
    \end{split}
\end{equation}
\end{linenomath*}
and potentially including regularising ICAR prior densities as in Section~\ref{sec:reg:trseq}. We compared eight models S$^\prime$1-S$^\prime$8 that were analogous to models S1-S8 in Section~\ref{sec:reg:trseq}. The models were fitted in Stan version 2.21, convergence diagnostics \citep{vehtari2019rank} were checked, and effective sample sizes were $\gg 10000$.  Table~\ref{table:sequence_recipient_postcheck} reports the proportion of hold-out data in 95\% PPCrI, the MAE, and ELPD for each model in 10-fold cross-validation. We chose to model deep-sequencing probabilities among newly infected participants across population strata with model S$^\prime$5. Figure \ref{fig:sequencing_recipient_posci} shows the marginal posterior densities of the regression coefficients under model S$^\prime$5, and Figure~\ref{fig:sequence_recipient_rate} shows the resulting deep-sequencing probabilities among recipients. We further compared the estimated deep-sequencing probabilities among recipients (approximated among newly infected participants regardless of ART use) to those among potential source cases (approximated among all infected participants that reported no ART use) in Figure~\ref{fig:sequencing_source_recipient_posci}. Deep-sequence sampling probabilities were estimated to be slightly lower among newly infected populations in older age groups when compared to source populations, but were overall very similar. Note that during the study period immediate start of treatment was not yet recommended by guidelines, and so infected populations were mostly ART naive at time of their first survey date, which explains why sampling probabilities in the two groups were overall similar. 

To summarise, the sampling probability of recipient cases was approximated by
\begin{linenomath*}
\begin{equation}\label{seq:sampling_of_recipients}
    \xi^R_a = \xi^p_a * \xi^{s^\prime}_a
\end{equation}
\end{linenomath*}
for all population sub-groups $a$.

\FloatBarrier

\subsection{Pairwise sampling probabilities of sources and recipients}\label{sec:pw_sampling_probs}
\begin{figure}[!t]
    \centering
     \includegraphics[width=0.85\textwidth]{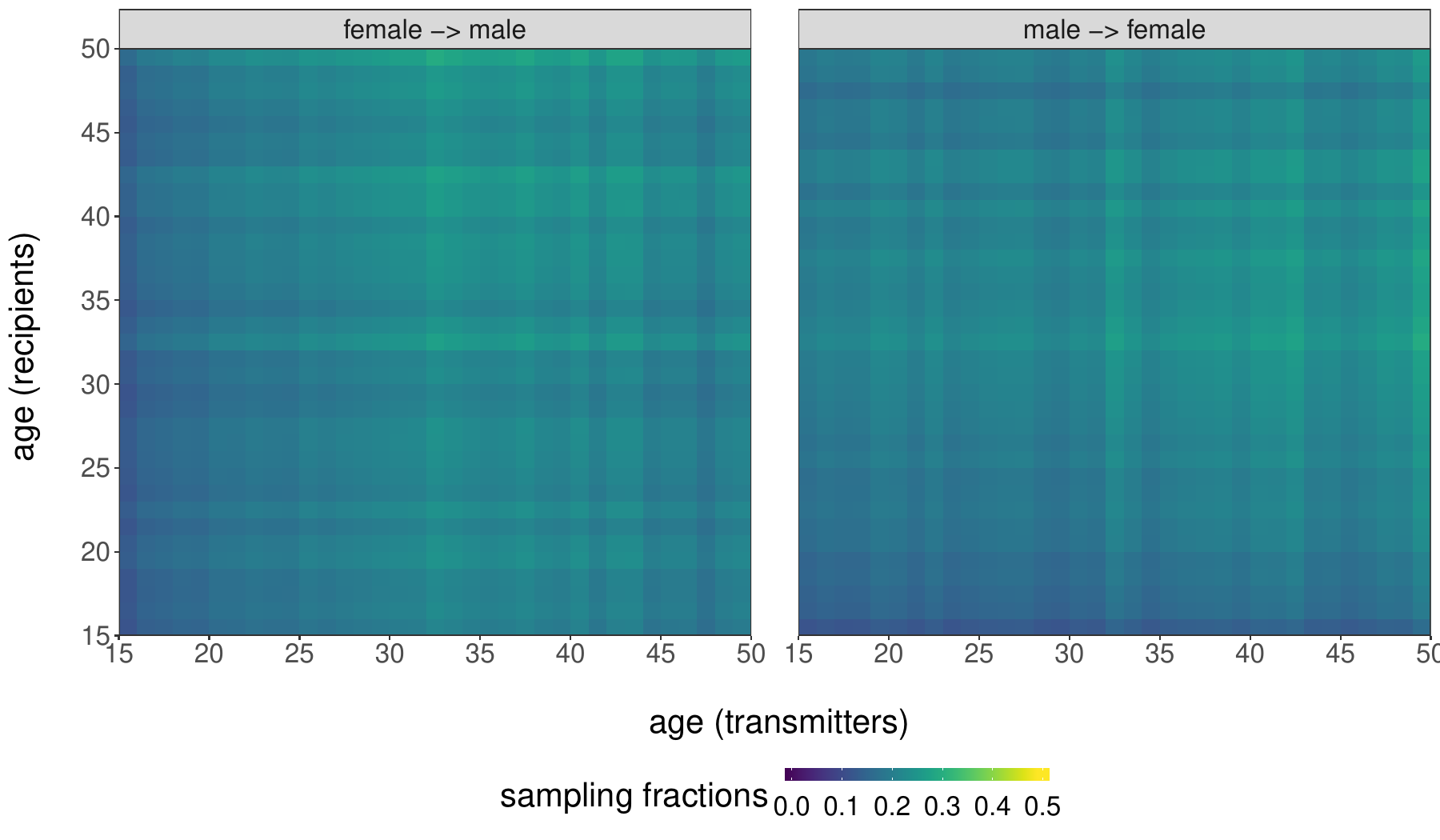}
    \caption{\textbf{Estimated sampling probabilities of transmission events in RCCS inland communities, by 1-year age bands.} Sampling probabilities of transmission events were calculated by multiplying the sampling probabilities of source and recipient populations. Shown are the posterior median estimates of the sampling probabilities by the age of source and recipient populations at the midpoint of the study period. Left panel: female to male transmission. Right panel: male to female transmission.}
    \label{fig:sampling_fraction_ii}
\end{figure}

\begin{figure}[!t]
    \centering
    \includegraphics[width=0.85\textwidth]{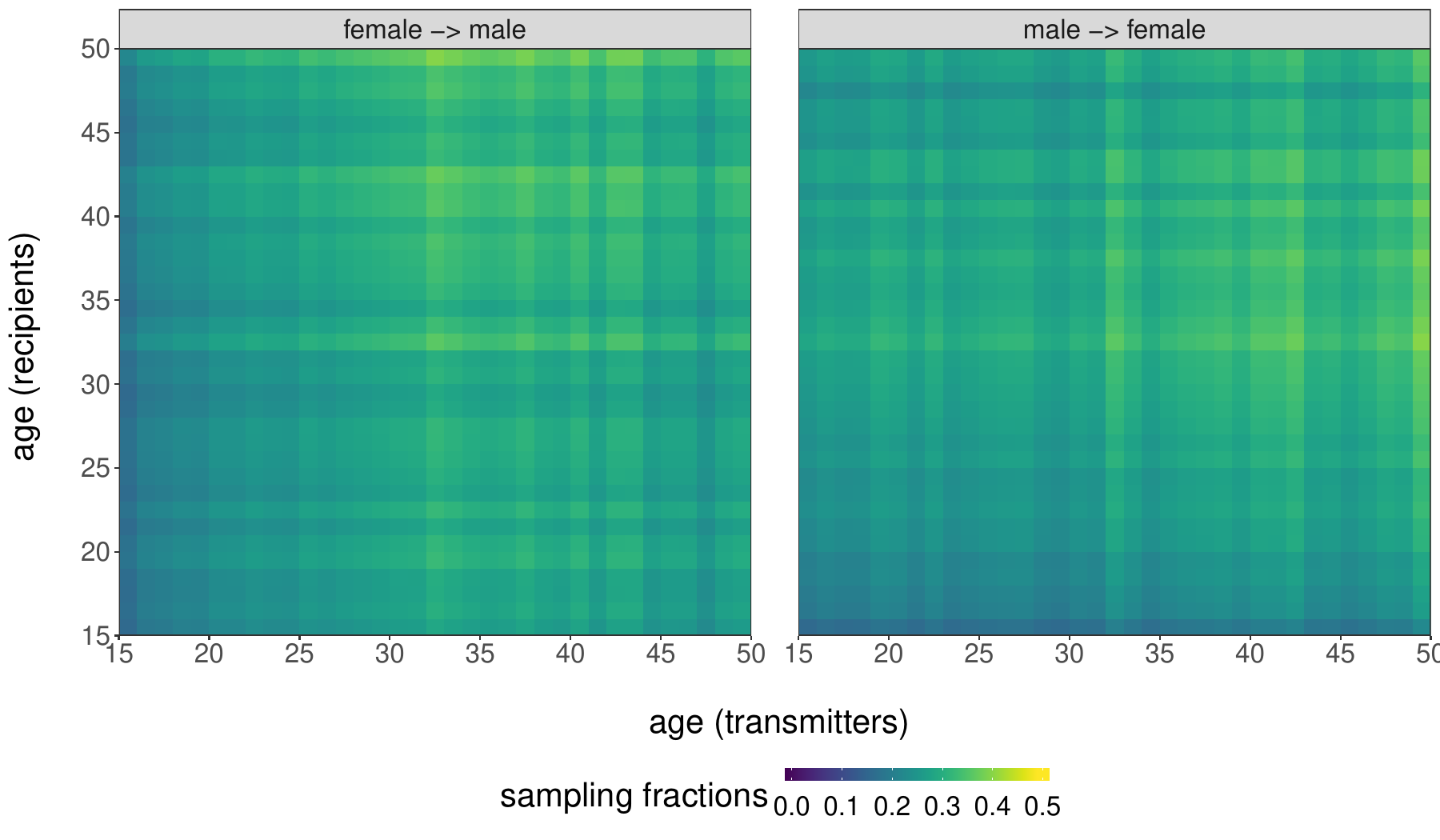}
    \caption{\textbf{Estimated sampling probabilities of transmission events in RCCS fishing communities, by 1-year age bands.} Shown are the posterior median estimates of the sampling probabilities by the age of source and recipient populations at the midpoint of the study period. Left panel: female to male transmission. Right panel: male to female transmission.}
    \label{fig:sampling_fraction_ff}
\end{figure}

We then used the sampling probabilities of source and recipient cases, $\xi^S_a$ and $\xi^R_b$ in \eqref{seq:sampling_of_sources} and \eqref{seq:sampling_of_recipients},
to calculate the sampling probability of transmission events from sub-population $a$ to sub-population $b$ as $\xi^S_a \times \xi^R_b$. Doing so we assumed that source and recipient cases are independently sampled. Figure \ref{fig:sampling_fraction_ii}  illustrates the resulting sampling probabilities of transmission events in inland communities, and Figure \ref{fig:sampling_fraction_ff} illustrates the sampling probabilities of transmission events in fishing communities.

\FloatBarrier

\section{Analyses using different age bands} \label{sec:supp:stratification}

In the main text, we stratified and analyzed data by $1$-year age bands to obtain detailed estimates of age-specific transmission flows. In this section, we present results on analyses using data that is stratified by $2$-year age bands, and $5$-year age bands. For the $5$-year age band analysis, we consider two stratifications, $15-19, 20-24,\dotsc$ and $(12.5-17.5], (17.5-22.5],\dotsc$. We used the same methodological approach as for $1$-year age bands, except that based on the fitted GP model in Equation~\eqref{eq:gp}, we also predicted transmission flows at $1$-year age bands from the model fitted to $2$-year or $5$-year stratified data. To illustrate the impact of using different age stratifications, we here focus on the proportion of transmissions from men that are of same age or younger than female recipients, up to $5$ years older, or $>5$ years older than their female counterparts.

\begin{figure}[!h]
    \centering
    \includegraphics[width=0.8\textwidth]{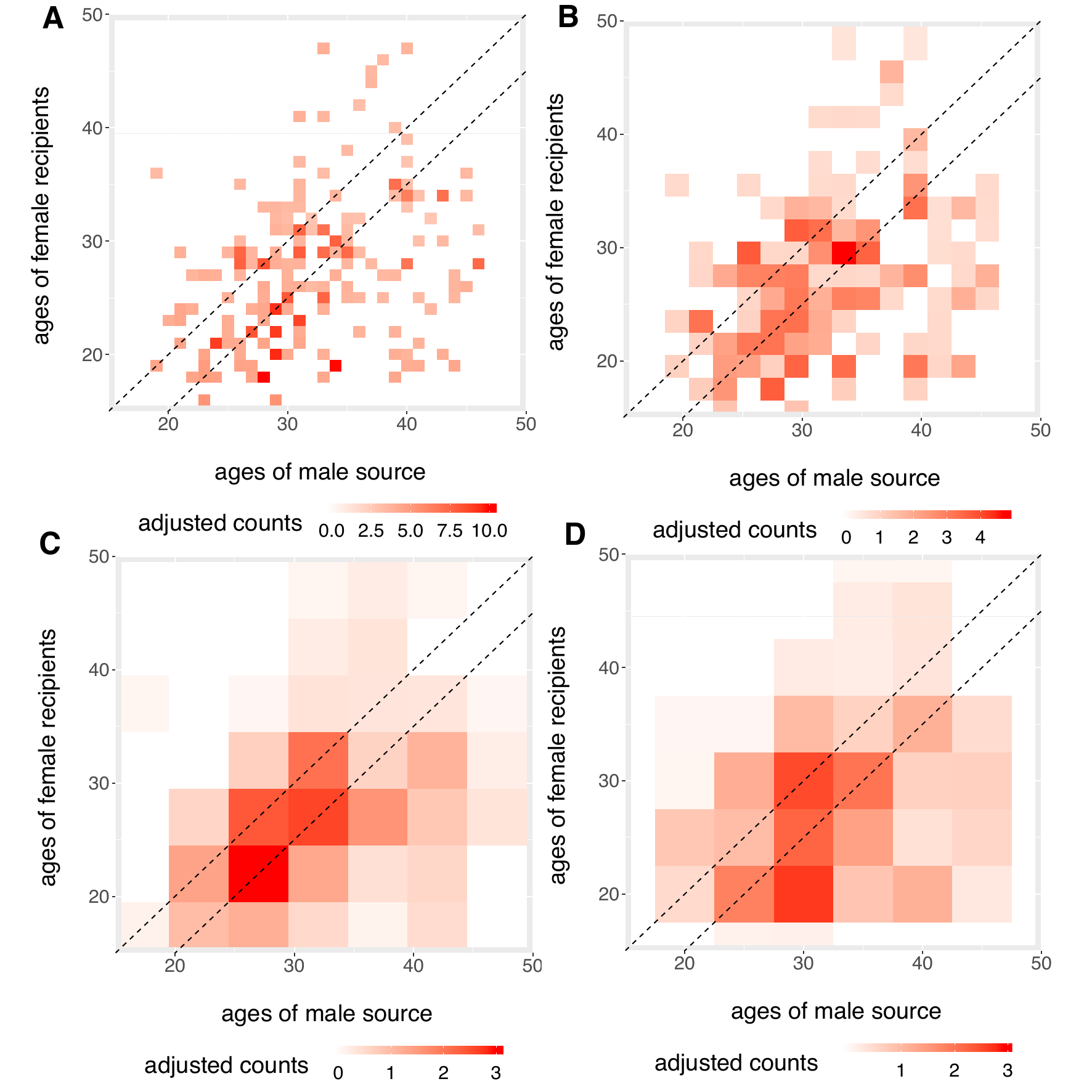}
    \caption{\textbf{Aggregated male-to-female transmission counts.} We compared male-to-female transmission counts (in colour) aggregated by 1-year (Panel A), 2-year (Panel B) and 5-year age bands (Panel C-D). These counts were adjusted by average sampling fractions per stratum. The two black dashed lines are reference lines, representing sources are as same age as recipients (top) and 5 years older than recipients.}
    \label{fig:counts_by_age_band}
\end{figure}

Figures~\ref{fig:counts_by_age_band}A-D show the observed data, the phylogenetically likely source-recipient pairs, respectively by $1$-year, $2$-year and $5$-year age stratifications. To facilitate comparison across the choice of age strata, the observed counts were divided by the average sampling fractions, and show the expected number of phylogenetically likely source-recipient pairs after adjusting for sampling differences. To characterise sources of infections in young and adolescent women, our primary interest is to estimate if male source partners are more than $5$ years older, i.e. for $17$ year old women, interest is in men aged $\geq 23$ or $<23$ and for $24$ year old women interest is in men aged $\geq 29$ or $<29$, and these cut-offs are indicated as a dashed lower diagonal line in Figure~\ref{fig:counts_by_age_band}. Intuitively, it is clear analyses stratified by $5$-year age strata are very coarse to address the primary analytic aims. 

\begin{figure}[!h]
    \centering
    \includegraphics[width=0.8\textwidth]{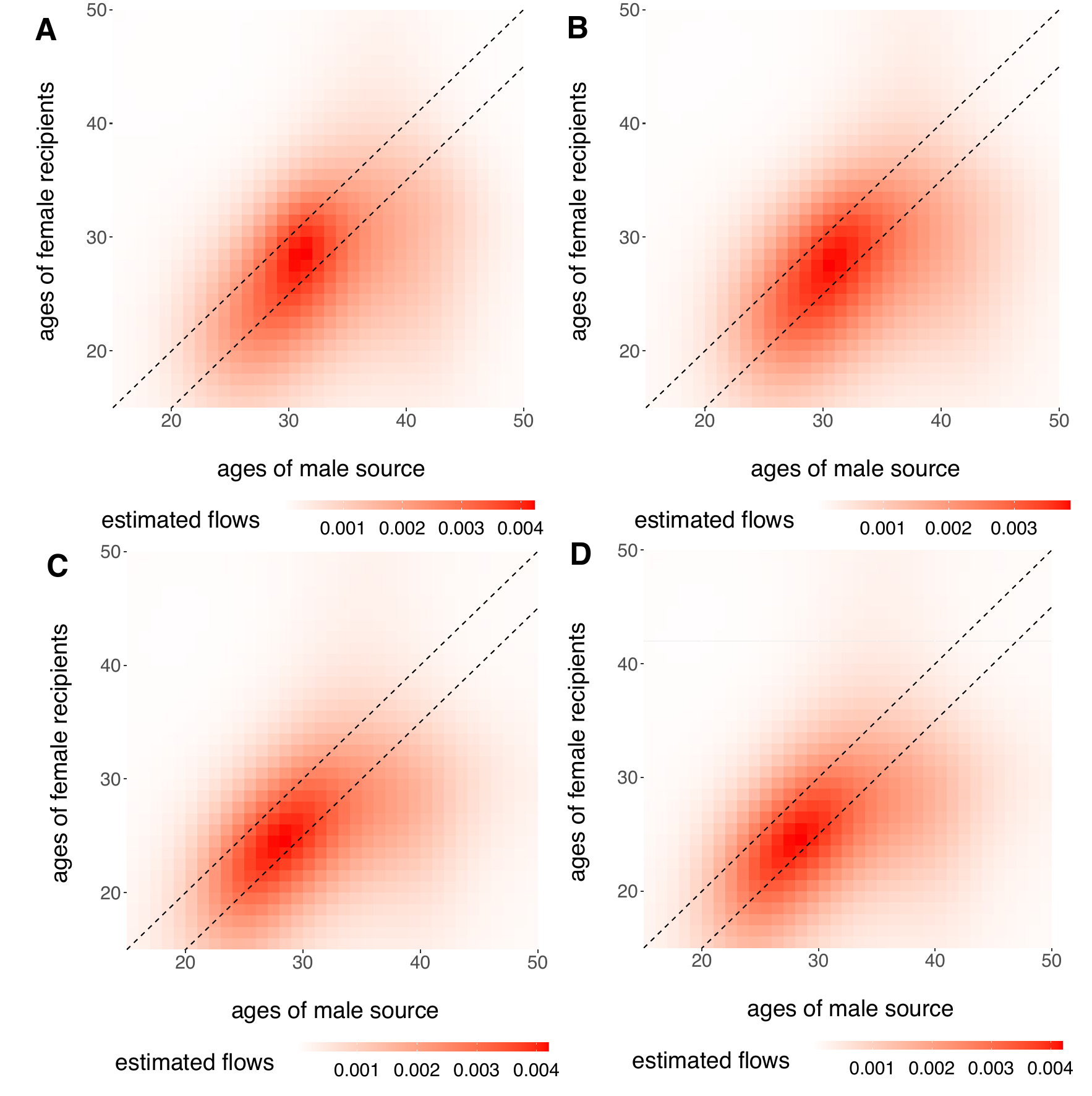}
    \caption{\textbf{Estimated male-to-female transmission flows.} Using the aggregated counts in Figure \ref{fig:counts_by_age_band}, we modelled the transmission intensities using the GP prior densities~\eqref{eq:gp}, and predicted flows by 1-year age bands. Panel A-D show the estimated transmission flows (in colour) from men (x-axis) to women (y-axis) using corresponding data in \ref{fig:counts_by_age_band}. The two black dashed lines are reference lines, representing sources are as same age as recipients (top) and 5 years older than recipients.}
    \label{fig:flows_by_age_band}
\end{figure}

\begin{figure}[!h]
    \centering
    \includegraphics[width=\textwidth]{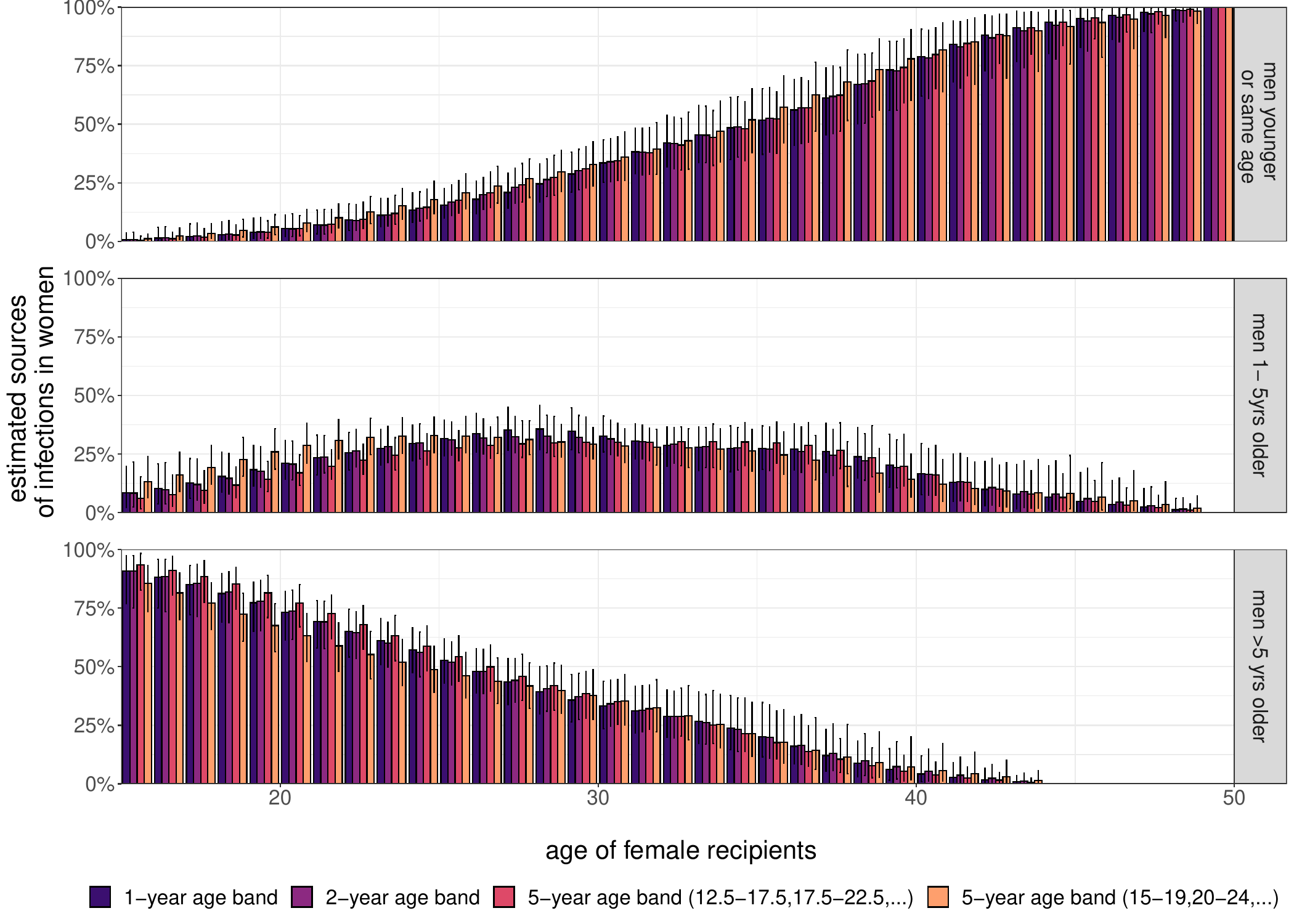}
    \caption{\textbf{Source of infections in women under 4 types of aggregation.} Using the aggregated counts in Figure \ref{fig:counts_by_age_band}, we modelled the transmission intensities using the GP prior densities~\eqref{eq:gp}, predicted flows by 1-year age bands and evaluate the proportion of infections from men in three age categories (younger or of the same age, aged 1-5 years older and aged $>5$ years older). The proportions of infections from 3 age categories (in facet) were shown against ages of female recipients in the bar plots, and estimations were based on 4 types of aggregations (in colour)}
    \label{fig:1yearage_by_age_band}
\end{figure}

Figure~\ref{fig:flows_by_age_band}A illustrates the estimated age-specific transmission flows at $1$-year age stratification (posterior median) from data stratified by $1$-year age bands, and Figures~\ref{fig:flows_by_age_band}B-D illustrate the predicted age-specific transmission flows at $1$-year age stratification (posterior predictive median) from data stratified respectively by $2$-year and $5$-year age bands. The figures show that when using coarser age stratifications, the inferred peak of male to female transmission flows shifts to younger ages, and falls more strongly into the diagonal band indicating transmission from men aged up to $5$ years older. We also find that the predicted transmission flows shift considerably depending on how exactly the start and end points of $5$-year stratifications are chosen. Figure~\ref{fig:1yearage_by_age_band} shows that using $5$-year age stratifications with different start and end points provide significantly different estimates into the sources of infections in young and adolescent women, while estimates using $1$-year and $2$-year age stratifications are very similar. 
 
These analyses suggest that statistical models that are able to borrow information across the exact ages of the individuals in likely transmission pairs are an important technique to estimate age-specific transmission flows, and in particular the sources of transmission in young and adolescent women. Comparing the findings from $1$-year and $2$-year age stratifications suggests further that analyses could have been performed by 2-year age bands at smaller computational cost.

\FloatBarrier

\section{Supplementary Figures and Tables}
\begin{table}
\caption{ Estimated sources of infections in women, Rakai, Uganda, 2009-2015.\label{tab:ageanalysis}}
\centering
\begin{tabular}{l l l l}
\hline\hline
{\bf Infected}    & \multicolumn{3}{c}{{\bf Estimated sources by age}} \\[1mm]
{\bf women} &  {\bf Men younger or }  & {\bf Men 1- 5yrs}  & {\bf Men >5 yrs} \\[1mm] 
&  {\bf same age}  & {\bf older}  & {\bf older} \\[1mm] 
Age & Posterior mean & Posterior mean & Posterior mean \\[1mm]
 & (95\% credibility interval) & (95\% credibility interval) & (95\% credibility interval)\\
[0.5ex]
\hline
15 & 0.6\% (0.1\% - 3.2\% ) & 7.6\% (2.1\% - 20\% ) & 91.8\% (77.6\% - 97.7\% ) \\ [1mm] 
16 & 1.1\% (0.1\% - 4.9\% ) & 9.7\% (3.7\% - 20.5\% ) & 89\% (75.2\% - 95.9\% ) \\ [1mm] 
17 & 1.7\% (0.3\% - 6.4\% ) & 12.3\% (5.9\% - 23.1\% ) & 85.6\% (72.9\% - 93.4\% ) \\ [1mm] 
18 & 2.5\% (0.6\% - 7.6\% ) & 15.6\% (8.8\% - 26.2\% ) & 81.5\% (69.2\% - 89.9\% ) \\ [1mm] 
19 & 3.5\% (1\% - 8.9\% ) & 19.4\% (12\% - 29.3\% ) & 76.7\% (65.1\% - 85.6\% ) \\ [1mm] 
 20 & 4.8\% (1.8\% - 10.5\% ) & 23\% (15.5\% - 32.6\% ) & 71.8\% (60.6\% - 80.7\% ) \\ [1mm] 
21 & 6.6\% (3\% - 12.6\% ) & 26.4\% (18.6\% - 35.3\% ) & 66.8\% (56.1\% - 76\% ) \\ [1mm] 
22 & 8.7\% (4.6\% - 15.2\% ) & 29.1\% (21.2\% - 37.8\% ) & 61.9\% (51.6\% - 71.4\% ) \\ [1mm] 
 23 & 11.3\% (6.4\% - 18.2\% ) & 31.3\% (23.5\% - 40.1\% ) & 57.1\% (47\% - 66.9\% ) \\ [1mm] 
24 & 14\% (8.6\% - 21.1\% ) & 33.2\% (25.4\% - 41.6\% ) & 52.5\% (42.6\% - 62.1\% ) \\ [1mm] 
25 & 17\% (11\% - 24.3\% ) & 34.9\% (27.6\% - 43.1\% ) & 47.8\% (38.2\% - 57.4\% ) \\ [1mm] 
26 & 20.2\% (13.8\% - 28.1\% ) & 36.1\% (28.6\% - 44.7\% ) & 43.4\% (34.2\% - 53.1\% ) \\ [1mm] 
27 & 23.9\% (16.6\% - 32.5\% ) & 36.6\% (28.6\% - 45.9\% ) & 39.2\% (30.3\% - 49.3\% ) \\ [1mm] 
28 & 28.1\% (20.2\% - 37.2\% ) & 35.8\% (28\% - 45.3\% ) & 35.8\% (26.7\% - 45.7\% ) \\ [1mm] 
29 & 32.8\% (24.5\% - 42.1\% ) & 33.8\% (26.6\% - 42.7\% ) & 32.9\% (24.2\% - 43.3\% ) \\ [1mm] 
30 & 37.7\% (28.6\% - 47.6\% ) & 31.2\% (24.3\% - 39.1\% ) & 30.8\% (22.4\% - 40.6\% ) \\ [1mm] 
31 & 42.2\% (32.1\% - 52.7\% ) & 28.6\% (21.2\% - 36\% ) & 28.9\% (20.6\% - 39.1\% ) \\ [1mm] 
32 & 46\% (34.8\% - 57\% ) & 26.6\% (18.8\% - 34.5\% ) & 27.2\% (18.6\% - 38.2\% ) \\ [1mm] 
33 & 48.9\% (36.4\% - 60.3\% ) & 25.6\% (17.5\% - 34.6\% ) & 25.1\% (16.6\% - 37.7\% ) \\ [1mm] 
34 & 51.6\% (36.9\% - 63.6\% ) & 25.3\% (16.9\% - 34.8\% ) & 22.5\% (13.9\% - 35.7\% ) \\ [1mm] 
35 & 54.6\% (39.3\% - 66.9\% ) & 25.5\% (17.1\% - 35.2\% ) & 19.4\% (11.1\% - 32.7\% ) \\ [1mm] 
\hline
\end{tabular}
\end{table}

\FloatBarrier

\end{bibunit}


\end{document}